# L'INFRA-MECANIQUE QUANTIQUE

Mioara Mugur-Schächter

http://www.mugur-schachter.net/ ; http://www.cesef.net/

**Abstract**. A *qualitative* representation of what is called 'microstates' is constructed quite independently from the mathematical formalism of fundamental Quantum Mechanics, by taking into accont exclusively the constraints imposed by *(a)* the cognitive situation in which a human being places himself if he decides to construct knowledge concerning microstates ; *(b)* the general requirements of human conceptualisation. The result – called *infra-(quantum mechanics)* – offers a semantic structure of reference that will permit to develop a unified coherent treatment of all the interpretation problems raised by the formalism of fundamental Quantum Mechanics.

## 1. Sur le processus d'émergence de la mécanique quantique

La façon dont la mécanique quantique s'est constituée comporte un certain caractère qui est unique dans l'histoire des théories physiques : elle a émergé – longuement, entre 1900 et 1935 environ – d'une vraie petite *foule* de contributions d'auteurs différents. Bohr, Plank, Einstein, de Broglie, Schrödinger, Heisenberg, Born, Pauli, von Neumann, Dirac. Et j'en oublie un bon nombre. Or ces contributions (dont quelques unes se sont même constituées d'une façon parallèle, pratiquement sans interactions) – qui toutes ont été essentielles, chacune vraiment originale et de grande envergure créative, mais aussi fortement dissemblables – se sont finalement assemblées dans un tout parfaitement cohérent. Pourtant il serait difficile d'attribuer cette mise en cohérence à une personne déterminée dont on puisse imaginer qu'elle l'a surveillée à l'intérieur de son esprit individuel, par un type de processus que chacun peut imaginer par introspection. Comme, par exemple, on est porté à attribuer à Newton la mise en cohérence des données connues à son époque concernant le mouvement des corps macroscopiques, ou à Maxwell la mise en cohérence des données connues à son époque concernant les phénomènes électriques et magnétiques, etc. Qu'est-ce qui assure donc la cohérence logique de cette foule de contributions qui se sont fondues dans le formalisme mathématique de la mécanique quantique ?

## 2. Une hypothèse

Le phénomène et la question mentionnés suggèrent une hypothèse : ceux qui se sont attelés à la tâche de représenter les microsystèmes et leurs états d'une façon qui puisse être tolérée à la fois par l'essence de la mécanique newtonienne et par la théorie macroscopique des champs électromagnétiques, se sont trouvés confrontés à une situation cognitive qui, à l'époque, était sans doute inusuelle à un point tel que l'effort nécessaire d'innovation dépassait de loin les facultés d'une seule intelligence. Et même les capacités d'un seul génie. Mais d'autre part cette situation cognitive singulière imposait plus ou moins implicitement des restrictions tellement contraignantes que celles-ci ont agi comme un moule commun qui a assuré un grand degré d'unité entre les résultats des différentes approches. C'est la situation cognitive qui a orchestré la construction de la mécanique quantique.

Placée sur un niveau supra individuel, intersubjectif, elle a remplacé d'une manière implicite le contrôle unificateur conceptuel-logique qui d'habitude fonctionne explicitement à l'intérieur d'un seul esprit novateur. Omniprésente d'une manière extérieure et neutre, elle a agi comme un organisateur et un co-ordinateur.

Le formalisme qui s'est constitué ainsi n'exprime pas explicitement la situation cognitive qui l'a déterminé. Toutefois, il a dû sans doute incorporer en état cryptique les contraintes qui l'ont modelé, puisqu'il est performant.

Mais cela est courant. Pour toute théorie mathématique d'un domaine du réel physique, les choses se passent plus ou moins ainsi. L'inhabituel, dans le cas de la mécanique quantique, doit donc consister dans la nature particulièrement nouvelle et contraignante de la situation cognitive impliquée. Celle-ci, après avoir fait obstacle à la conception du formalisme par un seul physicien, et après avoir ensuite orchestré la construction du formalisme par tout un ensemble de physiciens, doit être aussi, par l'*extériorité* dans laquelle elle s'est maintenue face à tout esprit individuel, la raison pour laquelle, jusqu'à ce jour, le formalisme quantique est ressenti comme si peu compréhensible, même par les physiciens et théoriciens qui l'ont longuement pratiqué et y ont réfléchi à fond. Parmi les fondateurs eux-mêmes, il serait difficile de trouver deux qui aient été entièrement d'accord sur les 'significations' incorporées dans le formalisme qu'ils ont contribué à créer.

Aucune autre théorie physique, pas même la relativité d'Einstein, n'a soulevé des débats aussi résistants concernant les significations. Ces débats subsistent et *évoluent* depuis plus de 75 ans, sans trouver des solutions qui soient acceptées de manière unanime. Le formalisme est là, cohérent et puissant. Mais aucun consensus n'a pu s'installer sur sa façon de signifier, ni, *a fortiori*, sur la question de savoir pourquoi cette façon est comme elle est et pas autre.

## 3. Un projet

### *3.1. Formulation du projet*

L'hypothèse formulée suggère à son tour un projet : faire abstraction du formalisme mathématique de la mécanique quantique et s'attacher à construire soi-même une représentation, uniquement qualitative, mais qui mérite clairement d'être appelée une description – et en termes *mécaniques* (position, vitesse, etc.) – de ce qu'on conçoit comme correspondant à l'expression 'états de microsystèmes'. Si l'on se place dans ces conditions primordiales de pénurie sévère, on sera forcé d'utiliser à fond le peu qui reste disponible et agit inévitablement. À savoir la situation cognitive justement. Et bien sûr aussi nos capacités opératoires, les traits qui caractérisent les modes humains de conceptualiser d'une manière *communicable* et *intelligible* (sans quoi ne peut exister aucun consensus intersubjectif, ni, *a fortiori*, 'scientifique'), et le *but* d'élaborer l'essence qualitative d'un formalisme cohérent et performant. En ces conditions, si l'hypothèse d'une détermination des spécificités du formalisme quantique, par les contraintes qu'impose la situation cognitive, est correcte, ce qui émergera devra être l'essence qualitative des algorithmes mathématiques de la mécanique quantique tels qu'on les connaît. Et si cette essence devenait connue, elle offrirait un solide élément de référence pour enfin *comprendre* les choix plus ou moins implicites qui ont conduit à la représentation mathématique des microétats, ainsi que la manière de signifier de cette représentation mathématique.

### *3.2 Nouveauté du projet*

On a énormément discuté la situation cognitive impliquée dans telle ou telle expérience particulière, souvent magistralement et très en détail. Einstein, Bohr, Schrödinger, de Broglie, Bell, Wigner, ont fait des analyses de cas très profondes, dans le lit desquelles se sont ensuite précipité des torrents de gloses, de spécifications, ou simplement de fantaisies. Mais toutes ces analyses font intervenir le formalisme mathématique de la mécanique quantique tout autant que la situation cognitive spécifique du cas considéré. Les problèmes et les analyses considérés ont en général renvoyé même plus aux algorithmes mathématiques impliqués, qu'à la situation cognitive sous-jacente. Ce mélange a piégé l'entendement. Il l'a empêché de s'extraire radicalement du formalisme mathématique et d'en pouvoir percevoir, isolément, les sources, la structure des racines qu'il a implantées dans la factualité physique, et les modes humains d'opérer et de conceptualiser qui ont agi pour structurer ce formalisme. Bref, il semble utile de construire *a posteriori* la genèse qualitative du formalisme mathématique de la mécanique quantique.

Le moment actuel y est peut-être beaucoup plus propice. Dans le passé, ceux pour qui le mot *comprendre* pointait malgré tout vers autre chose que l'application automatique d'un algorithme dont l'efficacité est établie, et qui ont recherché une approche qui puisse doter d'un caractère de nécessité les algorithmes quantiques et leur fonctionnement, sont tous restés isolés, même les plus grands comme Einstein, Schrödinger, et de Broglie.

Ce n'est que dans les années 1980 – exactement à partir du théorème de Bell [1964] concernant la question de 'localité'[1] – que les débats ont pris de la densité. Ce théorème impliquait des *expériences réalisables* qui pouvaient établir si oui ou non les microphénomènes sont 'non locaux', comme on acceptait que l'affirme le formalisme quantique. Tout à coup, tout le monde s'est mis à vouloir comprendre, ouvertement et avec une sorte de désespoir. La tension de bizarrerie introduite par ce qu'on appelait le caractère non local du formalisme quantique était brusquement ressentie comme insupportable. Cette nouvelle phase, imprévue, était comme la tombée d'un masque officiel. Ceux qui avaient soutenu qu'il n'était pas nécessaire de comprendre, dans le fond d'eux-mêmes avaient été convaincus qu'il n'y avait rien de vraiment *important* à comprendre : le formalisme n'est pas local parce qu'il est newtonien, non relativiste. D'accord. On y pense et puis on oublie. Car la réalité physique, elle, est certainement locale comme il se doit selon la relativité d'Einstein. Un point c'est tout. Tout est en ordre. Un formalisme utile à 99% est bon. Il faut être pragmatique. Mais les expériences ont été réalisées et elles ont *confirmé* (dans ce cas, spécifiquement) les prévisions du formalisme quantique 'non local'. Elles les ont confirmé là où l'on ne voulait pas et on ne s'attendait pas à ce qu'elles soient confirmées! Du coup, il devenait urgent de comprendre cette situation. Ainsi le problème de l'interprétation du formalisme s'est finalement trouvé officialisé, institutionnalisé par un succès inattendu du formalisme.

A partir de là a commencé une phase nouvelle de communauté des questionnements et celle-ci a conduit à des bilans globaux qui se poursuivent. Désormais s'est installée une attention plus ou moins générale aux questions d'interprétation. Seule la persistance des questionnements de départ et l'accumulation de questionnements nouveaux ont pu conduire – dans une génération renouvelée de physiciens et penseurs – vers la claire perception d'un problème *global* de la signification du formalisme quantique ; un problème porté et mûri par le cours du temps (Schlosshauer [2003]).

## 4. Une représentation qualitative des microétats construite indépendamment du formalisme quantique

### 4.1. Préalables

Je fais donc abstraction du formalisme quantique : je me trouve devant une table rase où les seuls éléments dont il est permis de faire usage sont les conditions cognitives qui contraignent l'approche, les impératifs introduits par les modes humains généraux de conceptualisation, et par le but de construire des connaissance concernant des microétats[2].

---

[1] Si un événement qui se produit en un point donné d'espace-temps, en un ici-maintenant donné *A*, influence un événement qui se produit en un point d'espace-temps *B* qu'aucun signal lumineux issu de *A* ne peut atteindre, alors on dit qu'il y a 'non localité' au sens de la relativité d'Einstein.

Je procéderai par une suite de questions-réponses. Dans chaque phase j'essaierai de faire apparaître : le but descriptionnel local caractéristique de cette phase ; la façon dont, dans la phase considérée, interviennent les modes humains de conceptualisation ainsi que des éléments historiques qui tissent (par continuité ou par opposition) ce qu'on veut construire de nouveau, à ce qui a été construit avant, sous condition d'*intelligibilité* ; comment, lors de chaque nouvelle avancée, s'entrelacent inextricablement concepts, opérations physiques, données factuelles, et mots et signes d'étiquetage qui assurent la communicabilité ; ce qu'on se donne et ce qu'on obtient qui préalablement était inconnu ; comment émergent progressivement des structures de penser-et-dire liées à des modes de faire ; comment on peut se heurter à un obstacle qui arrête toute progression et qui ne peut être contourné que par une *décision méthodologique* adéquate. Au bout du processus j'examinerai en quel sens ce qui émerge peut être regardé comme une description des microétats, nonobstant le fait que personne ne perçoive des microétats et que, en outre, ce qu'on aura pu produire ne soit *pas* un *modèle*, n'offre aucune image de ce que le terme général 'microétat' désigne, ni, d'autant moins, de la manière d'être du microétat particulier étudié.

Bref, j'essaierai de mettre au jour pas à pas le développement du processus par lequel peut émerger un système cohérent de manières de penser et d'exprimer concernant ce qu'on dénomme 'microétat', faisant unité organique avec un système correspondant de procédures physiques et conceptuelles.

Afin de réussir une communication sans flou, je serai obligée d'introduire quelques notations : des dénominations par symboles au lieu de mots, afin d'éviter des formulations verbales répétées à outrance et trop longues à chaque fois. Mais ces notations n'auront rien de mathématique. Notamment, les dénombrements de fréquences relatives, ainsi les probabilités d'un événement, seront seulement définis et dénotés, sans que l'on *calcule* avec eux. L'exposé restera rigoureusement qualitatif afin de mettre en évidence ce qui, véritablement, s'impose au niveau primordial des structures de faits-et-concepts.

### 4.2. Comment introduire un microétat en tant qu'objet de description ?

#### *4.2.1. Une opération de génération d'un micro-état*

Toute *description* implique une entité-objet[3] définie et des qualifications de cette entité-objet, et elle doit être *communicable*. Les descriptions scientifiques sont en outre soumises à l'exigence de permettre un certain consensus intersubjectif. Le but propre de la mécanique quantique est d'offrir des descriptions *mécaniques* d'*états* de micro-*systèmes*, de micro-états[4] (désormais nous écrirons la succession de ces deux mots sans tiret interne, sauf quand il faudra accentuer la différence entre leurs désignés). Ceux-ci sont des entités hypothétiques

---
[2] Toutefois les notes de bas de page contiendront souvent des références au formalisme mathématique de la mécanique quantique : cela, sans nullement modifier le niveau d'abstraction sur lequel se place l'élaboration entreprise dans le texte principal, et notamment sans altérer son caractère purement qualitatif, guidera la compréhension des physiciens et préparera au face-à-face entre le formalisme quantique et la représentation obtenue ici, qui est sa principale visée.
[3] Dans tout ce qui suit, l'expression 'entité-objet' est une abréviation de l'expression 'entité-(objet d'étude)' où 'objet d'étude' spécifie un *rôle descriptionnel*. (Par conséquent le pluriel s'écrit : entités-objet). Cette spécification est très importante d'un point de vue conceptuel : en général l'objet *d'étude* n'est pas aussi un 'objet' au sens de la pensée et le langage classique, ni au sens de la logique classique fondée sur la pensée classique.
[4] Les caractéristiques stables d'un micro-système – qu'on peut baptiser 'objectités' – sont étudiées par la physique atomique et la physique des particules. La mécanique quantique fondamentale présuppose connues les objectités des micro-systèmes qu'elle considère, et elle s'assigne la tâche spécifique d'étudier leurs états, des microétats, c'est à dire, leurs caractères *instables* qui dépendent de l'environnement. C'est dans ce sens que Dirac emploie 'état' et 'microétat', sens à distinguer de celui plus usuel et diamétralement opposé de *statique*.

qu'aucun humain ne perçoit. Nous voulons expliciter la stratégie descriptionnelle qui a conduit à la description quantique des microétats. La première question qui s'impose est donc :

*Q1 :* « Comment fixer un microétat en tant qu'objet d'étude ? »

Pourquoi ce mot, 'fixer' ? Parce qu'en l'absence de *toute* espèce de stabilité –récurrence ou reproductibilité – on ne peut pas commencer à conceptualiser. Ce qui est dépourvu de toute stabilité peut être étudié aussi, bien sûr, mais en tant que changements subsumés sous quelque chose de durable ou de récurrent ou de répétable que l'on place en position fondamentale. Il s'agit là d'une manière de hiérarchiser qui est inhérente au fonctionnement de l'esprit humain.

Les microétats qui ont la plus grande importance pragmatique sont conçus comme étant liés par des forces d'attraction au sein de microstructures stables, atomes, noyaux, molécules, etc., et là ils sont stables eux aussi. C'est dire que, bien que non perceptibles, ils sont conçus comme étant déjà disponibles pour être étudiés, en état naturel et stable. C'est pourquoi les toutes premières ébauches de pensée 'quantique' ont concerné les microétats liés.

Mais quelle que soit leur importance particulière, les microétats liés n'épuisent pas la totalité des microétats concevables : il convient d'admettre que l'espace fourmille de microsystèmes en états non liés, libres ou progressifs.

La question *Q1* concerne un microétat libre dont la manière d'être nous est totalement inconnue.

Puisqu'un microétat n'est pas perceptible, il est clair qu'on ne peut pas le fixer en tant qu'objet d'étude en le sélectionnant tout simplement dans un ensemble d'entités préexistantes, comme un caillou, par exemple, que l'on ramasserait par terre et que l'on poserait sur la table du laboratoire. Il faut trouver un autre procédé.

D'autre part, il est clair également que si nous – des humains liés par nos corps et nos capacités perceptives à un certain niveau d'organisation de la matière, celui qu'on dit macroscopique – voulons étudier des microétats non perceptibles par nous, alors nous aurons besoin d'appareils enregistreurs, c'est-à-dire, d'objets macroscopiques eux aussi, qui soient aptes à développer *sur eux*, à partir d'interactions supposées avec les microétats présupposés, des marques qui, elles, soient perceptibles par nous. Mais tant qu'on n'a pas introduit un procédé qui puisse fixer un microétat en tant qu'objet de l'étude à faire, il n'y a aucun moyen de savoir quelle marque correspond à quel microétat *:* une médiation introduite exclusivement par un appareil enregistreur ne peut suffire. Cet appareil manifesterait juste une foule de marques (visibles ou audibles, etc.), rien d'autre. Ces marques ne seraient pas des descriptions de microétats, elles seraient dépourvues de signification. Elles ne seraient que des événements connaissables mais tout à fait déconnectés du concept hypothétique de microétat. Or, de par sa définition courante, le concept de description concerne un objet spécifié. Ce problème qui consiste à fixer un microétat en tant qu'objet d'étude, est incontournable et primordial.

Pour amorcer une dynamique de solution, considérons une opération impliquant des objets et des manipulations *macroscopiques* et qui soit telle que, sur la base de certaines données dont nous disposons par la voie historique du développement de la physique, nous puissions imaginer que cette opération engendre un microétat. (Par exemple, on a été conduit à admettre qu'une plaque métallique chauffée suffisamment perd, par agitation thermique intensifiée, des électrons qui circulent librement sur sa surface. Ce procédé semble donc pouvoir être conçu comme engendrant des états libres d'électrons).

Mais on se demande aussitôt : Comment peut-on savoir que l'opération considérée engendre vraiment un tel microétat ? Et tout d'abord, quel sens cela a-t-il de parler de 'microétats' avant de savoir ce que c'est ?

Eh bien, on ne peut pas d'emblée savoir ce que c'est, justement. Mais on peut essayer de *construire* un sens pour cette affirmation de connaissance, sur la base de données préexistantes, de raisonnements et de conventions : il s'agit de construire une stratégie cognitive. On peut commencer cette construction en supposant, sur la base invoquée plus haut, que telle opération macroscopique met en jeu ce qu'on appelle un électron, et en détermine un certain état. D'ailleurs, on n'a pas de choix. Si l'on veut initier une étude des microétats, on doit investir de quelque façon, car de rien on ne peut rien tirer. Et pour cela on doit puiser dans le réservoir dont nous disposons. Or ce réservoir ne contient que des systèmes de connaissances pré-existants avec les représentations qui y sont affirmées, des opérations macrocopiques réalisables à l'aide d'appareils macroscopiques, et les modes humains généraux de conceptualisation. On investit donc du savoir conceptuel pré-constitué, vague et hypothétique – la supposition qu'un microétat déterminé mais inconnu de telle ou telle sorte de microsystème, a été produit par telle opération macroscopique réalisée à l'aide de tels appareils macroscopiques – afin de se donner une base pour tenter de construire à partir d'elle un savoir nouveau, défini et *vérifiable* concernant le microétat inconnu.

### *4.2.2. Étiquetage et communicabilité*

La question *Q1*, toutefois, subsiste toujours : en quel sens peut-on fixer ce microétat hypothétique et inconnu, en tant qu'objet stable d'une étude subséquente ? Le fait d'avoir reconnu qu'il faut créer ce microétat par une opération macroscopique, n'efface pas cette question. L'unique réponse – qui, à la réflexion, s'impose – est frappante : *En étiquetant le microétat à étudier par l'opération qui l'a engendré*.

Mais un tel étiquetage ne peut être utile que si l'opération de génération d'état mise à l'œuvre est *reproductible*. Si elle l'est, et seulement dans ce cas, alors nous pouvons en effet *convenir* de dire qu'à chaque fois qu'on réalise cette opération, elle fait émerger 'le microétat correspondant'. Sur la base de cette convention on peut maintenant étiqueter l'opération de quelque manière, et attacher la même étiquette au microétat correspondant.

Il est en effet strictement nécessaire d'étiqueter, car sinon on ne peut pas communiquer ce qu'on a obtenu, même pas à soi-même. Cependant que ce que nous voulons réaliser, est une procédure entièrement communicable, et en plus consensuelle, conduisant à une description consensuelle d'un microétat. En outre, sans étiquetage, sans notation symbolique, cet exposé écrit non plus ne pourrait continuer sans s'enfoncer dans des vertiges de mots. Convenons donc d'une notation. Par exemple, notons *G* une opération donnée de génération de microétat. Le micro-état correspondant pourra alors être symbolisé $me_G$, ce qui se lit : *le micro-état engendré par l'opération de génération dénotée G*.

L'opération *G* de génération de microétat est supposée être bien définie par la spécification de paramètres opérationnels macroscopiques, mais elle est quelconque (de même que lorsqu'on dit « pensez à un nombre », l'interlocuteur pense à un nombre défini, *3* ou *100*, etc., mais n'importe lequel). Si la nécessité se présente de considérer deux opérations de ce type, l'on pourra, par exemple, introduire les notations *G1* et *G2*.

Mais aussitôt on se demande : « En quoi ces précisions nous avancent-t-elle concernant le problème de fixer un microétat spécifié en tant que l'objet d'une étude subséquente ? Des notations ne sont pas des faits physiques. Ecrire $me_G$ ne peut pas fixer physiquement un microétat en tant qu'objet d'étude ». Or, si, justement. Car l'opération de génération dénotée *G*, elle, est un acte physique, et qui se trouve sous notre contrôle puisque c'est nous qui le concevons et l'accomplissons. Et les associations [(opérations)-(symbolisations)] ont une grande puissance d'organisation dans un processus de conceptualisation. On peut s'en rendre compte le plus clairement précisément lorsqu'on se trouve en situation de pénurie conceptuelle.

Pour y voir clair, procédons par contraste. Imaginons d'abord une entité physique macroscopique qu'il nous est loisible d'examiner directement, par exemple un morceau de tissu. Supposons la tâche de 'fixer l'état' de ce morceau de tissu. Comment s'y prendrait-on ? On l'examinerait et l'on enregistrerait quelque part (dans sa mémoire, sur papier, dans un ordinateur, etc.) des mots ou des notations indiquant les propriétés constatées et qui paraissent suffire pour l'individualiser : genre de substance, dimensions, forme, couleur, degré d'usure, défauts, etc. L'on assignerait aussi un nom à ce morceau de tissu, ou une étiquette, ou les deux à la fois, et ensuite on le rangerait en se disant que désormais il sera possible de le retrouver tel qu'il était, ou bien au pire de le reproduire.

Mais un procédé de ce type n'est évidemment pas applicable à un microétat qui n'est pas directement perceptible. Tout ce dont on dispose pour fixer ce microétat en tant qu'objet d'étude, c'est qu'il a été engendré par l'opération de génération d'état qui a été dénotée $G$ : *c'est le microétat généré par l'opération $G$ et étiqueté $me_G$*. Point. Rien de plus, au départ. Or – et cela est remarquable – déjà cela suffit comme bout de fil. En effet – selon notre rationalité humaine – l'entité étiquetée $me_G$, puisqu'elle a émergé par l'opération physique $G$, a dû émerger imprégnée de certaines marques physiques relatives à cette opération. Des marques non connues, mais dont on conçoit qu'elles ont singularisé cette entité factuellement à l'intérieur du continuum du réel physique. Désormais cette entité 'existe' donc, en ce sens qu'elle a cessé de se fondre dans le reste du réel physique. Elle existe d'une façon spécifique qui porte le sceau de l'opération de génération $G$.

En outre puisque l'opération dénotée $G$ est reproductible, le microétat dénoté $me_G$ est désormais 'capturé', en *ce* sens qu'on peut désormais le re-produire lui aussi autant de fois qu'on voudra.

Les traits minimaux de stabilité et de communicabilité sans lesquels on ne peut pas démarrer un processus de conceptualisation publique, viennent d'être acquis. Une association [(opération physique)-(symbolisations)] a permis de franchir ce premier pas. Bref : On peut fixer un microétat en tant qu'objet d'étude en le *créant* physiquement à l'aide d'une opération de génération reproductible, et en l'étiquetant par l'étiquette associée à cette opération.

### *4.2.3. Une décision méthodologique inévitable*

Mais à nouveau on doute. On se dit : « Si l'on veut aboutir à décrire des microétats il faut en effet partir d'une opération macroscopique accomplie à l'aide d'un appareil macroscopique, car nous sommes rivés au niveau macroscopique pour initier une action dans le réel microscopique ; il faut en effet poser l'hypothèse que ce qui résulte de l'opération accomplie est ce qu'on appelle un microétat, sinon on ne peut pas commencer la construction d'une démarche qui mérite le nom d'étude des microétats : un échantillon de réel purement factuel ne peut être hissé dans le réseau de conceptualisation qu'en l'incluant *a priori* dans un réceptacle conceptuel. En outre il faut aussi se référer à ce microétat hypothétique en l'appelant *le microétat qui correspond à l'opération de génération $G$* et en le dénotant de quelque façon qui inclue l'étiquette $G$, car on ne dispose d'aucune autre caractérisation, cependant que, pour communiquer, il faut bien nommer. Tout ceci paraît en effet nécessaire. Mais qu'est ce qui prouve que lorsqu'on effectue ce qu'on appelle reproduire l'opération de génération étiquetée $G$ c'est vraiment la même opération qui se réalise ? Et même s'il en était ainsi, qu'est-ce qui prouve en outre qu'à chaque fois qu'on réalise $G$, c'est le même microétat qui émerge ? ».

Eh bien, rien ne prouve ces deux 'mêmetés', ces deux invariances, celle du désigné de $G$ et celle, corrélative, du désigné de $me_G$. D'autre part elles sont cruciales. Car comment pourrait-on étudier un 'microétat' qui pour l'instant n'est que juste un nom d'un concept posé par nous à l'avance, dont le désigné n'est pas perceptible, et dont on ne saurait même pas si l'on arrive ou non à le re-créer ? Cela est au-delà des limites du concevable. Et il est évident que vouloir d'abord savoir si chaque réalisation de l'opération de génération $G$ aboutit ou pas à re-produire le même microétat, afin de décider sur la base de ce savoir si oui ou non on peut 'logiquement' se lancer à commencer de construire des connaissances concernant les microétats, c'est se laisser piéger dans un cercle vicieux et renoncer au but initial.

La pensée scientifique classique nous a profondément habitués à poser l'existence simultanée d'entités matérielles ou conceptuelles préconstituées, à partir desquelles, par des appositions de symboles (notamment verbaux) et des mises en successivités purement discursives appropriées, on peut assurer de la *déductibilité*. Mais ici il est évident qu'il ne s'agit pas de déduire et qu'on ne dispose pas d'un réservoir de simultanéités préconstituées, déjà disponibles. Nous sommes encore dans une phase radicalement initiale de pure construction, une construction primordiale, à la fois physique et conceptuelle, au cours de laquelle seul l'ordre approprié des actions cognitives peut engendrer les éléments qui, successivement, s'imposent comme strictement nécessaires pour assurer la possibilité d'une progression descriptionnelle. Or dans le cas qui nous occupe, l'arrêt de la constructibilité ne peut être évité que par ce mortier abstrait qu'est la *décision méthodologique* suivante.

> *Ce* qui émerge lors d'une réalisation quelconque de l'opération de génération telle qu'elle est spécifiée à l'aide de paramètres macroscopiques – les seuls dont nous disposions – sera dénommé par définition 'le microétat $me_G$ correspondant à $G$ ', quel qu'en soit le contenu factuel non connaissable. Et l'on posera que la relation entre '$G$' et '$me_G$' est une relation *de un-à-un*. Nous dénoterons cette relation par $G \leftrightarrow me_G$.

Sans cette décision méthodologique il n'est tout simplement pas possible d'assigner une définition claire à ce qu'on veut désigner comme 'le microétat $me_G$ correspondant à $G$ '. L'on ne pourrait donc pas s'y appuyer pour construire. Si l'on *veut* tenter de développer une démarche qui permette d'aboutir à des connaissances consensuelles concernant des 'microétats', alors il n'y a pas de choix : on est obligé de commencer par *admettre* pour les désignés des symboles $G$ et $me_G$, les 'mêmetés' corrélées posées plus haut, sans s'immobiliser dans la question de savoir si elles sont 'vraies' ou non. Car *on manque de tout accès direct de contrôle de la vérité de ces 'mêmetés'*. Cependant que, dans la phase actuelle de la progression amorcée, rien n'interdit cette décision, et elle seule peut permettre d'avancer. Nous verrons bien ce qui en découle. Plus tard, enrichis de la construction accomplie, il deviendra possible de discerner des conséquences de la décision méthodologique posée à sa base. Nous serons alors armés pour conclure de manière finale si nous devons revenir en arrière et abandonner cette décision, avec la construction fondée sur elle, ou si au contraire nous pouvons adhérer définitivement à sa pertinence. *Pas à sa vérité, mais à sa pertinence méthodologique*.

Pour l'instant nous construirons donc sur la base de la décision méthodologique qui vient d'être posée.

Enregistrons et soulignons cette première entrée en scène spectaculaire de la nécessité inévitable de décisions méthodologiques au cours d'un processus explicite de conceptualisation.

*4.2.4. Une catégorie particulière d'opérations de génération d'un microétat :
les opérations de 'génération composée'*

Il existe une catégorie particulière d'opérations de génération de microétats qui est liée à un fait d'expérience bien connu, celui que l'on désigne par l'expression 'interférence corpusculaire' (interférence avec des microsystèmes lourds, i.e. à masse non nulle).

Soit un microsystème lourd d'un type donné, disons un électron. Dès qu'on a spécifié une opération de génération $G1$ qui engendre pour un tel microsystème un microétat correspondant $me_{G1}$, et également une autre opération de génération $G2$ qui lui engendre un autre microétat correspondant $me_{G2}$, on peut spécifier pour ce type de microsystème, une opération de génération *composée* qui combine $G1$ et $G2$ et engendre un microétat tel que, en un certain sens détaillé plus bas, on peut le considérer comme étant un état d' "interférence" des microétats $me_{G1}$ et $me_{G2}$.

Historiquement, l'étude des états de microsystèmes à masse non-nulle a été abordée en connaissant déjà, d'une part le comportement des mobiles lourds tel que celui-ci est décrit par la mécanique newtonienne, et d'autre part le comportement macroscopique des ondes (sans masse), tel que celui-ci est décrit par l'électromagnétisme classique. Or – lorsqu'on les découvre avec ces acquis historique dans l'esprit – les microétats d'interférence corpusculaire frappent l'attention. Car dans ce cas, face aux repères constitués par ces acquis, on se trouve devant un système surprenant de similitudes et de différences quant aux manifestations observables. En effet, avec des ondes, une configuration – étendue – d' "interférence" se forme *d'un seul coup*, pas comme cela se passe avec des microétats, au fur et à mesure, *via* des impacts localisés qui sont observables individuellement. Cependant que selon la mécanique classique, les impacts isolés que l'on obtient par les réitérations d'un même état d'un mobile donné (un 'corpuscule', si ce mobile est très petit), devraient ne pouvoir jamais engendrer une configuration finale étendue ayant la même forme qu'une figure d'interférence obtenue avec une onde[5].

La possibilité des phénomènes d'interférence corpusculaire apparaît donc comme une spécificité surprenante de certains états de microsystèmes lourds. Cette spécificité a joué un rôle central dans la formulation mathématique de la mécanique quantique. Ici cette formulation mathématique est ignorée. Toutefois l'importance des phénomènes d'interférence corpusculaire reste majeure, également, face aux buts propres de la construction tentée ici. Car ces phénomènes comportent des implications concernant les opérations de génération de microétats, qu'il est vital de formuler explicitement. Arrêtons-nous donc un instant sur les phénomènes d'interférence corpusculaire. On peut le faire le plus simplement à l'aide d'un compte rendu de la célèbre 'expérience des trous d'Young', fait dans les termes que nous avons construits jusqu'ici.

\* Afin de fixer les idées supposons qu'on travaille avec des électrons. On utilise un écran opaque. En aval de cet écran on place une plaque couverte d'une substance sensible (où se forme une marque observable lorsque la surface de la plaque est atteinte[6] par un microsystème lourd). En amont de l'écran opaque on produit, par quelque opération de génération préliminaire $G_p$, un microétat préliminaire d'électron, et ensuite on laisse passer du temps. On constate alors que jamais, à la suite de la génération du microétat préliminaire, il n'apparaît une

---

[5] Les 'mobiles' au sens de la mécanique classique, lorsqu'ils rencontrent un obstacle partiel, produisent sur un écran placé derrière les obstacles des impacts distribués de manière à constituer une figure de "diffusion" (la définition peut être truvée dans les manuels de mécanique classique), pas une figure d' "interférence".

[6] Conformément à la note technique de la fin de l'introduction générale, les expressions verbales qui suggèrent des images importées de la pensée et le langage courants sont écrites sous la forme '…..' où les points indiquent l'expression verbale. Cette précaution permet de maintenir ces expressions sous contrôle, et finalement les éliminer.

marque observable sur la plaque sensible placée en aval de l'écran opaque. On exprime ce fait en disant que le microétat préliminaire ne traverse pas l'écran opaque.

\* On perce maintenant l'écran opaque d'un trou. Dénotons-le par le chiffre *1*. Dans ces nouvelles conditions on constate que, à la suite de chaque acte préliminaire de génération $G_p$ accompli en amont de l'écran opaque, si les paramètres sont convenables il se produit systématiquement un impact observable sur la plaque sensible placée en aval de l'écran opaque. On exprime ce fait en disant que le microétat préliminaire correspondant à $G_p$ passe par le trou *1*, devenant de ce fait un *nouveau* microétat. Si l'on élimine toute image suggérée par cette façon de dire, on reste avec l'affirmation que, dans le contexte expérimental considéré, l'existence du trou *1* agit comme une nouvelle opération de génération *G1* qui, à partir de l'état préliminaire produit par $G_p$, crée en aval de l'écran troué un microétat correspondant $me_{G1}$.

En répétant la même procédure un grand nombre de fois, on obtient sur la plaque sensible une certaine configuration étendue *c(1)* de marques observables – pas *une* seule même marque réitérée – qui est *compatible* avec la figure de *diffusion* que la mécanique newtonienne affirme, dans des conditions similaires, pour un corpuscule au sens classique.

\* Appliquons la même procédure décrite ci-dessus, mais en utilisant un écran opaque percé d'un trou *2* qui est placé à un *autre* endroit que celui où était placé le trou *1*. On observe alors de nouveau, à la suite de chaque opération préliminaire de génération $G_p$, une marque sur la plaque sensible placée en aval de l'écran. On exprime ce fait en disant que cette fois le microétat préliminaire passe par le trou *2* ; ce qui, en faisant abstraction de toute image suggérée par le langage employé, revient à considérer que, dans ce nouveau contexte expérimental, l'existence du trou *2* agit comme une opération de génération *G2* qui, à partir de l'état préliminaire $G_p$, produit en aval de l'écran troué un microétat correspondant $me_{G2}$, différent de $me_{G1}$.

En répétant $G_p$ un grand nombre de fois, on obtient cette fois sur la plaque sensible une nouvelle configuration étendue de marques observables, *c(2)*, consistant dans une figure de *diffusion* compatible avec la mécanique newtonienne et qui est déplacée face à celle obtenue avec le trou *1*.

\* On utilise maintenant un seul écran opaque percé de *deux* trous, *1* et *2*. En amont de cet écran opaque on répète un grand nombre de fois l'opération préliminaire de génération $G_p$. Dans ces conditions nouvelles on constate qu'à la suite de chaque réalisation de $G_p$ l'on obtient sur la plaque sensible placée en aval de l'écran troué *une* seule marque observable. Et – à condition que les trous *1* et *2* soient assez rapprochés – la configuration finale de l'ensemble des marques observables, *diffère* cette fois de la figure que l'on devrait obtenir selon la mécanique classique, à savoir une figure ayant le même aspect qu'une juxtaposition ('addition') des deux configurations de diffusion *c(1)* et *c(2)* mais où l'on ne conterait qu'une seule fois deux marques produites à un même endroit : la configuration finale inattendue que l'on obtient – dénotons-la *c(1+2)* – reproduit, par des impacts *individuels* et *successifs*, l'aspect de la figure d'*interférence* que, avec une onde, l'on obtiendrait d'un seul coup.

On peut exprimer cette situation par la représentation qui suit.

*(a)* On considère que lors de chaque réalisation de la procédure impliquée, l'existence simultanée des deux trous *1* et *2* agit comme un paramètre qui définit *une* seule nouvelle opération de génération où les deux opérations précédentes *G1* et *G2* 'se composent'. Dénotons cette nouvelle opération de génération par *G(G1,G2)* et baptisons-la *opération de génération composée*. Notons $me_{G(G1,G2)}$ le microétat correspondant généré en aval de l'écran opaque et nommons-le *microétat correspondant à une opération de génération composée*.

*(b)* Convenons de dire que, dans le microétat à génération composée $me_{G(G1,G2)}$, les deux microétats $me_{G1}$ et $me_{G2}$ que produi*raient*, respectivement, l'opération *G1* seule et l'opération *G2* seule, 'interfèrent' : l'introduction de cette façon de dire permet de faire une référence verbale aux figures d'interférence ondulatoire étudiés dans la théorie macroscopique des ondes électromagnétiques, mais qui apparaissent *d'emblée*, pas par l'apparition progressive d'impacts successifs observables individuellement.

On vient de voir comment s'est forgé dans ce cas un système de représentation comportant : des opérations physiques *effectuées* ($G_p$) par l'expérimentateur ou bien seulement *conçues* (*G1*, *G2*, *G(G1,G2)*) ; des objets et procédures macroscopiques (ceux qui interviennent dans l'expérience d'Young décrite plus haut) ; des faits observables (les marques individuelles sur la plaque sensible et les configurations globales constituées de ces marques) ; des façons de dénoter et de dire où interviennent des références à des conceptualisations accomplies auparavant ; et enfin, certains prolongements commodes de la pensée et du langage courants ('le microétat *traverse* l'écran', etc.), mais dont par la suite on peut faire abstraction.

On peut concevoir un nombre illimité de microétats distincts correspondants à des compositions distinctes d'une seule et même paire d'opérations de génération de départ, *G1* et *G2*. En effet on peut complexifier la manière de composer *G1* et *G2* : en posant un 'filtre' sur le trou *1* on peut affaiblir l'intensité de ce qui passe par le trou *1* relativement à ce qui passe par le trou *2*, quoi que cela puisse vouloir dire (car en fait on n'en sait strictement rien) ; ou *vice versa*. Il existe également des procédés pour 'retarder la progression de ce qui passe par l'un des trous, face à la progression de ce qui passe par l'autre trou'. L'ensemble des opérations que l'on exprime en termes de tels affaiblissements ou retardements relatifs, engendre une nombre illimité de micro-états distincts, mais qui tous, chacun *via* une opération composée correspondante *G(G1,G2)*, sont liés à un même couple de deux opérations élémentaires de génération de départ, *G1* et *G2*.

En outre, on pense que les constatations formulées plus haut pour le cas de deux opérations de génération *G1* et *G2*, peuvent être généralisées à tout nombre fini d'opérations de génération (de trous). En tout cas à ce jour on n'a pas signalé des restrictions concernant ce nombre. Acceptons donc l'hypothèse de la possibilité d'opérations de génération composées *G(G1,G2,…Gn)* où *n* est quelconque et appelons cette hypothèse *le principe de composabilité des opérations de génération de microétats*. Etant donnée une opération de génération *G(G1,G2,…Gn)* composée bien définie, le microétat à génération composée correspondant sera dénoté $me_{G(G1,G2,…Gn)}$.

### 4.2.5. Mutation du concept de définition d'une entité-objet-d'étude : indépendance de toute qualification

Notons que d'ores et déjà la dynamique de construction qui s'est imposée implique un caractère très inhabituel. Le produit de l'opération de génération *G* émerge en tant que morceau – imaginé – de pure factualité physique, conçu comme portant des spécificités physiques relatives à *G* mais qui sont tout à fait inconnues. Car dire que ce qu'on étiquette $me_G$ est le microétat engendré par *G*, ne renseigne nullement sur la façon d'être spécifique de l'entité (hypothétique) particulière elle-même qui – parmi toutes les entités subsumées sous le concept général de microétat – est la seule qui est étiquetée $me_G$. Dire cela c'est dire exclusivement comment l'entité dénotée '$me_G$' a été produite, pas comment elle est. De même que les expressions à sonorité modélisante (comme 'le microétat $me_G$ *passe* par tel trou', etc.) ne disent strictement rien concernant le comportement du microétat dénoté $me_G$, et en outre, on vient de le voir, elles peuvent être éliminées du langage final qui s'est constitué. En ce sens, et en ce sens seulement, aucun *modèle* défini n'est affirmé concernant le microétat particulier engendré.

Il reste toutefois que l'entité physique dont il s'agit, 'le microétat $me_G$ engendré par $G$', *est attrapé dans un réseau de conceptualisation préexistante*, comme un poisson dans un filet. Sans cela il ne serait tout simplement pas possible de le hisser et l'insérer dans le volume du conceptualisé et communicable : il *faut* parler, il *faut* écrire, et avec des mots qui *existent* et que chacun peut *comprendre*.

Il est important de noter que la servitude face à la conceptualisation humaine spontanée et au mode correspondant de parler et penser, est inévitable : ceux-ci constituent la structure de départ des élaborations spécialisées. On peut s'en éloigner indéfiniment, mais le long de cheminements qui y sont génétiquement connectés. *A posteriori*, dans le produit final, on peut occulter les traces de sa genèse ainsi que celles de la structure de penser et parler de départ, comme on le fait plus ou moins dans les langages techniques, et de manière radicale dans les langages codés. Mais le degré auquel une telle occultation est souhaitable est limité par le degré de communicabilité intelligible que l'on souhaite assurer (dans le cas de l'exposé présent l'intelligibilité aussi immédiate que possible est l'un des buts majeurs). Toutefois l'utilisation d'un réseau de concepts-et-mots ('microsystème', 'microétat', 'génération de microétat', etc.) qui relie à la conceptualisation courante, ne dit pas plus comment sont et se comportent les entités physiques désignées, que la structure du filet utilisé par le pêcheur ne dit comment est le poisson qui y a été attrapé, ou comment il bouge.

Le fragment de conceptualisation accompli, a donc contourné l'impossibilité, au départ, d'utiliser des prédicats conceptuels pour définir un microétat (comme, pour définir une chaise particulière ou un certain couteau, on dirait respectivement «la chaise marron qui se trouve.....etc. » ou « le couteau très coupant, à poignée rouge...etc. »), ou de montrer du doigt, ou de *pointer vers*... à l'aide de contextes verbaux. Il a contourné le fait que rien, aucun moyen n'est disponible pour que le microétat à étudier, étiqueté '$me_G$', soit individualisé au sein de la catégorie générale dénommée 'microétat' ; qu'il y soit individualisé d'une manière qui s'applique directement à lui et qui renseigne sur lui, *spécifiquement*. Malgré cette totale pénurie de façons usuelles de définir, nous avons produit une 'définition' pour le microétat à étudier : une définition a-conceptuelle, non-qualifiante à proprement parler.

Ce qui importe dans cette réussite est que dès que cette sorte de définition est acquise, elle permet de continuer le processus de construction d'une description, cependant que sans elle ce processus resterait bloqué.

Bien sûr, il existe d'innombrables autres entités-objet que l'on ne définit qu'opérationnellement. Par exemple, une robe est elle aussi un objet que (souvent) on ne définit que par une suite d'opérations. Mais pendant qu'on fabrique une robe, on perçoit tous les substrats sur lesquels s'appliquent nos actions de génération, ainsi que les processus que ces actions produisent. Et quand la confection est terminée, on voit la robe, on la porte, etc. Tandis que la définition d'un microétat par l'opération qui le produit descend à l'intérieur d'un domaine de factualité physique qui est de l'inconnu total. On n'y peut disposer d'aucun autre connecteur au domaine du connu qu'un réceptacle conceptuel préfabriqué – toute une classe nommée microétat – que l'on immerge mentalement dans ce domaine d'inconnu pour y accueillir comme dans un ascenseur conceptuel le microétat particulier étiqueté $me_G$, et le hisser dans le domaine du conçu et du dicible. L'opération de génération $G$ introduit comme objet d'étude un morceau de pure factualité physique, jamais encore qualifié auparavant, lui, spécifiquement. Elle l'introduit d'une manière qui est entièrement indépendante de toute action de qualification passée ou future de cette entité-objet particulière. Toute qualification à proprement dire, spécifique de l'entité-objet particulière dénotée $me_G$, qui puisse, non pas seulement la singulariser de manière a-cognitive au sein du factuel physique inconnu, mais aussi la caractériser cognitivement et de façon communicable au sein de la classe de tous les microétats, devra être réalisée *ensuite*. Il en découle que le processus de description qui vient d'être initié et qui, pour être accompli, requiert, comme toute description, l'introduction d'une entité-objet bien définie *et* des qualifications de celle-ci, est – en général – radicalement scindé en deux étapes *indépendantes* l'une de l'autre : une première étape de *génération* factuelle et d'étiquetage de l'entité-objet-d'étude – non qualifiante –, et une étape subséquente, qui reste à construire, de qualification de cette entité-objet (de son étude à proprement parler).

*Le concept de définition de l'entité-objet se sépare du concept de qualification de cette entité.*

Cela est foncièrement nouveau par rapport à la conceptualisation classique. La pensée courante et les langages qui l'expriment – le langage courant mais aussi la logique, les probabilités, toute la pensée scientifique classique – n'impliquent pas une structure de cette sorte pour initier un processus de création de connaissances. Les entités-objet-d'étude sont partout introduites par des actions qui, d'emblée, sont plus ou moins *qualifiantes* **:** des gestes ostentatoires qualifiants (là, ici, celui-ci, etc.), des référence à des contextes qualifiants, ou carrément des qualifications verbales**.** Ouvrons un dictionnaire. On y trouve « *chat* **:** un petit félin domestique, etc. ». Et dans les manuels scientifiques il en va souvent de même. Nous sommes profondément habitués à ce qu'une entité-objet et ses qualifications nous soient données *dans la même foulée*. Les grammaires définissent en général l'entité-objet d'une assertion descriptionnelle (d'une *proposition* au sens grammatical) en introduisant un nom d'*objet* au sens courant classique, 'maison', 'ciel', 'montagne' etc., suivi des nom(s) de *propriétés*, de prédicats (au sens grammatical) qui – eux – *définissent en le qualifiant* l'objet indiqué par le nom considéré. Tout cela est presque exclusivement verbal et suppose la *pré-existence* des entités-objet. Ceux-ci ne doivent qu'être sélectionnées dans le champ de l'attention, pour usage, étude, etc. Pas question de les créer physiquement. Et la sélection d'entités-objet, ce qui introduit dans le champ de l'attention telle ou telle parmi ces entités pré-existantes – que ce soit au sens descriptionnel ou au sens grammatical – ce sont des qualifications qui l'opèrent, des prédicats, non pas une opération physique non-qualifiante. La logique classique entérine cette façon de faire. Toute la pensée classique flotte dans le nuage de trompe-l'œil conceptuels que nous désignons par les mots 'objet' et 'propriété'.

Mais les microétats ne sont pas des 'objets' au sens classique. Ce ne sont que des entités physiques que nous voulons mettre dans le rôle d'entités-(objet-*de-description*). Et afin de réussir ce tour difficile, le référent courant du mot 'définition' doit être soumis à une mutation du sens qu'il comporte ; une mutation telle qu'elle vide ce référent de tout contenu sémantique propre en le séparant radicalement du référent du mot 'qualification'.

### 4.3. Qualifier un microétat

Sur la base acquise à partir de la question $Q$ il est désormais possible d'envisager d'étudier le microétat étiqueté $me_G$, c'est-à-dire d'essayer d'acquérir quelques données communicables sur sa manière d'être propre et particulière. Il s'est créé une ouverture vers l'acquisition d'un savoir nouveau et non-hypothétique, lié à des caractéristiques spécifiques au microétat $me_G$ créé par l'opération de génération $G$ considérée, et par aucune autre. Cette ouverture n'aura été utilisée que lorsqu'on aura réussi à en tirer des manifestations observables par le concepteur-observateur humain, des manifestations qui impliquent le microétat $me_G$ et dont on puisse dire, en quelque sens précisé, qu'elles *qualifient* $me_G$. Mais quelles sortes de manifestations? Comment les concevoir? Comment les réaliser ? Comment établir leur signification? Bref, on se trouve maintenant en présence de la question *Q2 :* Comment qualifier un microétat $me_G$ ?

Un chemin pour aborder la questions *Q2* s'ébauche lorsqu'on commence par examiner comment nous qualifions habituellement.

### *4.3.1. Comment qualifions-nous habituellement :*
### *une grille normée de qualifications communicables*

Lorsqu'on qualifie un objet on le fait toujours relativement à quelque point de vue, quelque biais de qualification, couleur, forme, poids, etc. Une qualification dans l'absolu n'existe pas. Supposons alors que l'on recherche une qualification de couleur pour une entité-objet macroscopique. Notons tout de suite que le mot couleur lui-même n'indique pas une qualification bien définie. Il indique une nature commune à tout un ensemble ou spectre de qualifications, rouge, vert, jaune, etc.. Il indique une sorte de dimension, ou terrain abstrait, bref un réceptacle ou support sémantique où l'on peut loger toute qualification qui spécifie une couleur bien définie. On pourrait alors dire, par exemple, que rouge est une 'valeur' (non numérique dans ce cas) que la dimension de couleur peut loger ou manifester, et que la couleur au sens général ne peut se manifester que par des valeurs de couleur. Ce langage peut paraître inutilement compliqué. Mais il apparaîtra vite que les distinctions introduites sont toutes nécessaires. Il en va de même pour ce qu'on appelle forme, poids, position, énergie, bref, pour tout ce qui indique un biais de qualification : Plus ou moins explicitement, mais toujours, une qualification fait intervenir deux paramètres de qualification, hiérarchisés : *(a)* une dimension de qualification et *(b)* un spectre de 'valeurs' porté par cette dimension.

Comment apprend-on quelle est la valeur de couleur d'un objet que l'on veut étudier du point de vue de la couleur ? On le regarde. Cela peut s'exprimer aussi en disant qu'on assure une *interaction de mesure* de couleur entre l'objet et notre *appareil* sensoriel visuel. L'avantage de cette façon de dire est qu'elle introduit d'emblée un langage qui pourra convenir aussi bien aux examens scientifiques délibérés, qu'aux examens plus ou moins spontanés de la vie courante. Cette interaction (biologique) de mesure de couleur produit une sensation visuelle que seul le sujet connaissant éprouve, donc peut connaître et reconnaître. Mais on admet que cette sensation visuelle se trouve en corrélation stable avec l'objet étudié, dans la mesure où cet objet et son état sont eux-mêmes stables cependant que l'état du sujet percepteur est stable également et 'normal'. En outre, par des apprentissages préalables qui impliquent des processus de comparaison et d'abstraction, le sujet finit par distinguer, dans sa sensation visuelle globale de l'objet d'étude, cette dimension sémantique particulière dont le nom public est 'couleur', ainsi que les valeurs particulières de couleur par lesquelles cette dimension se manifeste à lui. Une valeur de couleur, telle qu'elle est ressentie par le sujet, est essentiellement indicible quant à sa qualité, sa *qualia*, sa nature subjective intime. Mais par l'apprentissage de la correspondance entre cette *qualia* et le *mot* que les autres prononcent en regardant la même entité physique (correspondance qui est constante dans le cas d'un groupe d'observateurs normaux et en situation observationnelle normale) le sujet arrive à *étiqueter* lui aussi cette *qualia* non communicable, par ce même mot, son nom public qui, lui, est consensuel : 'rouge', 'vert', etc. Cela lui permet de s'entendre avec les autres sujets humains en ce qui concerne les valeurs de couleur[7]. Bref, dans le cas de notre exemple, le sujet annoncera par un mot consensuel que le résultat de la mesure (de l'estimation de valeur) de couleur qu'il vient de réaliser à l'aide de ses yeux, est telle valeur de couleur, disons la valeur 'rouge'.

Mais un aveugle peut-il répondre à une question concernant la couleur d'un objet ? Ce n'est pas impossible. Il peut mettre l'objet d'étude dans le champ d'un spectromètre de couleurs connecté à un ordinateur vocal qui annonce en noms publics de couleurs les résultats de l'analyse spectrale qu'il opère. Ainsi le résultat produit par une interaction de mesure de couleur entre l'objet d'étude et un appareil de mesure de couleur qui est différent des appareils sensoriels biologiques de l'aveugle, est perçu par l'aveugle *via* un appareil sensoriel biologique dont il dispose, son ouïe. Cela le met en possession – directement – de l'expression publique du résultat de l'interaction de mesure considérée. La perceptibilité sensorielle visuelle, dont il manque, a été court-circuitée.

En outre, nonobstant l'absence de *toute* perception sensorielle visuelle, l'aveugle peut néanmoins se construire progressivement une certaine 'perception intellectuelle' subjective de la signification des noms de couleur, 'rouge', 'vert', etc.. Cela est possible à l'aide de certains contextes (verbaux ou d'autres natures) qu'il est capable de percevoir. À partir de ces contextes, le total vide d'intuition qui, pour lui, se cache sous le mot publique 'rouge' qu'il a appris à utiliser, est osmotiquement pénétré d'une sorte de brume de *qualia*, peut-être associée à un mélange d'images fugaces, comme un 'modèle' vague et changeant.

Cette dernière remarque n'est pas dépourvue d'importance parce que l'expérimentateur scientifique est comparable à l'aveugle en ce qui concerne le type de signification qu'il peut associer à certaines qualifications qu'il réalise sans aucun autre support perceptif que des marques 'sans forme' ou des annonces par symboles recueillies sur des enregistreurs d'appareils qui sont extérieurs à son corps (pensons aux indications que l'on lit sur un écran de surveillance des phénomènes qui se passent dans un accélérateur du CERN) : à la différence des qualifications déclenchées exclusivement par ses perceptions sensorielles biologiques, les significations associées à de telles marques ou annonces ne parviennent plus à la conscience de l'expérimentateur sous la forme de *qualia* liée à l'entité-objet-d'étude, il ne les perçoit *que* par l'intermédiaire d'étiquetages publics de résultats obtenus opérationnellement : la dimension sémantique qui porte la 'valeur' impliquée par ces résultats, n'est plus sensible. Mais après coup, comme l'aveugle, l'expérimentateur se construit une certaine perception intellectuelle subjective de cette dimension sémantique. Quand il dit, par exemple : «j'ai mesuré une 'différence de potentiel électrique'», il associe à cette expression un certain mélange flou d'images portées par les formulations verbales qui ont formé ce concept dans son esprit et qui, au plan intuitif, le relient à son expérience sensorielle de départ.

On vient de détailler comment se constitue une *grille normée de qualifications communicables* de couleurs. Une conclusion analogue vaut pour toute autre sorte de qualification physique communicable et normée, de forme, poids, etc., et même pour des qualifications communicables et normées abstraites.

Le schéma général affirmé au départ pour une grille de qualification, peut maintenant être précisé : une telle grille comporte toujours *(a)* une dimension sémantique portant un ensemble de valeurs ; *(b)* une procédure d'interaction de 'mesure' mettant en jeu, soit d'emblée et exclusivement un ou plusieurs appareils sensoriels biologiques du concepteur humain, soit d'abord un appareil de mesure artificiel sur les enregistreurs duquel s'enregistrent des marques physiques qui, elles, sont directement observables par le concepteur *via* ses appareils sensoriels biologiques (il doit *toujours* y avoir un effet *final* d'un acte de mesure, que le concepteur humain perçoive directement par ses sens biologiques) ; et *(c)* une procédure de traduction de l'effet *final* perçu par le concepteur humain, en termes communicables et publiquement organisés désignant une, et une seule, parmi les *valeurs* portées par la dimension sémantique introduite.

---
[7] L'essence de ce procédé est la même que dans l'entière physique macroscopique, les relativités d'Einstein inclusivement : un ensemble d'observateurs perçoivent et examinent tous une même entité physique qui leur est extérieure ; ils étiquettent les résultats subjectifs de leurs observations selon des règles publiques conçues de façon à assurer certaines formulations qualifiantes finales qui sont invariantes d'un observateur à un autre.

Telle est l'essence du schéma qui fonctionne lors des actions de qualification de notre vie courante et, en général, lors des qualifications effectuées dans le cadre de la pensée classique. Au premier abord, les éléments de ce schéma et les phases de son édification peuvent n'apparaître que d'une façon confuse. Mais la présence de chaque élément et de chaque phase est toujours identifiable par analyse.

### *4.3.2. De la grille usuelle de qualifications communicables,*
### *à une 'condition-cadre générale' pour la qualifiabilité d'un microétat*

Considérons un microétat $me_G$ spécifié par une opération de génération *G*. On veut le qualifier. Il est clair d'emblée que pour ce cas le schéma classique ne fonctionne plus tel quel, ne serait-ce que parce qu'un microétat n'est pas disponible tout fait et stable, de manière à ce qu'on puisse lui 'appliquer' une grille qualifiante déjà disponible et stable elle aussi. Mais il peut y avoir d'autres difficultés, insoupçonnées. Il faut donc identifier explicitement, une à une, les spécificités qui apparaissent, et faire face systématiquement aux contraintes qu'elles comportent.

#### *4.3.2.1. Spécificités d'une opération de qualification d'un microétat:*
#### *émergence d'un concept de 'condition-cadre générale'*

Partons de l'exemple des couleurs considéré plus haut. Dans cet exemple on suppose que l'entité-objet préexiste et que, d'emblée, elle est apte à produire, lors d'une interaction avec nos yeux, l'effet de couleur dénommé 'rouge'. Ce résultat est supposé se produire en vertu d'une *propriété* dont on conçoit que d'ores et déjà elle est actuelle, réalisée en permanence *dans* l'entité-objet considérée, d'une façon intrinsèque, indépendante de toute interaction d'estimation de la couleur : la propriété d'émettre constamment des radiations électromagnétiques d'une longueur d'onde comprise dans l'intervalle étiqueté par le mot 'rouge'.

L'aveugle admet lui aussi que l'entité-objet dont il ne peut voir la couleur, préexiste avec constance, puisqu'il peut à tout instant la toucher, l'entendre tomber, etc. Et il admet également que cette entité-objet 'possède' de par elle-même une propriété qui, par interaction avec un appareil récepteur adéquat, produit le genre d'impression qu'on appelle 'couleur'. C'est sur cette double base qu'il soumet l'entité-objet, telle qu'elle est, à une interaction avec un spectromètre, c'est à dire avec un simple détecteur de couleur qui ne fait qu'enregistrer des effets de la propriété, préexistante dans l'entité-objet, d'émettre des radiations dans la bande du visible par l'homme normal.

Or dans le cas d'un microétat, la supposition de propriétés intrinsèques préexistantes dans l'entité-objet d'étude, ne vaut plus. Une telle supposition serait inconsistante avec la situation cognitive dans laquelle on se trouve dans la phase actuelle du processus de conceptualisation que nous avons entrepris ici. Concernant – spécifiquement – un microétat particulier qui est 'disponible' pour être étudié mais qui n'a encore jamais été qualifié, on ne sait *rien* à l'avance en dehors de la manière dont il a été généré. (J'emploie de manière répétée ces curieuses précisions – spécifiquement, particulier – afin de constamment distinguer le savoir nouveau que l'on veut gagner, du type de 'savoir générique d'accueil' porté par l'affirmation *posée a priori* que l'effet de l'opération de génération $G$ mise à l'œuvre est de la catégorie qu'on convient d'appeler 'un microétat'). Dans la phase initiale du processus de construction de connaissances concernant un microétat, celle de création de l'entité-objet-d'étude, nous n'assignons aucune propriété propre (si l'on peut dire) au microétat que nous avons étiqueté '$me_G$'. Nous n'avons même pas encore d'indices directs que ce microétat existe. On pose qu'il existe (et qu'il est en une relation de un-à-un avec l'opération $G$ qui l'a produit) mais on ne le sait pas.

Dans ces conditions, afin d'arriver à associer au microétat $me_G$ telle valeur de telle dimension de qualification, il faudra *construire* tout un chemin. Il faudra, en général tout au moins, commencer par *changer* ce microétat hypothétique de telle manière qu'il produise quelque effet *observable* ; et il faudra que cet effet observable puisse – de quelque façon bien définie – signifier une valeur précisée parmi toutes celles considérées comme possibles pour la dimension de qualification voulue (ce qui, accompli, étayera aussi l'hypothèse que le microétat $me_G$ existe). Et il faudra rester vigilant concernant *ce* que – exactement – la valeur ainsi obtenue qualifie. Bref, il faudra *créer*, conceptuellement et *physiquement*, une qualification du type recherché, et en outre il faudra construire un sens pour l'affirmation que cette valeur là peut être associée au microétat étudié.

L'affirmation de nécessité de vigilance signale d'emblée un piège vers lequel le langage pousse subrepticement la pensée, et auquel il faut échapper. Lorsqu'on dit « je veux qualifier ceci » (cette pierre, ce lac), ce qu'on escompte automatiquement est un renseignement sur la manière d'être *de* 'ceci' *même*. Or dans notre cas, puisqu'on doit changer le microétat, et peut-être même le changer radicalement, le résultat *final* observé impliquera un *autre* microétat, différent du microétat à étudier, celui qui a été initialement engendré par l'opération de génération $G$ en tant qu'entité-objet d'étude. En outre ce résultat ne qualifiera même pas *exclusivement* cet autre microétat changé, il ne qualifiera que, globalement, *l'interaction de mesure* qui aura produit le changement et la manifestation finale observée. En effet la valeur indiquée par la manifestation observée aura été produite par cette interaction de mesure considérée globalement, pas exclusivement et directement par ce que nous appelons le microétat $me_G$ à étudier. Il faudra surveiller de près cet ordre d'idées et notamment expliciter s'il existe un sens, dans ces conditions, dans lequel le résultat des qualifications construites peut être regardé comme une 'description du microétat étudié $me_G$' lui-même. Et si un tel sens existe, il faudra le délimiter avec rigueur.

Une autre remarque s'impose. Le formalisme mathématique de la mécanique quantique concerne spécifiquement les déplacements des microétats dans l'espace, puisqu'il s'agit d'une *mécanique*. Les dimensions sémantiques de qualification qui y interviennent sont indiquées par les mots 'position', 'quantité de mouvement', 'énergie', moment de la quantité de mouvement'[8]. Comment a-t-on pu arriver à associer de telles qualifications à des microétats non perceptibles et encore strictement non connus, dont on ne savait même pas si, et en quel sens, on peut dire qu'il se *déplacent* ? Cela, sans se fonder sur aucun modèle ? Car n'oublions pas que le modèle du genre bille en mouvement, que l'on associait à un microétat avant la construction de la mécanique quantique, par transfert de la mécanique classique, a échoué, et que c'est à la suite de cet échec qui a laissé un *vide de modèle* qu'on a conçu la nécessité d'une autre conception sur les microétats. Les acquis historiques qui au début 20$^{ème}$ siècle appelaient une nouvelle 'mécanique' des microétats, lançaient cet appel avec un mot qui s'était vidé de tout contenu établi.

---

[8] À ces qualifications fondamentales, ont été ensuite ajoutées d'autres (spin, parité).

Un nouveau départ avait été marqué par le célèbre modèle onde-particule proposé dans la thèse de Louis de Broglie (de Broglie, L., [1924] et [1963]). Mais ce modèle n'a pas clairement survécu dans la formalisation finale acceptée généralement. En tout cas il n'y a pas survécu avec la même signification que dans la thèse et les travaux ultérieurs de Louis de Broglie. Il n'en est pas vraiment absent non plus. Il y subsiste implicitement dans les écritures mathématiques et dans le langage qui accompagne ces écritures, où il a instillé une foule de traces. Mais celles-ci ont diffusé et ont perdu les marques de leur origine. À tel point que, depuis Bohr et Heisenberg et jusqu'à ce jour, on affirme couramment que la mécanique quantique actuelle n'introduirait aucun modèle, ni de microsystème ni de microétat.

Or cela est certainement inexact. Le but d'élaborer une mécanique des microétats a *dû* présupposer la signifiance de 'grandeurs mécaniques' pour ces entités inobservables et hypothétiques étiquetées par le mot 'microétats'. Dans la mécanique classique les grandeurs mécaniques n'ont été définies *que* pour des mobiles macroscopiques. Leur signifiance pour des 'microétats' également n'a pu être qu'un postulat posé *a priori*. Un postulat admis sur la base de tout un ensemble d'indices expérimentaux et conceptuels-historiques, mais un postulat *a priori* tout de même. Ce postulat ne pouvait être justifié que d'une manière constructive, par la réalisation effective d'une représentation des grandeurs mécaniques à l'intérieur d'un tout doté d'efficacité prévisionnelle, constitué par un système d'algorithmes mathématiques et d'opérations physiques et appareils associés à ces algorithmes. On a *décidé* d'induire des 'grandeurs mécaniques' conçues initialement dans une discipline macroscopique, dans une représentation nouvelle liée à des dimensions d'espace et de temps dont les ordres de grandeur dépassent à un degré gigantesque les seuils de perception des organes sensoriels biologiques de l'homme. On a construit selon cette décision et la construction s'est justifiée. Mais afin de pouvoir construire l'on a forgé des éléments formels tirés des expressions mathématiques des modèles classiques de mobiles qualifiables par des grandeurs mécaniques (à savoir des reformulations hamiltonienne et lagrangienne de la mécanique newtonienne). Ceux-ci ont été *adaptés* à des actions descriptionnelles d'un type foncièrement différent de celui des actions descriptionnelles classique[9]. Toutefois la forme mathématique de ces éléments formels nouveaux, qui leur sont attachées, et surtout *la structure de l'opération physique de mesure* affirmée comme correspondant à chaque opérateur, déclarent souvent ouvertement leur origine historique et leur rôle de franche modélisation par prolongement d'une dimension sémantique de qualification mécanique définie au niveau macroscopique et posée *a priori* comme pouvant acquérir un sens conceptuel-opérationnel pour des microétats aussi.

Afin de concrétiser nous donnons immédiatement un exemple d'une telle opération de mesure, présenté dans des termes qui relient à ce que nous avons déjà établi ici et anticipent sur les questions qui seront traitées plus loin.

---

[9] À savoir *via* la définition d'un opérateur différentiel associé à chaque grandeur mécanique classique, avec ses 'états propres' et ses 'valeurs propres'.

Soit la grandeur – vectorielle – $X \equiv$ 'quantité de mouvement', dénotée $\mathbf{p}$[10]. Selon Feynman un acte de mesure pour un microétat, d'une valeur $Xn$ du spectre de cette grandeur (dénotons-la $Xn \equiv \mathbf{p}_n$) doit être accompli par la méthode 'time of flight', de la manière suivante.

Soient, respectivement, $\delta E(G)$ et $\delta t(G)$ les domaines d'espace et de temps qui seront peuplés par une réalisation de l'opération $G$ de génération d'un exemplaire du microétat $me_G$ à qualifier. On place un écran sensible $\mathcal{E}$ – très étendu – suffisamment loin du domaine d'espace $\delta E(G)$ pour que ce domaine puisse être considéré comme quasi ponctuel par rapport à la distance $O\mathcal{E}$ entre $\delta E(G)$ et $\mathcal{E}$ mesurée le long d'un axe $Ox$ partant – en gros – de $\delta E(G)$ et tombant perpendiculairement sur $\mathcal{E}$ ; cependant que $\delta t(G)$ puisse être regardé comme négligeable par rapport à la durée moyenne qui s'écoule entre le moment $t_o$ de la fin de la réalisation de l'opération $G$ et le moment $t$ où l'on enregistre un impact sur $\mathcal{E}$.

*(a)* On accomplit effectivement une opération $G$ en notant le moment $t_o$ assigné à sa fin, i.e. celui assigné au début de l'existence du microétat $me_G$. (Notons que $t_o$ est une *donnée* de départ caractéristique de $G$, ce n'est pas un enregistrement obtenu par l'acte de mesure sur $me_G$ qui doit suivre).

*(b)* Si $G$ a comporté des champs, au moment $t_o$ *on les éteint*. Si entre le support d'espace-temps $\delta E.\delta t(G)$ et l'écran $\mathcal{E}$ il préexiste des champs extérieurs ou des obstacles matériels, *on les supprime*. Sur la base de ces précautions l'*évolution de mesure* assignée à l'exemplaire du microétat $me_G$ créé par l'opération $G$, est posée être 'libre' (dépourvue d'accélérations) (cette précaution ne peut se rapporter qu'à la présupposition, dans $me_G$, d'une 'quantité de mouvement' dont toute accélération modifierait la valeur vectorielle).

*(c)* Après quelque temps il se produit un impact $P_n$ sur l'écran $\mathcal{E}$. L'aiguille d'un chronomètre lié à $\mathcal{E}$ aqcuiert alors une position, disons $ch_n$, qui marque le moment $t_n$ de cet événement (l'ensemble des données qui concernent l'acte de mesure considéré ici et qui pourraient changer de valeur dans un autre acte de mesure accompli sur un autre exemplaire de $me_G$, sont indexés par un numéro d'ordre $n$). On dit que 'la durée de vol' (time of flight) – mais *de qui? de quoi?* – a été $\Delta t_n = t_n - t_o$.

*(d)* La valeur de la distance – vectorielle – $\mathbf{d}_n$ parcourue entre $\delta E(G)$ et le point d'impact $P_n$, est $\mathbf{d}_n = \mathbf{OP}_n$. Le carré de la valeur absolue de cette distance est $|\mathbf{d}_n|^2 = d_{xn}^2 + d_{yn}^2 + d_{zn}^2$ où $d_{nx} \equiv O\mathcal{E}$ est mesurée sur l'axe $Ox$ et $d_{yn}$, $d_{zn}$ sont mesurés sur deux axes placés dans le plan de $\mathcal{E}$ et qui, avec $Ox$, déterminent un système de référence cartésien droit.

*(e)* On définit, respectivement, la valeur $\mathbf{p}_n$ mesurée pour la grandeur $X \equiv \mathbf{p}$, et sa valeur absolue $|\mathbf{p}_n|$, selon les formules

$$\mathbf{p}_n = m(\mathbf{d}_n / \Delta t_n) \qquad |\mathbf{p}_n| = m(\sqrt{d_{xn}^2 + d_{yn}^2 + d_{zn}^2} / \Delta t_n)$$

où $m$ est la masse associée au microsystème dont on étudie le microétat $me_G$ (définie dans la physique atomique ou la théorie des particule élémentaires).

Ceci clôt l'acte de mesure considéré. Notons maintenant ce qui suit.

Dans le cas exposé plus haut les *manifestations physiques observables* produites par l'acte de mesure sont : le point $P_n$ et la position $ch_n$ de l'aiguille du chronomètre lié à l'écran $\mathcal{E}$. Ces manifestations ne sont *pas* directement des valeurs numériques, ni n'en 'possèdent'. Ce sont seulement des *marques physiques* perceptibles, disons $\mu_{1n}$ et $\mu_{2n}$, respectivement, produites par l'acte de mesure, sur les deux 'enregistreurs' de 'l'appareil' de mesure qui a été conçu pour accomplir cet acte de mesure (l'appareil étant constitué du chronomètre associé à l'opération $G$, de l'extincteur de champs extérieurs, de l'écran $\mathscr{E}$ et du chronomètre lié à l'écran).

---

[10] Ici l'écriture en gras '$X$' n'indique que la nécessité, dans un formalisme mathématique pour décrire des microétats, de définir une composition, pour la dimension sémantique considérée, de trois autres dimensions sémantiques ; une composition reliée à celle qui, pour un vecteur au sens classique, unit ses trois 'composantes').

Les *significations* associées aux manifestations observables enregistrées, ainsi que les valeurs numériques associées à ces significations, à la fois, sont définies par : la manière de *concevoir* un acte de mesure de la grandeur $X \equiv$ 'quantité de mouvement' assignée à un microétat et dénotée $p$ et par les relations posées $p=m(d/\Delta t)=mv$, $\Delta t=t-t_o$ et $|d|=\sqrt{(d_x^2+d_y^2+d_z^2)}$, $\Delta t_n=t_n-t_o$ et $|d_n|=\sqrt{(d_{xn}^2+d_{yn}^2+d_{zn}^2)}$ dont les trois premières définissent une fonction générale '$f$' de structure de '$p$', à savoir $p = f(m, d, \Delta t)$, et les autres permettent de calculer la valeur $p_n$ mesurée pour $p$ en cohérence avec la fonction de structure '$f$' et sur la base des deux manifestations physiques observables $\mu_{1n} \equiv P_n$ et $\mu_{2n} \equiv ch_n$.

On voit clairement sur l'exemple donné que la quantité de mouvement $p$ associée au microétat à étudier $me_G$, est définie de la manière *classique*, par $p=mv$ ; que les prescriptions pour calculer la valeur numérique $p_n$ sont exactement celles que l'on devrait suivre pour un mobile classique libre ; que cela *seul* justifie que l'évolution de mesure posée pour $me_G$ exige d'éteindre tout champ et d'éliminer tout 'obstacle', i.e. de supprimer tout ce qui – selon la mécanique *classique* – modifierait par des 'accélérations' la valeur $p$ 'de départ' et que c'est cette exigence qui constitue la base sur laquelle on conçoit (plus ou moins explicitement) qu'un acte de mesure 'time of flight' est *convenable* par ceci précisément qu'il change le microétat étudié $me_G$ d'une manière qui n'altère pas aussi la valeur à mesurer de $p$. Bref, l'entière procédure qui vient d'être exposée serait entièrement arbitraire – et même inconcevable – en l'absence d'un modèle macroscopique classique pour lequel on a décidé d'assigner, par prolongement, une signification, aussi, pour des microétats.

Tout ce qui vient d'être dit plus haut concerne le formalisme mathématique de la mécanique quantique. Ce n'était qu'une déviation faite afin de permettre de mieux comprendre ce qui, dans la démarche développée *ici*, peut ou ne peut pas être affirmé concernant les qualifications de microétats. Car la démarche que nous développons ici est soumise délibérément à des contraintes plus draconiennes que celles qu'a subies la construction du formalisme de la mécanique quantique, qui avait *un autre but*, celui de construire une *mécanique* des microétats. Ici l'on recherche la structure descriptionnelle qualitative qui découle des contraintes imposées par, *exclusivement*, les modes humains généraux de conceptualiser et la situation cognitive où se trouve un observateur-concepteur qui veut construire des *connaissances* concernant des microétats. Cette ascèse sert le but d'obtenir une structure de référence qui permette ensuite, par comparaison, de percevoir le statut conceptuel de chaque élément descriptionnel qui intervient dans le formalisme quantique. L'utilisation – ici – de modèles de prolongement de la mécanique classique, déborderait notre règle de jeu. Mais cela n'entraîne nullement que l'insertion dans le formalisme quantique, de modèles de prolongement de la mécanique classique, soit répréhensible de quelque point de vue : *sans de telles insertions il n'y aurait simplement pas de mécanique quantique*.

Bref, en ce qui concerne la qualification de microétats, on ne peut viser ici qu'à formuler, en termes qualitatifs, des 'conditions-*cadre* générales' valides uniformément à des qualifications quelconques applicables à des microétats.

*4.3.2.2. La condition-cadre-générale pour qualifier un microétat*

Soit une grandeur qualifiante. Nous avons mis en évidence que, par définition générale, une telle grandeur doit introduire une dimension sémantique *X* et un spectre *{X1,X2,X3,…Xj,…}* formé de l'ensemble de toutes les valeurs *Xj* de *X* qui sont prises en considération[11], exprimables à l'aide de nombres[12]. Comme l'indique l'indexation par des entiers *1,2…j…*, on suppose que ce spectre est *discret*[13]. Le signe *Xj* indique une valeur de *X* bien définie mais quelconque, tandis que *X3* indique exclusivement la valeur dénotée ainsi.

Nous dirons que la paire *(X, {X1,X2,X3,…Xj,…})* constitue *une grille de qualification applicable à des microétats* dénotée *gqX(me)*, si et seulement si elle comporte les éléments suivants :

***gqX1(me)***. Une définition *conceptuelle-formelle* de *X*, dénotons-la *déf(X)*, qui soit spécifie une structure 'élémentaire' de *X* posée sans faire intervenir d'autres dimensions sémantiques, soit spécifie la forme de la dépendance de *X* d'autres dimensions sémantiques, *Y,Z,…*, *via* une fonction *déf(X)=f(Y, Z,…)*. Bref, *déf(X)* est une spécification de la structure sémantique de *X*. Notamment, elle justifiera le *nom* assigné à la grandeur *X* (dans le cas de grandeurs mécaniques, 'position', 'quantité de mouvement', 'énergie totale', etc.).

***gqX2(me)***. La spécification, pour la paire *(X,{X1,X2,X3,…Xj,…})*, d'une opération *physique* correspondante dénommée *interaction de mesure (des valeurs numériques) de X* et dénotée *Mes(X)* (lire : 'mesure de *X*') qui soit applicable à un microétat et soit cohérente avec *déf(X)*.

Toute interaction de *Mes(X)* doit :

* Se réaliser entre le microétat à étudier $me_G$ et un *appareil A(X)* approprié pour *changer* le microétat $me_G$ d'une manière qui, chaque fois, aboutisse à un groupe de *m* manifestations physiques observables $\mu_k$, *(j=1,2,…m)* (*m* : un entier (petit en général)) produites sur/par les enregistreurs de l'appareil *A(X)* utilisé (marque sur un élément de *A(X)* sensible à des micro-impacts, son émis par un compteur d'interactions microscopiques, position d'aiguille d'un chronomètre, etc.[14]). L'appareil *A(X)* est regardé comme partie intégrante du processus de mesure *Mes(X)* ; il accomplit le processus de *Mes(X)* soit exclusivement *via* des structures matérielles macroscopiques qui créent des champs de force ou agissent comme des obstacles, soit *via* de telles structures *et* des entités microscopiques *messagères*[15]. Mais dans tous les cas, à l'aide d'enregistreurs macroscopiques adéquats (écran, boite à déclics, milieu sensible, chronomètres, etc.), l'appareil *A(X)* expose *sur/par ses enregistreurs* les *m* effets observables $\mu_k$, *(k=1,2,…m)* qui résultent de l'interaction de *Mes(X)*. Pour cette raison nous parlerons de manifestations *liées* au microétat étudié (par l'interaction de mesure *Mes(X)*) mais TRANSFEREES sur les enregistreurs de *A(X)*.

---

[11] Pas qui 'existent', ce qui serait un concept non-effectif, mais qui *sont prises en considération* (par exemple, un mètre de couturier qui ne porte pas des marques entre deux traits successifs distancés l'un de l'autre de 1 mlimètre, ne prend pas en considération des valeurs de longueur spécifiées avec une précision supérieure à celle qui correspond au milimètre).

[12] Le symbole *X* joue le rôle que jouait dans l'exemple classique le nom générique de couleur, et les symboles *Xn* jouent le rôle des noms de valeurs de couleur (rouge, bleu, etc.). Mais ici, puisque les conditions recherchées visent les contenus sémantiques d'une théorie de physique mathématique, les valeurs *Xj* de la dimension de qualification *X* doivent être exigées comme spécifia*bles* (pas spécifiées) en termes *numériques*. Cela est nécessaire si l'on veut identifier des conditions-cadre générales suffisamment restrictives. Mais cela n'entache nullement le caractère qualitatif de l'approche.

[13] Pour simplifier, mais aussi pour d'autres raisons beaucoup plus essentielles qui deviennent claires dans le cadre de la méthode générale de conceptualisation relativisée (Mugur-Schächter [2006]) et dans (Mugur-Schächter [2008]).

[14] Le microétat lui-même ne 'manifeste' rien, toute ce qui est observable est effet d'une interaction de *Mes(X)* manifesté sur/par un enregistreur de *A(X)*.

[15] Lorsque le microétat étudié est lié (pas libre et progressif), on est contraint à des procédures de qualification *indirecte*, par exemple émission de microsystèmes 'messagers' qui, eux, interagiront avec le microétat étudié.

\* Comporter un codage qui, à chaque ensemble *donné* $\{\mu_{kq}, (j=1,2,...m)\}$ de $m$ manifestations physiques observables toutes indexé par un même indice $'q'$ et produites par un acte de $Mes_q(X)$ indexé par ce même indice, fasse correspondre une valeur numérique $Xj$ de $X$ et une seule, en cohérence avec $def(X)$[16]. Dénotons ce codage par $\mathcal{C} \equiv (\{\mu_{kq}, j=1,2,...m\} \leftrightarrow Xj)$[17].

L'exigence d'un codage $\mathcal{C} \equiv (\{\mu_{kq}, j=1,2,...m\} \leftrightarrow Xj)$ est d'importance majeure. Si elle n'était pas respectée la valeur numérique $Xj$ de $X$ produite par l'acte d'interaction de $Mes_q(X)$ resteraient tout simplement non identifiable cet l'acte de mesure ne servirait à rien. Car un ensemble $\{\mu_{kq}, j=1,2,...m\}$ de manifestations enregistrées par un acte de mesure sur un microétat n'indique en général d'aucune manière quelle valeur numérique $Xj$ il convient de lui assigner[18].

Notons maintenant un fait remarquable. On pourrait croire au premier abord que tout codage $\mathcal{C} \equiv (\{\mu_{kq}, j=1,2,...m\} \leftrightarrow Xj)$ doit être établi sur la base des particularités imposées sur les actes de $Mes(X)$ par la grandeur $X$ considérée. Or, sur la base du fait que toute manifestation physique se produit nécessairement dans l'espace et le temps, on peut concevoir une possibilité de principe de définir le codage $\mathcal{C}$ d'une manière applicable *uniformément* à tout acte de $Mes(X)$. En effet imaginons que les opérations physiques comportées par un acte de $Mes(X)$ quelconque seraient *par construction* telles qu'elles engendrent – à des distances et dans une durée finies – une correspondance quasi-biunivoque entre : d'une part la région globale d'espace-temps (ou seulement d'espace *ou* de temps) où se produisent *toutes* les $m$ manifestations observables $\{\mu_{kq}, k=1,2,...m\}$ qui closent l'acte de $Mes_q(X)$, et d'autre part une valeur $Xn$ bien définie de la grandeur mesurée $X$ et *une seule*. Autrement dit, imaginons que, par construction, tout acte de $Mes(X)$ introduirait au bout d'une durée finie, une *séparation* entre des régions d'espace-temps (ou seulement d'espace *ou* de temps) dont *chacune*, globalement, est réservée aussi exclusivement qu'on veut à un seul ensemble $\{\mu_{kq}, k=1,2,...m\}$ de manifestations observables, donc à une seule valeur $Xj$.

Dans ce cas *la région d'espace-temps (ou bien d'espace ou de temps) où se produit cet ensemble de manifestations, considérée globalement, coderait pour une valeur correspondante $Xj$ de $X$*[19].

Les conditions exigées par la définition $gqX(me)$ d'une grille de qualification applicable aux microétats constituent ensemble une *condition-cadre générale* que nous dénoterons $CCG \equiv [gqX(me)]$. Le mot 'cadre' souligne que $CCG$ ne dit strictement rien concernant la manière de qualifier un microétat par, spécifiquement, une grandeur de telle ou telle nature particulière.

La transposition du concept habituel de grille de qualification, au cas des micoétats, a séparé une sorte de vide à l'intérieur de la structure que nous sommes en cours de bâtir ; un vide délimité par la condition-cadre *générale CCG* qui ne spécifie que le concept de 'qualification quelconque'. Strictement rien n'est dit concernant la manière de peupler ce vide par une qualification de telle ou telle nature donnée, mécanique ou autre. Car à ce sujet ni les conditions cognitives dans lesquelles on se trouve, ni les impératifs opérationnels-conceptuels qui interviennent, ne donnent aucune indication.

---

[16] L'exemple donné dans le paragraphe précédent concernant la grandeur 'quantité de mouvement' peut aider à vite comprendre cette condition.
[17] Les opérations par lesquelles un codage $\mathcal{C}$ s'accomplit dans le cas général peuvent être imaginées facilement sur la base de l'exemple 'time of flight'.
[18] Dans le cas d'une mesure macroscopique aussi, interviennent toujours certains codages. Mais si l'entité-objet d'étude est directement perceptible et l'opération de mesure l'est aussi, le codage est appliqué d'une manière plus ou moins implicite et triviale. Cependant que dans le cas d'une opération $Mes(X)$ effectuée sur un microétat non percevable concernant lequel en outre la définition de la grille de qualification est entièrement abstraite, l'utilisation explicite d'un codage bien défini est impérative.
[19] Les mesures de spin par la méthode de Stern et Gerlach possèdent cette caractéristique d'une manière qui frappe. Et si l'on analyse un acte de mesure 'time of flight', on trouve que dans ce cas aussi la condition mentionnée est remplie.

D'autre part, le formalisme quantique a rempli ce vide – directement en termes mathématiques – sous les contraintes du but de bâtir une mécanique des microétats qui assure une continuité historique avec les concepts et les dénominations utilisés dans la mécanique classique.

Lors d'un futur face-à-face entre la structure qualitative qui s'élabore ici, et le formalisme quantique, il sera intéressant de mettre en évidence les résultats spécifiques de la confrontation entre la condition-cadre générale *CCG* et la théorie quantique des mesures de grandeurs mécaniques assignées à des microétats.

### *4.3.3. Conclusions sur la transposition aux microétats du concept de grille normée de qualification*

Il a été donc possible d'établir les contraintes générales à imposer à un acte de *Mes(X)* d'une grandeur *X quelconque* pouvant qualifier un microétat ; cela sans avoir eu besoin ni d'un modèle de microétat ni de la spécification effective de la définition de telle ou telle grandeur *X* considérée. Mais en dehors de la formulation de ces contraintes générales, le processus d'élaboration poursuivi dans ***3.3.2*** implique deux faits conceptuels notables, l'un concernant le concept de 'propriétés' et l'autre concernant la relation entre qualification et modèle. Dans ce qui suit nous mettons en évidence ces faits conceptuels.

#### *4.3.3.1. Qualifications de microétats et 'propriétés'*

Afin d'établir les contraines-cadre générales *CCG* il n'a pas été nécessaire de supposer l'existence permanente, pour un microétat, de 'propriétés'. En fait, à la faveur de la structure d'exigences des éléments *gqX1(me)*, *gqX2(me)* comportés par la définition d'une grille de qualification *gqX(me)*, une véritable mutation s'est introduite subrepticement en ce qui concerne les propriétés *de* l'entité-objet étudiée. Par exemple, il est clair que si l'on veut finalement réaliser pour les microétats, des qualifications en termes de grandeurs mécaniques, il est obligatoire d'utiliser un certain concept de *position*, car c'est l'un des deux concepts de base, avec la vitesse, de ce qu'on appelle une mécanique. Ce concept veut dire 'ici' ou 'là', à tel endroit localisé de l'espace physique. Il répond à la question 'où ?'. Or il est apparu que, concernant un microétat, un tel renseignement ne peut être obtenu que *via* une manifestation physique observable *transférée* sur l'enregistreur d'un appareil. La condition-cadre générale *CCG≡[gqX(me]* n'exclut pas que cette marque ou déclic puisse se traduire, notamment, en termes d'une valeur bien définie de la grandeur *X≡position*. Au premier abord il peut même paraître que le seul fait que la marque se soit produite à endroit d'espace bien défini *suffit* pour permettre de parler en termes de position en relation avec des microétats[20]. Mais en fait ce n'est pas les cas. Car rien ne prouve que *ce* qui, lorsque l'enregistrement de la marque s'est produit, a agi de telle manière qu'il se soit engendré une marque quasi ponctuelle sur l'écran sensible d'un appareil, existait dès *avant* l'émergence de l'enregistrement ; ou que cela ne se trouvait pas ailleurs, etc. : si l'on ne pose *a priori* vraiment aucune ombre de modèle, il n'existe aucune base pour exclure que l'enregistrement final d'une marque observable localisée ait été *créé* de toutes pièces par l'interaction de *Mes(X)* où *X* est dénommé 'position'. Cette éventualité a été évoquée et discutée (de Broglie, L., [1956]). Par exemple, Einstein a soutenu qu'elle impliquerait la possibilité de processus *physiques* de localisation quasi instantanée à l'intérieur de l'étendue assignée au 'microétat' – qui peut être illimitée – ce qui, pensait-il, pouvait être contraire à la théorie de la relativité (en fait, en rigoureuse absence de tout modèle, cette objection ne peut être reçue).

---

[20] En mécanique quantique on parle souvent, et même pratiquement toujours, de la position de la 'particule' ou de la position du 'système' (pas du microétat), et l'une comme l'autre de ces façons de parler conduit à une cohue de confusions entre 'microétat' et *modèle* de micro-système. En physique des particules élémentaires, par contre, la 'position' n'est pas traitée comme une 'grandeur'.

Cet exemple permet d'estimer l'ampleur de la mutation que subit le concept classique de *propriété* '*de*' une entité-objet, en la stricte absence de toute connaissance présupposée concernant ces fragments de pure factualité que nous avons dénommés microétats.

Puisqu'il est nécessaire d'admettre qu'en général une interaction de mesure *change* le microétat à étudier $me_G$ tel qu'il a été créé par l'opération de génération correspondante $G$, la manifestation observable produite par un acte de *Mes(X)* n'est plus liée nécessairement à quelque concept de propriétés préexistantes *de* l'objet d'étude. En toute rigueur, on doit raisonner en admettant que les manifestations observables pourraient être *entièrement* créées par l'interaction de l'appareil avec l'entité-objet strictement inconnue dénotée $me_G$. Créées en tant que propriétés émergentes *des enregistreurs de l'appareil*, pas du microétat étudié. La seule affirmation que l'on puisse faire avec certitude est que ces manifestations observables expriment une propriété acquise par les enregistreurs de l'appareil, en conséquence de l'interaction de *Mes(X)* : le microétat-objet-d'étude n'est pas qualifié *isolément* par les processus de mesure *Mes(X)*.

Les considérations qui précèdent permettent de mesurer quel abîme conceptuel a été franchi, dans l'implicite obscur, par la manière de concevoir sur la base de *modèles classiques* les opérations physiques a accomplir *sur un microétat* lors d'un acte de mesure d'une grandeur 'mécanique' $X$ dont on pose à l'avance la signifiance (comme dans le cas de la méthode du temps de vol).

### *4.3.3.2. Qualification de microétats et modèle classique*

L'entière élaboration d'un concept de qualification de microétats, et notamment les considérations sur la mutation du concept de propriété que ce concept comporte, impliquent une conclusion non triviale qui concerne la relation entre qualifications de microétats et modèles classiques.

Il est apparu d'abord que les qualifications mécaniques spécifiées par le formalisme quantique, n'ont pu être définies qu'à partir de modèles classiques prolongés pour le cas des microétats par la voie d'éléments descriptionnels mathématiques et de définitions correspondantes d'opérations physiques de mesure. Il est apparu ensuite que les conditions-cadre générales *CCG*, parce qu'elles refusaient toute référence à un modèle donné, ont dû laisser un vide de spécification en ce qui concerne la nature de la grandeur qualifiante et, par voie de conséquence, en ce qui concerne la structure particulière de l'opération de mesure appropriée pour qualifier un microétat en termes de valeurs de telle ou telle grandeur qualifiante nommément. On sent que l'on est là en présence de deux symptômes de l'action d'un même fait conceptuel. Ce fait peut être exprimé par les remarques suivantes.

La structure des fonctionnements biologiques et psychiques de l'être humain, comme aussi la structure de ses intuitions et de sa pensée-et-langage, se sont forgées par des interactions directes avec du réel macroscopique, *via* les appareils biologiques neurosensoriels humains. Ces structures sont si profondément et inextricablement inscrites dans l'être même de chaque humain normal, qu'il est vain de vouloir les 'dépasser entièrement' par des entraînements de familiarisation avec des concepts et structures auxquels ont abouti des élaborations mathématiques ou méthodologiques réalisées dans les théories scientifiques modernes. Les injonctions qui exigent un tel dépassement total sont irréalistes. Il subsiste *nécessairement* un cordon ombilical plus ou moins évident qui rattache tout construit conceptuel au réseau d'intuitions et de conceptualisation issu de nos interactions sensorielles directes avec du réel macroscopique. Il n'est pas pensable de *qualifier* du non percevable d'une façon qui ne fasse intervenir strictement aucun élément induit par la conceptualisation naturelle macroscopique, et qui néanmoins soit *compréhensible*. On peut avoir des surprises totales face à des faits d'observation incompréhensibles qui apparaissent spontanément, comme lors de la découverte des rayons X. Mais toute qualification d'entités physiques inobservables qui est organisée *délibérément*, qu'elle soit recherchée en tant que données informatives primordiales (comme en mécanique quantique fondamentale) ou en tant que modélisation explicative à vérifier par la suite (comme dans la théorie des particules élémentaires), est organisée sur la base d'éléments tirés de la conceptualisation modélisante classique. Sinon cette qualification n'émerge pas.

### *4.3.3.3. Mot de conclusion sur la qualification de microétats*

Notons que la démarche développée ici, parce qu'elle part sur une base plus fondamentale que celle sur laquelle a été construite la mécanique quantique, et entièrement non restrictive en ce qui concerne le type de qualification souhaité, a permis de percevoir, par des contrastes, l'abîme franchi concernant le concept de propriété lorsqu'on décide de qualifier des microétats, d'une part, et d'autre part aussi la nécessité inamovible d'une relation avec la conceptualisation classique, si l'on veut conduire les qualifications jusqu'au bout et leur assigner une expression compréhensible.

### **4.4. De quelques définitions fondamentales**

Nous venons de parler d'opération de qualification par un acte de *Mes(X)* opéré sur un seul exemplaire d'un microétat $me_G$ engendré par une opération de génération d'état *G*, et dont l'effet, une fois décodé, se traduit en termes d'une valeur $Xj$ de la grandeur mesurée *X*. Cela implique de manière évidente qu'un seul acte de mesure sur un seul exemplaire du microétat étudié $me_G$, ne peut produire que l'enregistrement d'une seule valeur d'une grandeur dynamique définie pour un microétat. Voulons-nous admettre cette implication ? La réponse est *non*, car un micro-état n'est pas la même chose qu'un micro-système [21]. Et un micro-état peut concerner un ou plusieurs micro-systèmes, selon le substrat physique sur lequel a été appliquée l'opération de génération *G*.

En outre, nous avons signalé que selon le principe de composabilité des opérations de génération de microétats (cf. *3.2.4*), deux opérations de génération *G1* et *G2* qui peuvent opérer séparément sur deux exemplaires distincts du microétat initial utilisé comme matière première, peuvent toujours s'appliquer également toutes les deux sur un seul exemplaire de ce microétat. Comment compte-t-on ces divers 'un' et 'deux' ? Quelles présuppositions sont ici à impliquer afin de rester en accord avec l'emploi de ces termes pour accompagner les écritures du formalisme quantique?

Lorsqu'on considère cette question on arrive tout de suite à expliciter l'organisation conceptuelle suivante, que nous adoptons ici explicitement.

Soit donc ce qu'on appelle un micro-état. Ce micro-état comporte nécessairement quelque micro-système, ou plusieurs, dont il est le micro-état.

---

[21] Dans ce paragraphe, pour clarté de lecture, nous écrivons 'micro-système' et 'micro-état', avec tiret à l'intérieur. Ensuite nous reviendrons à l'écriture en bloc, sauf dans les cas ou l'on accentue de nouveau la distinction entre "système" et "état".

*Définitions [(un micro-système) et (un micro-état de un micro-système)]*

Soit un micro-état tel qu'une seule opération de mesure accomplie sur un seul exemplaire de ce micro-état ne peut produire qu'une seule valeur observable. On dira que ce micro-état met en jeu ce qu'on appelle un micro-système et que par conséquent c'est un micro-état de un micro-système.

*Définition [un microétat de n microsystèmes]*

Soient maintenant *n>1* micro-systèmes, c'est à dire deux ou plusieurs micro-systèmes dont on sait qu'on peut engendrer pour chacun séparément un micro-état au sens de la définition précédente. Mais soit *Gn* une seule opération de génération qui a engendré *à la fois* tous ces *n* micro-*systèmes* eux-mêmes, *avec* leurs *n* états[22] ; ou bien qui a engendré *à la fois* seulement tous les *n* micro-*états* de ces *n* micro-systèmes[23]. *Dans les deux cas on dira que le micro-état correspondant est un micro-état de n micro-systèmes, et nous le dénoterons par $me_{Gn}$*.

*Définition [mesure complète sur un micro-état de n micro-systèmes]*

Une opération de mesure *complète* accomplie sur un seul exemplaire d'un micro-état $me_{Gn}$ de *n* micro-systèmes, produit *n* valeurs observables distinctes, une valeur pour chaque micro-système impliqué, les grandeurs auxquelles ces valeurs sont liées pouvant être identiques ou différentes.

*Définition [mesure incomplète sur un micro-état de n micro-systèmes]*

Une opération de mesure accomplie sur un seul exemplaire du micro-état $me_{Gn}$ et qui produit moins de *n* valeurs observables, est *une mesure incomplète* sur $me_{Gn}$ (c'est-à-dire qui ne tire pas de $me_{Gn}$ une qualification pour chaque micro-système impliqué et par conséquent ne qualifie pas cet exemplaire de ce micro-état, par cette mesure, aussi complètement qu'on peut le concevoir conformément à la définition précédente).

*Remarque*

En conséquence des définitions explicitées plus haut, un acte de mesure *Mes(X)* opéré sur un seul exemplaire d'un micro-état $me_G$ de *n* micro-systèmes ne peut – par construction des concepts – produire que tout au plus *n* valeurs observables $X_j$. Cette conséquence (d'apparence triviale, mais qui est importante) ne peut être niée qu'en niant soit la pertinence de la conditions-cadre générale, soit celle des définitions explicitées plus haut.

*Définition [un micro-état à génération composée]*

Soit un micro-état de un micro-système ou de *n>1* micro-systèmes, indifféremment. Si ce micro-état a été engendré par une opération de génération *G(G1,G2,..Gk)* qui a composé – sur un *autre* micro-état de départ utilisé comme matière première afin de générer le micro-état considéré – les effets de deux ou plusieurs opérations de génération *G1,G2,..Gk* (où *k* est un entier) dont on sait que chacune aurait pu agir séparément, alors on dira que le micro-état considéré est *un micro-état à génération composée*[24].

---

[22] Il s'agit alors d'une 'création' au sens de la physique des particules.
[23] Il s'agit alors d'une 'interaction' au sens employé dans les exposés de mécanique quantique fondamentale.
[24] Cette dernière définition, déjà introduite avant pour le cas d'un micro-état d'un seul micro-système, place maintenant tout micro-état à génération composée (i.e. lié au principe de composabilité des opérations de génération), dans le contexte de l'ensemble des définitions de ce paragraphe 3.4.

**4.5. Du caractère *primordialement* statistique des qualifications 'mécaniques' d'un microétat**

Dans ce qui suit immédiatement nous supposerons qu'il s'agit d'un micro-état d'un seul micro-système. (Le cas d'un micro-état de $n>1$ micro-systèmes sera évoqué de nouveau plus tard).

Afin de placer désormais notre *langage* dans le même domaine de pertinence que celui auquel se réfère le formalisme de la mécanique quantique, nous parlerons constamment de grandeurs mécaniques. Mais ce cas particulier est soumis *exclusivement* à la condition-cadre générale qui laisse un vide de spécification par rapport aux traitements mathématiques qui, dans le formalisme quantique, ne concernent *que* des grandeurs mécaniques. Ainsi, en fait, les considérations qui suivent ne sont nullement restreintes à des grandeurs mécaniques, elles sont valides pour tout acte de qualification d'un microétat, par un aspect physique[25].

Considérons l'exemplaire du microétat $me_G$ produit par *une* seule réalisation d'une opération de génération $G$. Supposons qu'il est soumis à une opération de mesure d'une grandeur mécanique (dynamique) particulière bien précisée. Notons $B$ cette grandeur (afin de la distinguer clairement, aussi bien d'une grandeur définie mais quelconque qui est dénotée $X$, que d'un appareil de mesure qui est dénoté $A(X)$). Nous considérons donc un acte de *Mes(B)*. Pour éviter des restrictions arbitraires *a priori* il faut admettre que, en général tout au moins, l'acte de *Mes(B)* doit *changer* le microétat $me_G$ de départ, et de telle manière que l'on obtienne un groupe $\{\mu_k, k=1,2,...m)\}$ de $m$ marques physiques observables relevées sur les enregistreurs de l'appareil $A(B)$. Via le codage approprié, ces marques interviendront dans le calcul de l'une parmi les différentes valeurs possibles du spectre de $B$, disons $B4$. Au bout de cette séquence *[G.Mes(B)]* de réalisation d'une opération de génération $G$ suivie d'un acte de *Mes(B)* :

\* Les manifestations observées – *sur l'appareil* – qui signifient la valeur $B4$ de la grandeur $B$, incorporent une inamovible *relativité au processus Mes(B)* qui a permis d'obtenir ces manifestations.

\* L'exemplaire individuel d'un microétat $me_G$ qui a été soumis à l'acte de *Mes(B)* en général n'*existe plus* tel qu'il avait été engendré par l'opération de génération $G$. En général ce microétat de départ a été d'abord changé par l'évolution de *Mes(B)*, et en outre, souvent, sa transformée finale reste capturée dans l'un ou l'autre des objets macroscopiques qui constituent les enregistreurs de l'appareil $A(B)$.

Cette dernière circonstance oblige, si l'on veut *vérifier* le résultat $B4$, d'engendrer d'autres exemplaires du microétat $me_G$ et d'autres successions *[G.Mes(B)]* d'une opérations $G$ et un acte de *Mes(B)*. Or les sciences physiques accordent une importance majeure à la condition de vérifiabilité des résultats annoncés : c'est cette condition qui garantit la possibilité d'un consensus intersubjectif, sans quoi il n'y aurait pas ce qu'on appelle objectivité. Bref : afin de faire face à la condition centrale de vérifiabilité, il faudra faire usage de tout un ensemble de réalisations d'une succession *[G.Mes(B)]* et de l'entier *ensemble* correspondant d'exemplaires du microétat $me_G$.

Imaginons alors que l'on ait effectivement répétée un grand nombre de fois la succession *[G.Mes(B)]*. Si à chaque fois l'on retrouvait le résultat $B4$ que l'on avait trouvé la première fois, on se dirait : « j'ai trouvé une petite loi : si un microétat $me_G$ engendré par l'opération de génération $G$ est soumis à un acte de *Mes(B)*, l'on obtient le résultat $B4$ ».

---

[25] Les qualifications purement formelles-conceptuelles, comme celle de 'parité', échappent au domaine de pertinence des considérations qui suivent.

On pourrait se demander ensuite si *toute* mesure de toute autre grandeur *C*, *D*, etc., effectuée de manière répétée sur des exemplaires du microétat $me_G$, produit stablement une et même valeur de la grandeur mesurée, disons *C17* pour *C*, *D154* pour *D*, etc.. Et s'il s'avérait qu'effectivement c'est le cas, on se dirait : « j'ai trouvé une nouvelle loi plus importante que la précédente : un microétat $me_G$ introduit un groupe bien déterminé de valeurs observables des grandeurs *B,C,D,…* considérées, une valeur et une seule pour chacune de ces grandeurs mécaniques re-définie pour des microétats ».

Mais il se trouve qu'en fait les choses se passent autrement. Lorsqu'on répète une succession *[G.Mes(B)]* qui comporte un acte de *Mes(B)* accompli sur un microétat $me_G$ engendré par une opération de génération *G*, en général on n'obtient *pas* à chaque fois une même valeur de la grandeur *B*. En général – nonobstant le fait qu'à chaque fois il s'agit de la 'même' opération *G* et du 'même' acte de *Mes(B)*, on obtient une fois telle valeur de *B* et une autre fois telle autre valeur. Et lorsque le nombre d'essais s'accroît, l'ensemble des valeurs obtenues ainsi tend à couvrir progressivement tout le spectre *{B1, B2,....Bj,...}* de *B*. Et même s'il arrive que, pour le microétat $me_G$, ce soit à chaque fois la même valeur de *B* qui apparaît, disons *B4* – et l'expérience montre la possibilité d'un tel cas –, alors on trouve *toujours* d'autres grandeurs différentes de *B* pour lesquelles, face à $me_G$, les résultats sont dispersés : *aucun* microétat ne s'associe avec un ensemble de résultats de mesure qui soit dépourvu de dispersion pour *toutes* les grilles de qualification mécanique définies pour des microétats : si un microétat $me_G$ donné est tel qu'il conduit à un ensemble de résultats de mesure dépourvu de dispersion pour une parmi ces grilles – ce qui n'est pas le cas général – alors il existe *toujours* d'autres grilles de qualification pour lesquelles $me_G$ conduit à un ensemble dispersé de résultats de mesure. Ceci est un fait d'expérience, une donnée factuelle. Il n'y a pas de stabilité *générale* de la valeur produite par des mesures opérées sur un microétat.

Dans ces conditions, il est clair d'emblée qu'une valeur donnée du spectre de la grandeur *B*, disons *B4*, peut apparaître, par un acte de *Mes(B)*, pour une infinité de microétats différents produits par des opérations de génération différentes. Une valeur d'une grandeur *X* n'est donc jamais spécifique à un microétat donné.

*Ainsi les faits nous éjectent sur un niveau statistique*.

Or le caractère statistique auquel on est confronté ici est *primordial*, en *ce* sens qu'on ne peut pas l'assigner à quelque ignorance. Car la structure de connaissances concernant des microétats dont on surveille la genèse *ici*, émerge par construction *foncièrement première*. Elle émerge d'un inconnu supposé être total ; elle émerge en dehors de – sans aucun lien avec – la pensée classique pré-constituée concernant les déplacements de 'mobiles', avec ses modèles et ses postulats spécifiques, notamment le postulat déterministe corrélatif d'une formulation *de base* des lois de mouvement qui est marquée d'un caractère individuel (non statistique) et de certitude présupposée.

### 4.6. La forme qualitative de la description d'un microétat

Ce n'est donc que sur un niveau statistique qu'on peut encore rechercher un invariant observationnel lié à un microétat donné. Or sans invariants il n'y a pas de lois, pas de connaissance générale, pas de science prévisionnelle. Comment procéder ?

De ce qui vient d'être dit il découle que pour chaque paire *(G,X)* il faudra, pour tout *X*, accomplir un grand nombre *N* de fois la succession d'opérations *[G.Mes(X)]* correspondante, enregistrer à chaque fois la valeur $X_j$ du spectre *{X1,X2,…Xj......}* de *X* qui a été obtenue, et rechercher un invariant impliquant l'ensemble des *N* valeurs obtenues. Ce sera alors forcément un invariant non individuel ; et si c'est un invariant de la statistique constatée ce sera un invariant probabiliste, car on ne connaît pas une autre sorte d'invariant sur le niveau de conceptualisation statistique.

Soit donc un grand nombre *N* de répétitions de la succession *[G.Mes(X)]*. Soit *{n(G,X1)/N, n(G,X2)/N,…n(G,Xj)N…… n(G,Xk)/N}* l'ensemble des fréquences relatives enregistrées (où : *n(G,Xj)* désigne le nombre des réalisations de la valeur *X(j)* pour le microétat *me$_G$*, *N* est le nombre total d'essais faits, et *n(G,Xj)N* est la fréquence relative du résultat *Xj* parmi les résultats des *N* essais *[G.Mes(X)]* accomplis). Cet ensemble de fréquences relatives est 'la distribution statistique' des *Xj*.

On dira que la situation s'avère être *probabiliste* si et seulement si, lorsqu'on mesure la distribution statistique *{n(G,X1)/N, n(G,X2)/N,…n(G,Xj)N.... n(G,Xk)/N}}* pour des nombres d'essais *N* qui s'accroissent autant qu'on veut, l'on constate que ces distributions *convergent* vers une distribution limite idéale. Celle-ci – si elle se manifeste – sera dénommée une *distribution de probabilité* et on dira qu'elle constitue *la loi factuelle de probabilité*

*{p(G,Xj)}={ p(G,X1, p(G,X2), …. p(G,Xj) …… p(G,Xk)}*

où *p(G,Xj)=lim.N→∞ [(n(G,Xj)/N]* dénote par définition la limite de convergence de la fréquence relative *n(G,Xj)/N*[26], supposée exister.

### *4.6.1. Spécificité face à me$_G$ d'une loi de probabilité p(G,Xj)* versus *grandeurs 'incompatibles'*

#### *4.6.1.1. Questions*

Lorsqu'on dit qu'on recherche la description du microétat *me$_G$* il est sous-entendu que l'on cherche un ensemble de qualifications de *me$_G$* qui soit spécifique à *me$_G$*, c'est-à-dire tel qu'aucun autre microétat généré par une opération de génération différente de celle, *G*, qui produit *me$_G$*, ne puisse faire apparaître exactement le même ensemble de qualifications. Ce qu'on cherche en fait ainsi est une re-définition de *me$_G$* qui soit à la fois conceptualisée et vérifiable par l'expérience et qui puisse *in fine* remplacer sa 'définition a-conceptuelle' de départ par l'opération *G* correspondante, qui ne nous donne rien à connaître concernant spécifiquement le microétat *me$_G$*.

Or comment savoir si la loi de probabilité *{p(G,Xj)}* obtenue est spécifique du microétat étudié *me$_G$* ?

*Si* elle l'était, alors d'ores et déjà, à elle seule, elle pourrait peut-être être regardée comme une 'description' du microétat *me$_G$*, nonobstant le fait qu'elle ne concerne pas *me$_G$* isolément, mais seulement les interactions de *me$_G$* avec l'appareil *A(X)* utilisé pour accomplir les *Mes(X)*. Mais si, même avec cette réserve de taille, cette distribution de probabilité n'est pas spécifique au microétat *me$_G$*, alors il est clair d'emblée qu'en aucun cas il ne s'agit là d'une 'description' de *me$_G$* : l'appeler ainsi trahirait radicalement la définition du concept. Or aucun fait ni aucun argument n'entraîne avec nécessité qu'une distribution de probabilité *p(Xj)* établie pour une paire *(G,X)* donnée (i.e. pour un microétat *me$_G$* et avec une grandeur *X* donnés), ne peut se réaliser également pour un paire *(G',X)* où *G'≠G*, c'est à dire avec le même *X* mais pour un autre microétat *me$_{G'}$*≠ *me$_G$*. Il faut don chercher un autre critère de spécificité observationnelle de *me$_G$*.

---

[26] Ici, provisoirement, j'emploie les manières courantes de parler. Mais en fait les concepts de 'convergence probabiliste' et de 'loi factuelle de probabilité' soulèvent des problèmes conceptuels très sérieux. Ceux-ci sont exposés et traités dans Mugur-Schächter [2006].

Cela conduit à la nouvelle question suivante. Que se passe-t-il si l'on considère deux grandeurs dynamiques *X* et *Y* différentes, au lieu d'une seule? Ne pourrait-on pas, à l'aide d'une paire de deux lois de probabilités *{p(G,Xj)}* et *{p(G,Yq)}* définir avec certitude une spécificité observationnelle face à *me$_G$* ? Ou bien, puisque peu de choses peuvent être certaines lorsque d'emblée il s'agit de probabilités, du moins agrandir la chance *a priori* d'avoir identifié une spécificité observable de *me$_G$* ?

*4.6.1.2. Incompatibilité ou compatibilité mutuelle de deux aspects de qualification d'un microétat*[27].

Quand il s'agit de microétats, deux grandeurs *X* et *Y* ne sont vraiment 'différentes' que lorsqu'elles sont mutuellement incompatibles, c'est-à-dire, lorsqu'elles s'excluent mutuellement en ce sens précis *qu'un acte de mesure de Mes(X) est conçu comme changeant le microétat étudié autrement qu'un acte de mesure de Mes(Y)*. Ce qui évidemment est impossible à réaliser sur *un* seul *exemplaire* du microétat à étudier. Deux grandeurs qui sont incompatibles au sens spécifié, ne peuvent donc pas être mesurées par un même acte de mesure opéré sur un et même exemplaire du microétat à étudier, si celui-ci – comme il est supposé ici – est un microétat d'un seul microsystème[28, 29].

Si deux grandeurs mécaniques *X* et *Y* ne sont *pas* incompatibles au sens spécifié, alors, par opposition aux conditions impliquées par la définition d'incompatibilité mutuelle, le mode de changement de changement d'un seul exemplaire du microétat étudié *me$_G$*, qui se réalise lors d'une interaction de *Mes(X)*, *peut* (avec un choix convenable d'appareil) être le même que celui qui se réaliserait pour ce même exemplaire lors d'une interaction de *Mes(Y)* : c'est la définition de la compatibilité mutuelle de deux grandeurs qualifiantes définies pour des microétats. Donc en cas de compatibilité, rien n'empêche de procéder de la façon suivante : un seul exemplaire du microétat à étudier est soumis au type unique de changement qui convient – à la fois – comme processus de changement par un acte de *Mes(X)* et comme processus de changement par un acte de *Mes(Y)*. On peut donc introduire une nouvelle notation : *un acte de Mes(XY)*. Cette unique interaction de *Mes(XY)* ne peut évidemment produire qu'un unique ensemble de marques physiques observables sur les enregistreurs de l'appareil *A(XY)*. Mais l'unicité de cet ensemble de marques physiques observées permet néanmoins de qualifier le microétat étudié, à la fois, par une valeur correspondante *Xj* de *X* et par une autre valeur correspondante *Yq* de *Y*. En effet, ces deux valeurs peuvent, elles, être distinguées l'une de l'autre, si la définition conceptuelle de la grandeur *X* introduit un codage en termes d'une valeur *Xj* de *X* qui est différent du codage introduit par la définition conceptuelle de la grandeur *Y*, mais qui est un codage *relié* à celui-ci, calculable à partir de celui-ci[30].

---

[27] Les considérations qui suivent ne s'appliquent qu'au cas d'un micro-état de un micro-système (cf. *3.4* et *3.9.1*).

[28] C'est l'essence de ce qui, en mécanique quantique, est lié au *principe de complémentarité*. Mais souvent (sinon toujours) la condition d'unicité de *l'exemplaire* du microétat de *un* microsystème que l'on considère, est oubliée lorsqu'on parle de complémentarité quantique. Cela conduit à beaucoup de confusions car c'est précisément cette condition d'unicité de l'exemplaire mis en jeu lors d'un seul acte de mesure, qui est essentielle (et qui en outre est douée d'universalité, comme il apparaîtra dans le dernier chapitre).

[29] Il apparaîtra dans le paragraphe *3.9.1* que cette définition de l'incompatibilité mutuelle de deux grandeurs définies pour un microétat, posée ici pour le cas d'un micro-état de *un* micro-système, *cesse* d'être valide dans le cas d'un micro-état de *deux ou plusieurs* microsystèmes : ce qui décide finalement de la 'compatibilité', est la possibilité, ou non, de mesurer les grandeurs considérées *simultanément sur un seul exemplaire du micro-état considéré*.

[30] Par exemple, imaginons que *X* est la re-définition conceptuelle pour le cas d'un microétat, de la grandeur classique 'quantité de mouvement' selon une seule dimension d'espace (dont en mécanique classique la valeur s'écrit alors *p=mv* où *m* est la masse du mobile et *v* dénote sa vitesse selon la dimension d'espace considérée), cependant que *Y* est la re-définition conceptuelle de la grandeur classique 'énergie cinétique' selon une seule dimension d'espace (dont en mécanique classique la valeur s'écrit $T=(p^2/2m)=(mv^2/2)$). En transposant ces définitions conceptuelles classiques, dans les termes du formalisme quantique (sur la base d'une construction abstraite qui prolonge sans l'avouer les modèles de la mécanique classiques), leur dépendance conceptuelle et formelle se maintient et elle conserve sa forme. En ces conditions il est évident que selon le formalisme quantique il suffit qu'un acte de *Mes(XY)* produise, à partir d'un seul exemplaire du microétat étudié, un unique ensemble de manifestations physiques observables commun à *X* et à *Y*, puisque le codage comporté par la re-définition conceptuelle de *X* permet d'associer à cet ensemble de marques un *sens* en termes d'une valeur de la quantité de mouvement *p*, cependant que le codage comporté par *Y* consiste à simplement calculer ensuite aussi un autre sens, énergétique, $T=(p^2/2m)$, en effectuant le

Ceci revient à dire que, dans ces conditions, on peut considérer que les noms '*X*' et '*Y*' ne désignent en fait *pas* deux dimensions de qualification qui sont distinctes *physiquement*. Qu'ils ne désignent que deux utilisations conceptuelles-formelles différentes mais corrélées, d'une seule dimension physique de qualification. Disons, en employant une image, que d'un point de vue physique opérationnel les deux grandeurs compatibles qui interviennent dans *Mes(XY)* peuvent être regardées comme deux *'directions de qualification colinéaires'* que l'on peut superposer dans une seule dimension de qualification.

### *4.6.2. Spécificité probabiliste face au microétat étudié :*
### *la forme qualitative de la description d'un microétat*

Il ne semble pas impensable que l'on obtiennent la même loi de probabilité *{p(G,Xj)}* pour deux microétats distincts (produits par deux opérations de génération différentes). Cela ne semble pas pouvoir être fréquent. Néanmoins on aurait des réticences à en affirmer l'impossibilité. Si alors, afin d'agrandir le degré de vraisemblance d'avoir établi une spécificité de *me$_G$*, on considéré deux grandeurs qualifiantes au lieu d'une seule, il résulte des considérations du paragraphe précédent que l'effet recherché ne se produira pas si les deux grandeurs choisies sont mutuellement compatibles, car dans ce cas elles n'introduisent qu'une seule dimension physique de qualification.

Par contre, sur la base de ces considérations il paraît suffisamment sûr d'admettre que deux groupes non compatibles de grandeurs mutuellement compatibles, agissent comme deux 'directions de qualification' distinctes qui, en s''intersectant', fournissent une spécificité du microétat étudié *me$_G$* ; c'est à dire, qu'aucun autre microétat engendré à l'aide d'une autre opération de génération différente de l'opération *G*, ne conduit exactement au même couple de deux lois de probabilité liées à ces deux grandeurs non compatibles, que le couple de lois trouvées avec *G* et *me$_G$*.

D'autant plus, l'ensemble de *toutes* les lois de probabilité obtenues avec une opération *G* fixée et *tous* les groupes non compatibles de grandeurs compatibles qui sont re-définies pour les microétats, peut être tranquillement considéré comme spécifique à *me$_G$*. En *ce* sens, et en ce sens seulement, il semble possible de poser que l'ensemble de lois de probabilité évoqué – bien qu'il ne concerne pas le microétat *me$_G$* isolé des interactions de mesures qui l'ont engendré – constitue néanmoins *la 'description'* (la re-définition conceptualisée) communicable, vérifiable et consensuelle, 'de' *me$_G$* (cependant que l'opération de génération *G* ne faisait que singulariser le microétat correspondant *me$_G$* de l'intérieur du réel physique, d'une manière purement factuelle, a-conceptuelle, ne comportant aucune connaissance liée à *me$_G$* spécifiquement).

La description que nous venons d'obtenir a une forme *foncièrement relative* à :

* l'opération de génération *G* qui agit dans toutes les successions réitérées *[G.Mes(X)]*, pour tout *X* ;
* l'effet *me$_G$* de l'opération *G* ;
* la grille de qualification constituée par l'*ensemble* des grandeurs mécaniques redéfinies pour des microétats, en conformité avec la condition-cadre *CCG≡[gqX(me]*.

---

carré de la valeur de *p* fournie par le codage lié à *X*, et en divisant ensuite ce carré par *2m*. Bref, on peut procéder selon la méthode 'time of flight' pour obtenir les marques physiques observables, et ensuite, faire les deux calculs mentionnés. On voit sur cet exemple pourquoi le changement, l'évolution physique imposée à l'exemplaire considéré du microétat étudié par une interaction de mesure *X*, peut être le même que le changement imposé par une interaction de mesure *Y*, si *X* et *Y* sont 'compatibles' : dire que deux grandeurs *X* et *Y* sont 'compatibles' ne veut dire *que* ceci, précisément, que *X* et *Y* ne diffèrent l'une de l'autre que conceptuellement, par la manière d'associer à un ensemble donné *unique* de marques physiques observées, les deux valeurs *Xj* et *Yq* différentes qui correspondent à cet ensemble unique selon deux codages différents mais reliés que l'on peut transformer l'un dans l'autre.

Posons le langage qui suit. Appelons *vue mécanique* la grille de qualification mécanique d'un microétat constituée par *l'ensemble $\{V_X\} \equiv V_M$* de toutes les *vues-aspect mécanique* $V_X$ correspondant chacune à une seule grandeur mécanique $X$ et où $V_X$ est une renotation de *Mes(X)*. On peut alors indiquer de manière synthétique la *description* mécanique du microétat $me_G$, par le symbole

$$D_M/G, me_G, V_M/ = \{p(G,Xj)\}$$

qui désigne l'ensemble *{p(G,Xj)}* de toutes les lois de probabilité obtenues pour $me_G$ et pour toutes les grandeurs mécaniques $X$ redéfinie pour des microétats. Si en particulier l'on a $V_M \equiv V_X$ l'on dénotera la description correspondante par $D_M/G, me_G, V_X/=p(G,Xj)$ et l'on dira qu'elle ne qualifie l'entité-objet $me_G$ que face à l'unique vue-aspect mécanique $V_X$. On pourra donc écrire :

$$D_M/G, me_G, V_M/ \equiv \{D_M/G, me_G, V_X/\}$$

Telle est la forme – *qualitative* et *générale* – des descriptions de microétats (notamment mécaniques) à laquelle conduit la démarche développée ici. C'est une forme descriptionnelle purifiée de toute adhérence d'éléments mathématiques cryptiques, et *intégrée*. La genèse de cette forme decriptionnelle – rappelée dans son symbole $D_M/G, me_G, V_M/$ – met en évidence explicitement toutes les trois relativités mentionnées plus haut[31]. Cette genèse a été exposée à tous les regards et par cette voie la fonctionnalité de chaque élément descriptionnel se trouve définie, aussi bien en ce qui concerne sa place dans la succession génétique, que par son contenu et son effet propres.

Les descriptions de microétats offertes par le formalisme quantique consistent elles aussi précisément en l'ensemble *{p(G,Xj)}* des lois de probabilité qui correspond à un microétat $me_G$ donné et à l'ensemble des grandeurs mécaniques (représentées chacune par un opérateur différentiel)[32]. Mais *dans le formalisme quantique la forme intégrée d'une telle description n'est spécifiée nulle part*, et la *genèse* opérationnelle-conceptuelle de la description reste cachée. Chacun s'en fait une intuition *via* la manipulation des algorithmes ; ceux-ci permettent de définir chaque élément de l'ensemble *{p(G,Xj)}* en combinant 'le ket d'état associé au microétat étudié', avec d'autres éléments du formalisme.

D'autre part, nous l'avons déjà souligé, les descriptions de la mécanique quantique sont toutes individualisées par la définition mathématique exacte de la grandeur mécanique $X$ qui intervient et de l'opération de *Mes(X)* correspondante. La relation entre la forme descriptionnelle générale $D_M/G, me_G, V_M/$ définie ici et la forme descriptionnelle impliquée dans les algorithmes de la mécanique quantique, est comparable à la relation entre un buste antique où les yeux sont vides d'un regard individualisé, et la tête de l'homme qui a servi de modèle à ce buste. Mais cela n'altère pas le fait que les algorithmes quantiques pointent, sur le plan qualitatif, vers ce *même* type descriptionnel : le type de forme descriptionnelle impliqué dans le formalisme quantique, est identifié.

---

[31] Au premier abord on pourrait douter de la nécessité de mentionner séparément la relativité à $me_G$, étant donné que l'on a déjà affirmé la relativité à l'opération de génération $G$, et que, par décision méthodologique, $G$ et $me_G$ sont posés être en relation de un-à-un. Mais la nécessité de considérer la relativité à $me_G$ indépendamment de celle à $G$, s'impose clairement dès qu'on réfléchit à la question (Mugur-Schächter [2006]). En effet, les résultats des actes de *Mes(X)* ne dépendent - directement - que de $me_G$ déjà réalisé, cependant que $me_G$ ne dépend directement que de $G$. On ne peut économiser aucune des trois relativités fondamentales de la description quantique $D_M/G, me_G, V_M/$ d'un microétat : à $G$, à $me_G$, à la vue qualifiante $V_M$.
[32] En effet un œil averti discerne tout de suite dans *le désigné {p(G,Xj)}* du symbole $D_M/G, me_G, V_M/$, le contenu de la description quantique d'un microétat. Chacun des algorithmes mathématiques de la mécanique quantique incorpore rigidement un *fragment* de cette forme descriptionnelle (tout en comblant par des choix implicitement *modélisants* le vide que, dans la construction qualitative accomplie ici, la condition-cadre générale laisse *subsister* en ce qui concerne la représentation formelle des grandeurs mécaniques dans des termes appropriés à la qualification d'un microétat).

Ainsi, d'ors et déjà, l'hypothèse faite au départ selon laquelle l'essence épistémologique du formalisme quantique serait déterminée par la situation cognitive dans laquelle on se trouve lorsqu'on veut décrire des microétats, semble se confirmer. Il reste toutefois à parfaire cette confirmation.

En premier lieu, il n'y a pas de doute qu'il convient d'affirmer qu'une description $D_M/G,me_G,V_M/$) est *probabiliste*, puisqu'elle ne consiste en rien d'autre qu'un ensemble de lois de probabilité. Mais elle est probabiliste en un sens nouveau et il importe d'expliciter à fond en quoi cette nouveauté consiste.

En second lieu, la genèse d'une description $D_M/G,me_G,V_M/$ a été indiquée d'une manière très étalée, argumentée pas à pas : l'on ressent le besoin d'une reformulation plus synthétique de cette genèse, qui permette de percevoir en un seul regard la structure globale de ses spécificités, de façon à pouvoir jauger les spécificités et les perspectives nouvelles que celles-ci ouvrent. L'examen de l'organisation d'espace-temps des descriptions quantiques conduit précisément à une telle reformulation synthétique.

### 4.7. Problèmes concernant la forme descriptionnelle $D_M/G,me_G,V_M/$ et réponses

Dans ce qui suit[33] on assistera à un processus remarquable. On percevra en pleine lumière la genèse et la nature profonde de quelques problèmes fondamentaux que le formalisme quantique suscite depuis sa constitution : le problème de complétude, le problème ontologique, le sens exact du qualificatif 'essentiel' que l'on associe aux probabilités qui concernent des microétats et la question de la relation entre celles-ci et le postulat de déterminisme, la question de 'la coupure quantique-classique' et celle, corrélative, de la modélisation des microétats. Pour des raisons historiques, les problèmes énumérés ont été perçus en relation inextricable avec le formalisme mathématique de la mécanique quantique – en tant que problèmes d' 'interprétation' *de ce formalisme* – et selon cette optique ils perdurent à ce jour.

Or il apparaîtra que face à la démarche purement qualitative développée ici, ces problèmes émergent de nouveau, indépendamment du formalisme quantique et pourtant dans des termes *quasi identiques* à ceux qui se sont imposés relativement au formalisme. Il apparaîtra également que lorsqu'ils sont rapportés à la forme descriptionnelle $D_M /G,me_G,V_M/$ ces problèmes *s'élucident* d'une façon simple et évidente que l'on pourrait qualifier de directe en ce sens que la solution découle, sans intermédiaire, de la genèse de la forme $D_M /G,me_G,V_M/$.

#### 4.7.1. Le désigné du symbole $D_M/G,me_G,V_M/$ peut-il être regardé comme une "description" du *microétat $me_G$* ?

On vient de voir que la connaissance qu'il a été possible de construire concernant le comportement mécanique d'un microétat, ne consiste en général pas en descriptions individuelle*s*. Elle consiste en lois de probabilité $p(G,X_j)$ concernant l'émergence des valeurs $X_j$ de telle ou telle observable dynamique *X*. En outre, les manifestations observables que ces lois concernent ne peuvent *pas* être regardées comme étant liées à des propriétés que le microétat étudié $me_G$ engendré par l'opération de génération *G* aurait 'possédées' d'emblée, avant toute évolution de mesure, d'une façon déjà actuelle, réalisée, et réalisée pour lui isolément, de façon indépendante de tout acte d'observation. Ces lois de probabilité *{p(X$_j$)}* n'offrent *aucun* renseignement concernant la façon d'être du microétat *me$_G$* lui-même, indépendamment de nos actions cognitives sur lui. Il en est ainsi à tel point qu'il est même possible de reformuler la représentation construite plus haut, en termes strictement opérationnels-observationnels-prévisionnels : une fois qu'une description $D_M$/$G,me_G,V_X$/ incluse dans l'ensemble {$D_M$/$G,me_G,V_X$//} symbolisé $D_M$/$G,me_G,V_M$/, a été établie, si d'abord l'on opère de la façon dénotée *G* et ensuite l'on opère de la façon dénotée *Mes(X)*, on sait à l'avance qu'on a telle probabilité *p(G,X$_j$)* d'observer telle groupe de manifestations physiques *de l'appareil employé*, codés *X$_j$* en termes conceptualisés. En ces conditions peut-on affirmer que $D_M$/$G,me_G,V_X$/ et $D_M$/$G,me_G,V_M$/ sont des descriptions *du microétat me$_G$* lui-même? Car on peut se dire ce qui suit.

---

[33] Certains passages de ce paragraphe interviennent aussi dans Mugur-Schächter [2006] mais dans d'autres contextes.

« Penser à la manière d'exister du microétat *me$_G$* lui-même n'est qu'une intrusion philosophique dans la pensée et le discours scientifique. L'ensemble des lois de probabilité *p(G,X$_j$)* liées aux observables dynamiques *X* et associées à une même opération de génération *G* donnée, constituent un invariant opérationnel-observationnel-prévisionnel relatif à *G* et aux processus de *Mes(X)* et qui est spécifique à *G*, et cela *suffit*. On peut même, à la limite, se débarrasser en fin de parcours de toute trace de pensée hypothétique, comme on se débarrasse de tous les éléments de l'échafaudage quand la bâtisse est achevée. Même l'expression 'le microétat correspondant à l'opération de génération *G* ' peut être regardée comme un simple appui verbal qui est utile mais qu'il faut se garder de réifier. On se retrouve finalement devant une sorte de pont entre, d'une part des opérations physiques, et d'autre part des observations codées en termes de valeurs d'une grandeur mécanique et des prévisions probabilistes tirées de ces observations. Le microétat étudié ne fait que hanter cette construction comme un fantôme inutile. En *tout* cas, il ne s'agit nullement d'une description de ce microétat lui-même ».

Ce problème a plusieurs visages que je vais maintenant spécifier à tour de rôle. D'abord je caractériserai les aspects que, face $D_M$/$G,me_G,V_M$/, l'on peut désigner par les mêmes dénominations déjà employées en relation avec le formalisme mathématique : le problème de *complétude* et le problème *ontologique*.

Une fois que ceci aura été accompli, je montrerai qu'à l'intérieur de la démarche développée ici *la décision méthodologique d'affirmer la relation de un-à-un G↔me$_G$, élimine a priori ces deux problèmes*. Cela permettra de percevoir clairement aussi les réponses à autre deux questions qui, face au formalisme quantique, ont soulevé un grand nombre de débats, à savoir la question de la nature des probabilités quantiques, et la question d'une définition claire de 'la coupure quantique-classique'.

Au bout de ce cheminement, une solution globale à tous les questionnements abordés, apparaîtra avec évidence : dans tous les cas examinés, uniformément, il s'agit d'incompréhensions engendrées par le fait que la forme descriptionnelle $D_M$/$G,me_G,V_M$/ est ignorée, donc aussi la genèse de cette forme. En conséquence de cela, sans s'en rendre compte, l'on soumet les descriptions de microétats à des exigences de la conceptualisation *classique* qui tout simplement n'ont pas de *sens* face à des descriptions du type désigné par le symbole $D_M$/$G,me_G,V_M$/. L'on fabrique ainsi des questions illusoires dans lesquelles on s'enlise.

*4.7.1.1. Le problème de 'complétude' de la forme descriptionnelle $D_M$/$G,me_G,V_M$/*

Le débat sur la 'complétude' *du formalisme quantique* – pas sur celle de la forme descriptionnelle qualitative $D_M$/$G,me_G,V_M$/) – a conduit à des *théorèmes d'impossibilité* dont les plus importants sont le théorème de von Neumann affirmant l'impossibilité de paramètres cachés compatibles avec le formalisme quantique (J. von Neumann [1955]) et le théorème de Wigner affirmant l'impossibilité de définir une probabilité conjointe de position et de quantité de mouvement qui soit compatible avec le formalisme quantique (E.P. Wigner [1971]). Ces théorèmes semblaient trancher ce qu'on avait dénommé le problème d'incomplétude de la mécanique quantique, sans pour autant le faire comprendre[34]. En fait, l'un comme l'autre de ces deux théorèmes faisaient usage, dans la démonstration, du formalisme quantique lui-même, ce qui est circulaire. Finalement ces deux théorèmes ont été invalidés (Mugur-Schächter [1964], Bell [1966], Mugur-Schächter [1977], Mugur-Schächter [1979]). Mais les invalidations ne font que remettre en circulation le problème d'une représentation plus 'complète' des microsystèmes 'eux-mêmes'. Elles n'éliminent nullement ce problème.

Or le contenu du paragraphe précédent fait pressentir que ce problème de 'complétude' renaît également face à la forme descriptionnelle qualitative $D_M/G,me_G,V_M/$ qui a émergé à l'*extérieur* du formalisme quantique ainsi que de toute autre formalisation préexistante. Cela suggère que les racines du problème de complétude se trouvent *sous* le formalisme quantique, dans la situation cognitive même qui y est impliquée. En effet lorsqu'on reconsidère la forme descriptionnelle $D_M/G,me_G,V_M/$, on est porté à revenir sur sa genèse dans un état d'esprit critique. On se dit :

« Il est clair que si l'on veut se mettre en possession d'une qualification du microétat $me_G$, en termes d'une valeur $Xj$ d'une grandeur mécanique $X$ redéfinie pour des microétats, il faut réaliser une succession *[G.Mes(X)]* où une réalisation de $G$ soit immédiatement suivie d'un acte *Mes(X)* de mesure de $X$ effectué sur l'exemplaire de $me_G$ engendré par cette réalisation de $G$. Mais si, afin de vérifier qu'on a bien procédé, l'on répète la réalisation d'une succession *[G.Mes(X)]*, en général on ne retrouve plus la même valeur $Xj$ la grandeur $X$. Ceci est une donnée expérimentale indéniable qu'on a exprimée en disant que 'la situation est statistique'. Donc afin d'avoir une chance d'accéder à une caractérisation invariante de $me_G$ en termes de valeurs $Xj$ de $X$ (alors nécessairement probabiliste), on doit réaliser un très grand nombre de répétitions de la succession *[G.Mes(X)]*. Or après l'enregistrement d'une valeur $Xj$ de $X$, l'exemplaire du microétat $me_G$ qui avait été mis en jeu par la réalisation de $G$, n'existe plus ; il a transmuté en un *autre* microétat. Donc *chaque séquence [G.Mes(X)] brise la continuité du processus global de constitution de la connaissance que l'on acquiert sur le microétat $me_G$*. Ce microétat doit être recréé pour chaque nouvel acte de mesure. Si je travaillais avec un dé macroscopique je pourrais réutiliser indéfiniment le mêmes dé, sans avoir à le recréer à chaque fois, et le processus de constitution d'une connaissance probabiliste ne serait pas brisé ainsi, ce serait clairement un tout. Car je pourrais concevoir que la dispersion statistique n'est due qu'à la non identité des jets, d'ailleurs permise à l'avance. Tandis que dans le cas d'un microétat il apparaît un problème qui est grave : comment peut-on être certain que l'opération de génération $G$ recrée vraiment le *même* microétat à chaque fois qu'elle est réalisée, comme on l'a admis – peut-être trop vite – par la décision méthodologique de poser une relation *de un-à-un*, $G \leftrightarrow me_G$ ? Car c'est plutôt du contraire de ce que pose cette relation qu'on se sent incliné à être certain. En effet l'opération $G$ est définie par des paramètres macroscopiques dont il est impossible par principe de dominer tous les aspects microscopiques. Donc ce qui selon les paramètres macroscopiques semble être un ensemble de répétitions de la même opération $G$, en fait, au niveau microscopique, est sans aucun doute tout un ensemble d'opérations de génération mutuellement différentes, qui engendrent tout un ensemble de microétats eux aussi mutuellement différents. En plus, un processus de *Mes(X)* est lui aussi défini seulement à l'aide de paramètres macroscopiques en dessous desquels se cache sans doute tout un ensemble de réalisations microscopiques différentes de ce processus. Dans ces conditions, la description des microétats que nous avons élaborée est certainement *incomplète* car elle escamote les différences qui existent entre les exemplaires de microétats engendrés par les répétitions d'une opération de génération *G*, comme elle escamote également les différences qui existent entre les réalisations distinctes de l'opération *G* elle-même, et celles qui existent entre les réalisations distinctes d''une' *Mes(X)*. On affirme des répétitions 'identiques' des séquences *[G.Mes(X)]*, mais en fait celles-ci sont fluctuantes dans *tous* leurs éléments, dans l'élément *G*, l'effet $me_G$ de *G*, et l'acte de *Mes(X)*. Nous avons partout indiqué fallacieusement par un symbole invariant, tout un *ensemble* caché d'entités physiques mutuellement distinctes. Nous aurions dû introduire explicitement tout un ensemble d'opérations de génération différentes au niveau microscopique mais correspondant toutes à un même groupement de paramètres macroscopiques, et étiqueter *cela* par le symbole *G*. Et à l'intérieur de cet ensemble il aurait fallu symboliser les différences, les dénoter. Et de même pour $me_G$ et pour *Mes(X)*. Ce n'est que de cette façon qu'on aurait pu espérer de construire une description véritablement complète des microétats »[35].

---

[34] Les conclusions y étaient exprimées en termes absolus et définitifs, comme s'il était concevable de déduire une impossibilité absolue et définitive ! (A l'intérieur de quel système formalisé ? Un système qui, dès tel moment donné, établirait *tout* ce qui est possible, à jamais ?).

Toutefois, parvenus en ce point on peut se troubler car on peut se dire également :

« Mais même de cette façon, ce qu'on obtiendrait finalement ne serait toujours pas une description des microétats eux-mêmes ! Car malgré toutes les précautions conceptuelles et notationnelles mentionnées, nous n'apprendrions finalement toujours rien, strictement rien, sur les microétats eux-mêmes. Tout ce que nous apprendrions concernerait toujours seulement les manifestations observables produites par des processus de mesure. Car lors d'une succession *[G.Mes(X)]* donnée nous ne saurons pas comment *choisir*, dans les ensembles microscopiques que nous venons de concevoir, la variante microscopique de l'opération de génération d'un microétat qui s'est réalisé effectivement, ni la variante microscopique de microétat qui s'est réalisée, ni la variante microscopique d'un acte de mesure *Mes(X)* qui s'est réalisée en fait. Non, il n'y a rien à faire. D'une part, toute la façon de parler et de noter qui a été développée – avec des singuliers partout, ' l' 'opération de génération *G*', *'le'* microétat *'correspondant'*, *'le'* processus de *Mes(X)* – induit tout simplement en erreur. Elle masque l'incomplétude de l'approche. Et d'autre part, si l'on supprime ce masque, si à la place de ces singuliers qui faussent on introduit partout les ensembles microscopiques qui s'imposent à la raison, on reste bloqué dans une impossibilité de choix en conséquence de laquelle toutes les distinctions imaginées ne servent à rien. *Nous sommes condamnés à l'ignorance*. Dans ce sens-là la description accomplie est effectivement 'complète'. Nous sommes piégé dans la même difficulté que celle qui nous nargue dans le cas du formalisme quantique. Et la complétude qu'affirme l'orthodoxie concernant le formalisme quantique est bien vraie, puisqu'elle se reproduit au niveau de la démarche sous-jacente pratiquée ici. C'est cette conclusion qui s'impose : même si cette complétude n'est pas démontrable déductivement, elle est néanmoins *vraie*. Mais cette sorte de 'complétude' est une prison insupportable ! Il faut absolument trouver le moyen d'en sortir et d'accomplir une représentation vraiment complète des microétats ».

Voilà l'essence du discours qui naît dans les esprits, et comment l'assertion de 'complétude' de la forme descriptionnelle $D_M/G, me_G, V_M/$ s'érige comme un mur à la fois inacceptable et indestructible.

Ces questions d'identité, ou pas, lors des répétitions d'une succession *[G.Mes(A)]*, sont très insidieuses. D'une part elles incluent un noyau dur qu'on ne peut pas ignorer. Elles touchent aux limites de la pensée.

---

[35] Notons qu'un grand nombre de spécialistes de la mécanique quantique parlent effectivement en termes d' 'ensembles'.

L'effet du choc est viscéral. Ce noyau dur est doté d'une grande force de fascination car on supporte mal de véritablement sentir que la pensée se heurte à une limite de ses capacités. Non pas se le dire ou l'entendre dire, mais le sentir. Dans le même temps cette question introduit une foule de glissements que l'on sent être fallacieux. Cela aussi on le sent de façon intime. Et cela aussi inquiète, tout en augmentant la fascination. C'est ainsi que 'le problème de complétude', en effet, se manifeste aussi face à la forme descriptionnelle qualitative $D_M/G, me_G, V_M/$. Et face à celle-ci, il acquiert même une texture plus concentrée et une intensité psycho-intellectuelle plus grande que face au formalisme quantique, parce que, en l'absence d'un formalisme le regard n'a pas où se disperser en suspicions qui occupent et détendent. D'autre part, cette fois le problème de complétude apparaît face à une genèse explicite que l'on peut retracer, et cela l'expose à un contrôle critique.

Mais suspendons ce contrôle, le temps d'énoncer aussi le problème ontologique.

### 3.7.1.2. Le problème du contenu 'ontologique' du concept de microétat

On peut suivre aussi un autre cheminement qui est intimement relié au précédent mais où l'accent tombe plutôt sur la question 'ontologique': comment *est* un microétat lui-même, vraiment, indépendamment de toute opération cognitive humaine accomplie sur lui ? Cependant que cette fois la question de complétude de *notre représentation* reste en retrait. Ce cheminement, lui aussi, a émergé d'abord relativement au formalisme quantique. Mais de nouveau on le retrouve relativement à la forme descriptionnelle $D_M/G, me_G, V_M/$. On se dit .

« Essayons malgré tout d'admettre toutes les 'identités' affirmées par décision méthodologique. Mais alors pourquoi un microétat, toujours le même, identiquement reproduit par des opérations de génération identiques et soumis à chaque fois à une évolution de *Mes(X)* qui est toujours la même, d'une seule et même observable *X*, conduirait-il en général à des valeurs *Xj* différentes, au lieu d'engendrer toujours la même valeur ? Ne serait-ce pas parce qu'un microétat est une entité dont la nature est *essentiellement* aléatoire ? ».

Mais aussitôt on réagit :

« Que peut vouloir dire, exactement, 'une nature essentiellement aléatoire' d'un microétat ? Et pourquoi, dans les conditions considérées, un microétat aurait-il une nature plus aléatoire qu'une opération de génération *G* ou qu'un acte de mesure *Mes(X)* ? Ne s'agit-il pas en fait exclusivement de l'incapacité opératoire, de notre part à nous, de reproduire, à partir de contraintes macroscopiques, exactement la même opération de génération, le même état microscopique, et aussi, exactement le même acte de mesure ? Car l'idée qu'un microétat serait de par lui-même 'essentiellement' aléatoire – ou alors peut-être plutôt *intrinsèquement* aléatoire ? ou *aléatoire en soi* ? – paraît vraiment très obscure. En outre, s'il ne s'agit en effet que d'une incapacité opératoire humaine, qu'est-ce qui me donne le droit de retourner une telle incapacité de l'homme, en affirmation ontologique de caractères qui seraient – eux – *essentiellement aléatoires*, c'est à dire aléatoires *dans les faits même* ? Et d'ailleurs qu'est-ce que cela veut dire 'la même opération *G*' ou 'le même microétat' ou 'le même processus de mesure' ? 'Même', *de quel point de vue ?* Dans l'absolu ? Mais n'est-ce pas là un non-sens ? ».

Les significations des mots glissent et se tortillent comme des anguilles et elles échappent à l'entendement. La pensée rebondit indéfiniment contre un mur insaisissable qui l'use, la déchire et l'enlise. Alors on renonce à penser. On se tait jusque dans l'âme et on attend, avec une sorte de foi impuissante.

Voilà en quoi consiste 'le problème ontologique' que suscite la forme descriptionnelle $D_M/G, me_G, V_M/$. Ce même problème – *tel quel* – émerge également face au concept de microétat employé dans les exposés du formalisme mathématique de la mécanique quantique.

On voit déjà en quel sens la question de complétude et la question ontologique sont distinctes mais reliées : La question de complétude ne concerne directement que notre représentation des microétats, cependant que la question ontologique ne concerne directement que la façon d'être assignée aux microétats eux-mêmes.

Mais ces deux questions, l'une comme l'autre, présupposent que cela posséderait un sens définissable de vouloir "savoir comment les microétats sont vraiment en eux-mêmes, indépendamment de toute action cognitive humaine, d'une manière absolue". Elles présupposent également toutes deux, bien que de manière plus vague, que cela posséderait un sens spécifiable de vouloir savoir *d'emblée* cela, *avant* de se lancer dans le processus de construction exposée ici. Enfin, elles présupposent aussi que cela posséderait un sens que de vouloir réaliser une sorte de saturation absolue de la description d'une entité-objet ; d'arriver à savoir 'tout' ce qui concerne cette entité-objet. Voilà l'essence multiple, floue et obscure des deux problèmes liés, de complétude et ontologique ; l'essence distillée, rendue indépendante de la formalisation mathématique des descriptions de microétats. Ces problèmes, qui ont d'abord été perçus face au formalisme mathématique de la mécanique quantique, émanent en fait d'en dessous de ce formalisme. La source de ces problèmes, qui implique directement et explicitement la forme descriptionnelle $D_M/G, me_G, V_M/$, peut aussi s'énoncer de la façon suivante.

*Être* et *description* se confondent dans un seul absolu dont on présuppose qu'on peut l'englober entièrement dans la *connaissance* du réel 'tel qu'il est en lui-même'. Le concept de description est alors illusoirement exonéré du tribut inévitable qu'il doit payer à des opérations de qualification, qui *seules* peuvent engendrer du *connu*, mais en le *séparant* de l'être-en-soi, inconnaissable.

Les trompe-l'œil conceptuels coagulés par cet absolu impossible sont projetés sur l'horizon de la connaissance où ils rejoignent le trompe-l'œil du 'vrai-en-soi'. Ces différentes variantes en trompe-l'œil de notre refus viscéral de réaliser que tout *connu* – qui est *description*, qui comporte du *qualifié* – est confiné à l'intérieur du domaine du relativisé aux qualifications accomplies, s'agitent vainement dans un tourbillon éternel de néants de sens, comme dans un enfer de Dante des concepts qui ont péché.

### 4.7.1.3. Retour sur la relation de un-à-un $G \leftrightarrow me_G$

Par les questions de complétude et ontologique qu'elle soulève irrépressiblement, la description $D_M/G, me_G, V_M/$ d'un microétat pousse la pensée naturelle à un corps à corps avec l'affirmation kantienne métaphysique de l'impossibilité de connaître le-réel-physique-en-soi. Or les questions de complétude et ontologique sont fondées toutes les deux sur la mise en doute de la pertinence de la relation de un-à-un dénotée $G \leftrightarrow me_G$. D'autre part, on l'a fortement souligné, seule l'acceptation *méthodologique* de la relation $G \leftrightarrow me_G$ permet d'aboutir à la construction de la forme qualitative $D_M/G, me_G, V_M/$ de la description d'un microétat. Et un œil averti discerne dans le désigné *{p(G,Xj)}* du symbole $D_M/G, me_G, V_M/$, un équivalent intégré et à genèse épistémologique explicite, des descriptions de microétats offertes par les algorithmes quantiques. Alors que faut-il finalement penser de la relation $G \leftrightarrow me_G$ ?

Il me semble que la question peut être tranchée définitivement par le dialogue imaginé qui suit[36].

---

[36] Cf. aussi la définition *D4* dans l'exposé du noyau de la méthode de conceptualisation relativisée, dans Mugur-Schächter [2006], p. 64 : là ce même dialogue intervient à un niveau *général*, non lié au cas particulier des microétats.

**L (le lecteur)**. Malgré tout, je me demande si vraiment on est *obligé* de poser la relation de un-à-un $G \leftrightarrow me_G$. Il existe peut-être une autre solution.

**M (moi)**. Tout d'abord, rien n'est obligatoire dans une construction. Ceci est convenable, cela ne l'est pas. Point. En l'occurrence, si l'on imaginait au départ que l'opération $G$ peut produire tantôt une chose et tantôt une autre, on aurait des difficultés pour parler de ce que $G$ produit. Et aussi pour y réfléchir, ce qui est beaucoup plus grave. Alors pour quelle raison devrait-on éviter d'introduire une organisation de langage-et-concepts qui évite ces difficultés ?

**L**. Pour ne prendre aucun risque de découvrir plus tard que l'on a affirmé quelque chose de faux.

**M**. Faux ? Mais face à quoi ? La question est là : *face à quoi* ? Forcément face à quelque *examen* futur pour qualifier le microétat $me_G$, n'est-ce pas ? Or ici, il ne s'agit pas d'une relation entre le microétat $me_G$ et les résultats d'examens futurs pour le qualifier. Il s'agit exclusivement de la relation entre l'opération de génération $G$ et son effet $me_G$. Lorsqu'on glisse subrepticement d'un problème à un autre, on étrangle l'entendement dans un nœud.

**L**. D'accord, mais ce qu'on admet maintenant peut entraîner des effets concernant ce qui se manifestera plus tard.

**M**. Magnifique ! Finalement je trouve que cet échange est le plus utile que l'on ait pu concevoir afin de rendre intuitive la relation $G \leftrightarrow me_G$. Vous êtes en train de m'offrir l'occasion d'étaler sans un pli devant les yeux publics, l'un de ces glissements incontrôlés qui sécrètent des faux absolus et faux problèmes où l'entendement reste piégé comme une mouche dans une toile d'araignée.

Donc exprimons-nous jusqu'au bout : vous craignez que le fait de poser d'emblée une relation de un-à-un entre l'opération de génération $G$ et le microétat $me_G$ qu'elle produit, puisse avoir des implications qui se révéleront fausses face aux résultats de quelque examen futur de $me_G$. Et cette crainte vous fait préférer de laisser ouverte la possibilité que cette relation ne soit pas de un-à-un, plutôt que de l'exclure prématurément par une assertion dictatoriale qui pourrait se trouver démentie par la suite. C'est bien cela ?

**L**. Tout à fait cela.

**M**. Alors faisons une expérience de pensée. Imaginons un examen dénoté *Ex.1* du microétat $me_G$ qui serait tel que, chaque fois que l'on réalise $G$ et l'on soumet l'effet $me_G$ de $G$ à l'examen *Ex.1*, l'on obtienne invariablement le même résultat. Que diriez-vous dans ce cas *concernant la relation entre $G$ et $me_G$* ? Qu'il est désormais *démontré* qu'il s'agit en effet d'une relation de un-à-un ? Vous pouvez répondre « oui », vous pouvez répondre « non », ou bien vous pouvez répondre « pas encore démontré ». Cela épuise les possibilités.

Supposons d'abord que vous répondiez « oui ». En ce cas, imaginons maintenant un autre examen dénoté *Ex.2* qui est différent de *Ex.1* et qui est tel que lorsqu'on répète $G$ plusieurs fois et à qu'à chaque fois on soumet l'effet $me_G$ obtenu, à l'examen *Ex.2*, l'on constate tantôt un résultat, tantôt un autre, donc en fin de compte tout un ensemble de résultats différents. Cela vous paraît-il impossible, étant donné que l'effet du premier examen *Ex.1* s'est avéré être stable ?

**L**. Non, pas nécessairement, en effet.... On peut imaginer par exemple que l'opération $G$ est définie de façon à produire à chaque fois une bille sphérique de dimensions données, mais dont on laisse la matière varier d'une réalisation de $G$ à une autre. En répétant alors $G$ et en soumettant à chaque fois le produit de $G$ à un examen de forme, on obtiendrait un ensemble de résultats identiques, cependant qu'avec un examen de poids on obtiendrait un ensemble de résultats dispersés….. Si l'on n'essaie pas de restreindre *G* à l'avance convenablement, on ne peut pas éliminer la possibilité que vous venez d'envisager.

**M**. Restreindre *G à l'avance* pour que *tout* examen futur, disons *Ex.j, j=1,2,......,…* conduise à un ensemble de résultats identiques si l'on répète des séquences *[G.Ex.j]* ? Cela avec un *j* quelconque, de l'ensemble d'examens quelconque considéré ? Cela vous paraît-il concevable ? *Il me semble que vous ne distinguez pas clairement entre une restriction qui pèserait sur l'opération de génération G et une restriction concernant les examens futurs que l'on pourrait accomplir sur les résultats de G*. Mais progressons systématiquement. Donc vous admettez que lorsqu'on répète l'opération de génération *G*, telle qu'elle a été spécifiée de par sa définition, le microétat qui en résulte pourrait manifester à chaque fois des résultats identiques lorsqu'il est soumis à l'examen *Ex.1*, cependant que l'examen *Ex.2*, lui, produirait des résultats non-identiques. Que diriez-vous en ce cas *concernant la relation entre G et $me_G$* ? Qu'il est désormais *démontré* qu'elle n'est pas une relation un-à-un ?…..J'ai l'impression que vous hésitez ? Pourquoi ?

**L**. Parce que je commence à concevoir qu'il se pourrait que le comportement du microétat produit par *G*, face à des examens futurs sur celui-ci, ne puisse jamais imposer une conclusion quant à la relation entre *G* et le microétat produit par *G*.

**M**. Donc, finalement, nous sommes en train de converger. Néanmoins allons jusqu'au bout systématiquement. Examinons maintenant la troisième réponse possible de votre part. Supposons donc qu'à ma première question concernant l'examen *Ex.1* vous ayez répondu « non, cela ne démontre pas encore que la relation entre G et son effet dénoté $me_G$ soit une relation de un-à-un ». Dans ce cas, je vous demanderais : *Quand* admettrez-vous qu'il *est* démontré que la relation entre l'opération *G* de génération du microétat et le microétat que *G* produit, est une relation de un-à-un ? Quand vous aurez vérifié l'identité des résultats pour tous les examens futurs ? Mais que veut dire 'tous' ici ? Tous les examens que l'on connaît, ou bien tous ceux que l'on connaît plus ceux que l'on imaginera jusqu'à la fin des temps ? Sur quelle base pourrait-on affirmer quoi que ce soit concernant cette 'totalité' ouverte, indéfinie d'effets d'examens futurs ?……

Je prends la liberté de considérer que le dialogue imaginaire qui précède a valeur d'une preuve ; qu'il a pu imposer désormais la nécessité, en général, de décisions méthodologiques lorsqu'on entreprend une construction, et aussi, en l'occurrence, la nécessité de la décision méthodologique de poser la relation de un-à-un $G \leftrightarrow me_G$. La justification – pas la *preuve*, mais la *justification* – ne pourra venir qu' *a posteriori*. Car cette relation de un-à-un non seulement est nécessaire *a priori* pour pouvoir construire une représentation qualitative de la description d'un microétat, mais en outre la nier, on vient de le voir, serait dépourvu de toute conséquence acceptable d'un point de vue conceptuel, comme aussi de toute conséquence utile.

Or l*a relation de un-à-un $G \leftrightarrow me_G$ élimine a priori les deux problèmes reliés de la complétude de la forme descriptionnelle $D_M/G,me_G,V_M/$ et du contenu ontologique du concept de microétat.*

Ce qui a fait obstacle à la compréhension de la situation conceptuelle lorsque les problèmes de complétude et ontologique ont été soulevés en relation avec la formalisme mathématique de la mécanique quantique, a été l'idée fausse induite par la présence du formalisme, qu'il s'agirait de questions de nature formelle, à résoudre par démonstrations de théorèmes. Cependant que lorsqu'ils sont rapportés à la forme descriptionnelle $D_M/G,me_G,V_M/$, ces célèbres et persistantes questions de la complétude descriptionnelle et du contenu ontologique d'un 'microétat', qui ont fait couler tant d'encre, simplement s'évanouissent. Dès qu'on tient compte de la genèse de la forme descriptionnelle qualitative $D_M/G,me_G,V_M/$, dès qu'on s'imprègne du *néant* conceptuel duquel la forme $D_M/G,me_G,V_M/$ a émergé *via* des contraintes cognitives et méthodologiques inévitables, on comprend intuitivement qu'*avant* les descriptions *transférées* des microétats il n'y a *RIEN* en tant que connaissances sur des microétats. On heurte ainsi la limite au delà de laquelle seule une *construction* est concevable. Et l'on réalise qu'une construction impose certaines contraintes inéluctables *d'ordre* dans l'action épistémologique, des contraintes qui priment tout et qui notamment éliminent les prudences logiques comme dépourvues à la fois de possibilité et de sens.

*4.7.1.4. Descriptions primordiales 'transférées'.*
*Probabilités classiques et probabilités primordiales*

Elaborons plus la remarque qui clôt le paragraphe précédent. Ella a trait à une autre obscurité qui subsiste. Les questions de complétude et ontologique sont *spécifiques* au type de représentation induit par la situation cognitive dans laquelle on se trouve lorsqu'on veut construire des connaissances concernant des microétats. En effet, les théories physiques classiques ne conduisent pas à ces questions. Pourquoi ? Qu'est ce qui pousse la description des microétats au bord du gouffre métaphysique, quand la physique classique vaque tranquillement au loin sur des plateaux lisses ?

La réponse, dans ce cas aussi, se dessine déjà dans l'exposé de la genèse de la forme descriptionnelle $D_M/G,me_G,V_M/$. Les données observables comportées par cette forme descriptionnelle consistent exclusivement en marques physiques *transférées* par les interactions de mesure, sur des enregistreurs d'appareils. Ces marques émergent éparpillées dans l'espace, et aussi dans le temps, et elles ne sont pas assignables – isolément – au microétat $me_G$ étudié, elles ne caractérisent que les interactions de mesure entre $me_G$ et l'appareil de mesure mis en œuvre, globalement.

Ces marques transférées caractérisent l'élaboration d'une toute *première* strate de construction de connaissances. Dans l'ordre de la genèse des connaissances elles sont *primordiales*.

En effet une description $D_M/G,me_G,V_M/$ consiste d'abord à extraire de la factualité physique a-conceptuelle, un fragment de réel physique qui, *a priori*, a été étiqueté '$me_G$' mais qui n'a jamais encore été qualifié auparavant et qui est inobservable directement ; et ensuite, à forger – sans percevoir ce fragment – des qualifications absolument initiales qui lui soient liées. Une description $D_M/G,me_G,V_M/$ apporte une toute première sorte de 'connaissance' spécifique et communicable associable à ce fragment de factualité qui vient d'être arraché au réel physique a-conceptuel. Or il se trouve que cette toute première sorte de connaissance émerge irrépressiblement avec un caractère statistique-probabiliste : si ce caractère ne se manifeste pas concernant une grandeur $X$ donnée, il se manifeste toujours pour d'autre grandeurs $X$ (cf. 3.5). Ceci est un fait. Le caractère statistique-probabiliste émerge en tant qu'une donnée première. Tout cadre pré-élaboré, théorique ou même seulement conceptuel, est absent lorsque ce caractère se manifeste. Les probabilités incorporées aux descriptions de microétats sont des probabilités *primordiales*. Ce sont des probabilités dépourvues de *tout* substrat de conceptualisation déjà accomplie précédemment. Les descriptions transférées $D_M/G,me_G,V_M/$ appartiennent à une toute *première* strate de conceptualisation dont le type descriptionnel qui est entièrement ignorée par la conceptualisation classique.

Je souligne encore que le mot 'primordiales' employé dans ce contexte ne désigne qu'un statut épistémologique, celui de *primauté dans l'ordre de la constitution d'une chaîne de connaissances*. Ce mot est pur de toute connotation ontologique. Je refuse le terme courant de 'probabilités essentielles' précisément parce qu'il comporte une forte connotation ontologisante qui suggère des *'propriétés intrinsèques'*.

Quant au cadre conceptuel qui correspond à la forme descriptionnelle $D_M/G,me_G,V_M/$ et qui s'élabore dans le même temps que cette forme elle-même, tout simplement il rejette un postulat déterministe, comme viennent de le montrer les monologues et le dialogue imaginés plus haut. Ce cadre rejette comme dépourvue de sens la question de savoir *pourquoi* une description transférée $D_M/G,me_G,V_M/$ émerge probabiliste ; *a fortiori*, il rejette aussi un postulat qui 'expliquerait' ce 'pourquoi'.

Cependant que dans la physique classique, au contraire, toute description probabiliste est seconde dans l'ordre d'élaboration des connaissances *:* elle se *définit* sur du préalablement conceptualisé, sur du conceptualisé en termes de ces *modèles 'intrinsèques' que sont les 'objets' au sens courant* (Mugur-Schächter [2006 pp.118-127]) conçus comme étant dotés en permanence de 'propriétés' propres, si l'on peut dire. *Dans la conceptualisation classique ces modèles 'objectifiants' sont traités comme des **données premières***.

La conceptualisation classique est déterministe dans son principe même. Elle présuppose par postulat la possibilité d'une représentation strictement causale des faits physiques et elle accomplit une représentation de base qui est soumise *a priori* à l'exigence de réaliser cette possibilité. On n'y démarre pas par l'accomplissement opérationnel *physique* de successions *[G.Mes(X)]* et la construction de statistiques et de lois de probabilité. On démarre avec un crayon, du papier, et de la réflexion abstraite qui s'applique sur des modèles préformés sur la base de perceptions directes et traités comme des données premières. La genèse de ces modèles qui roulent dans la pensée courante se perd dans les brumes de l'évolution biologique qui l'a enfermée dans les boites noires des réflexes neurosensoriels. Selon la conceptualisation classique explicite, qui s'appuie directement sur des modèles-'objets', le caractère probabiliste d'une description (par exemple celle des effets des jets d'un dé, où celles de la théorie cinétique des gaz) serait *éliminé* si l'observateur-concepteur appliquait en toute rigueur la théorie fondamentale déterministe qui constitue le cadre conceptuel-mathématique de base qui est accepté (la mécanique newtonienne, ou celle d'Einstein, ou l'électromagnétisme de Maxwell, ou quelque composition de ces théories). Par postulat, tout caractère probabiliste s''explique' par quelque ignorance ou quelque abstraction faite de données qui, en principe, sont disponibles dans la représentation rigoureuse de base. Le fait que ce postulat est entaché de non effectivité n'a pas encore fait scandale (Longo [2002], Mugur-Schächter [2002C])). À ma connaissance, personne, à ce jour, n'a organisé une mise à jour globale de la question du déterminisme classique. La science classique se meut encore, sereine, dans un univers conceptuel qui émane du déjà conceptualisé dans les langues et la pensée courantes. C'est un univers conceptuel à substrat lissé par un déterminisme décrété dont le caractère arbitraire se cache dans le flou de la pensée courante. Mais le postulat déterministe lui-même, tout entier, n'est qu'un *modèle* général affirmé concernant l'agencement foncièrement *inobservable* de tout phénomène physique.

Or si l'on éliminait, par principe déterministe, le caractère probabiliste d'une description de microétat, qui est primordiale, qui ne consiste *que* en marques transférées sur des enregistreurs d'appareils, qu'en resterait-il ? *Rien* : cette description toute entière n'est rien de plus que les lois de probabilité construites en relation avec des fragments de substance physique a-conceptuelle, sur la base de groupes de marques physiques transférées par des interactions de mesure sur des enregistreurs d'appareils macroscopiques, et qui, irrépressiblement, émergent dispersés. Point.

Le seul problème que ces distributions de probabilités primordiales soulèvent véritablement est celui de leur *constructibilité*, et celle-ci exige de poser la relation de un-à-un $G \leftrightarrow me_G$. Dans l'ordre *génétique* de la construction de connaissances, on ne peut fourrer aucune étape préalable en *dessous* des descriptions transférées des microétats. Aucune connaissance *préalable* qui puisse les 'expliquer' d'une manière plus définie, plus 'complète', plus 'singulière' au sens grammatical, ne peut être trouvée 'avant' ou 'en dessous' de ces descriptions, comme on choisira de dire : il n'y a aucune place où l'on puisse loger une telle étape, puisqu'il s'agit de descriptions foncièrement premières. C'est pour cette raison que l'on est obligé de se lancer dans la postulation méthodologique d'une relation $G \leftrightarrow me_G$ de un-à-un.

Cela ne veut pas dire que l'on doive rester à jamais dépourvus d'"explication". Cela veut dire seulement que toute explication conçue en termes de qualifications *séparées* des processus de *Mes(X)* définis pour les microétats et en termes 'intrincisés', résorbés à l'intérieur du concept de microétat, sera à construire *après* et *sur la base* des descriptions transférées, primordialement probabilistes, des microétats.

L'ordre qui régit la genèse de connaissances est *distinct* et différent de l'ordre déductif qui explique à l'intérieur d'un volume de connaissances constituées. Sur la direction du développement des conceptualisations, il ne faut pas confondre l'ordre génétique factuel de construction, qui est exclusivement progressif, et d'autre part l'ordre déductif-explicatif qui s'institue *a posteriori* et qui revient en arrière pour placer des modèles explicatifs à un niveau *virtuel* placé en dessous des débuts absolus de la genèse constructive des conceptualisations, que rien ne peut repousser en arrière.

C'est la confusion de ces deux ordres distincts qui engendre le problème illusoire de complétude, le problème ontologique, le problème du statut des probabilités microscopiques face au postulat déterministe. Cette confusion engendre une sorte de vertiges de l'entendement qui immobilisent dans l'incompréhension.

Dans ces conditions, tirons les conséquences jusqu'au bout.

Lorsque la microphysique moderne fondamentale n'offre que des descriptions primordialement probabilistes, cependant que la physique macroscopique postule directement un déterminisme ontologisé, assigné entièrement aux faits mêmes, et qui se trouve explicitement en difficulté à la fois logique et métaphysique, il paraît approprié d'introduire le postulat factuel suivant.

> **Toute** *phase initiale d'un processus de construction de connaissances nouvelles – conscient ou réflexe, factuel ou abstrait, à objets microscopiques ou macroscopiques ou cosmiques – possède uniformément un caractère primordialement statistique où peuvent se loger des invariants statistiques, que l'on peut dénommer 'probabilistes'.*

Selon un tel postulat chaque discipline déterministe de la physique classique serait à regarder d'emblée comme un construit *modélisant* qui – dans l'ordre génétique de construction de connaissances – est *superposé* à la phase initiale sous-jacente du processus de construction de connaissances impliqué. Ce construit posé explicitement comme étant modélisant, préserverait tous les avantages pragmatiques du postulat déterministe classique, sans assigner un caractère déterministe au mode d'être général des faits physiques eux-mêmes, et sans que, sur cette base insoutenable, l'on soit conduits à incriminer comme 'incomplètes' les représentations génétiquement premières qui n'englobent pas le caractère déterministe. Cela nous mettrait dans un schéma de pensée simplifié et cohérent.

*4.7.1.5. 'La coupure classique-quantique'*

Nous venons de mettre en évidence que l'ensemble des descriptions transférées $D_M/G,me_G,V_M/$ de microétats appartient à une strate primordiale de conceptualisation où la structure descriptionnelle est radicalement distincte de celle des descriptions classiques. Nous avons déjà rappelé que les descriptions au sens classique de ce mot s'appliquent à des modèles-'objets' élaborés *précedemment* (sans doute sur la base de descriptions transférées primordialement probabilistes qui se sont construites de manière réflexe ou bien seulement implicite). Ces modèles-'objets' qui préexistent sont dotés par la modélisation qui les a générés, de 'propriétés' indépendantes de toute action cognitive. Et selon la pensée courante, les langages naturels, ainsi que selon la logique, les probabilités, et l'entière *science* classique, les 'propriétés' – rendues abstraites et dotées d'une existence propre permanente – sélectionnent parmi les modèles-'objets' préexistants, des objets-d'étude qu'elles qualifient *dans la même foulée*.

On mesure la distance qui sépare les descriptions au sens classique, des descriptions transférées $D_M/G,me_G,V_M/$ : il s'agit clairement de deux phases distinctes des processus de conceptualisation.

Ces remarques conduisent trivialement à la conclusion suivante.

La mystérieuse 'coupure quantique-classique' est la coupure entre : *(a)* les descriptions transférées de microétats, $D_M/G,me_G,V_M/$, qui émergent primordialement probabilistes et ne disent strictement *rien* sur la manière d'être d'un microétat considéré *séparément* de toute action de mesure accomplie sur lui[37], et *(b)* les descriptions au sens classique, fondées sur des modèles qui affirment des propriétés intrinsèques à l'entité qualifiée[38].

La raison pour laquelle la définition du contenu de cette coupure oppose à ce jour même une telle résistance, est que la forme qualitative intégrée $D_M/G,me_G,V_M/$ des descriptions de microétats est restée non connue d'une manière explicite.

*4.7.1.6. Modélisations associées à la forme $D_M/G,me_G,V_M/$*
*et niveaux descriptionnels*

Les travaux de Louis de Broglie [1924], [1956] et de David Bohm [1952] et ses élèves – très particulièrement Peter Holland [1997] – offrent une modélisation qui 'complète' le formalisme quantique le rendant déterministe. Cette modélisation a été largement critiqué[39] mais elle a aussi un nombre de plus en plus non négligeable d'adeptes.

Quoi qu'il en soit, soulignons de nouveau que toute modélisation individuelle (non probabiliste) du concept de microétat doit, de manière explicite, être construite sur la *base* de la description primordiale $D_M/G,me_G,V_M/$. Or il convient de concevoir qu'une telle construction est à accomplir selon certaines règles qui doivent être spécifiées : la foncière stratification de la conceptualisation, qui vient de se faire jour, est désormais à respecter très attentivement, avec toutes ses conséquences (Mugur-Schächter [2006]).

---

[37] Le fait que dans les descriptions impliquées par les algorithmes quantiques, les grandeurs mécaniques considérées et les mesures correspondantes sont entièrement spécifiées, ne change nullement le fait que ces descriptions ont elles aussi le même statut descriptionnel, celui de DESCRIPTIONS PRIMORDIALES TRANSFEREES.
[38] Dans Mugur-Schächter [2006] – à l'intérieur de la méthode de conceptualisation relativisée – 'la coupure *quantique*-classique' apparaît comme un cas particulier incorporé dans une coupure *générale* [(descriptions transférés)-(descriptions classiques)]. Dans le volume de l'ensemble des conceptualisations tel qu'il existe à tout moment donné, cette coupure générale est dotée d'une certaine épaisseur dont la méthode mentionnée spécifie l'entière *structure*, ainsi que la *relation* qui unit la strate des descriptions transférées à la strate des descriptions classiques.
[39] Personnellement, le trait qui me paraît inacceptable dans la modélisation Bohm-Holland – par ailleurs incontournable par certains autres traits – est qu'elle ne distingue pas explicitement entre le niveau de description individuelle où doit se loger tout modèle individuel de microétat, et le niveau de description statistique-probabiliste. Mais je tiens que cette indistinction peut certainement être levée en tenant compte de la structuration du connu en deux strates ( Mugur-Schächter [2006], Mugur-Schächter [1991] pp. 1434-1448). La critique formulée s'applique moins au modèle de Louis de Broglie. Toutefois, là encore, les "champs quantiques" sont définis à l'aide de l'amplitude de l' 'onde' qui représente des *probabilités*, pas un microsystème individuel.

Mais d'ors et déjà, du chemin accompli jusqu'ici il se dégage une conclusion. Modéliser des données nouvelles veut dire les rattacher à du connu d'une manière qui les rende facilement intelligibles et communicables. La connaissance qui se constitue dans l'esprit intuitivement, sur la base des processus neurosensoriels dont est doté un corps humain normal, conservera donc indéfiniment un rôle d'attracteur vers ses propres formes. D'autre part, les buts d'obtenir des données nouvelles (comme lorsqu'on veut créer de la connaissance concernant du réel microphysique) ou de créer de la cohérence logique et du consensus de plus en plus large concernant des données macroscopiques (comme dans le cas des relativités d'Einstein), conduisent en certains cas à construire des voies de représentation qui s'éloignent indéfiniment des représentations qui se constituent par des processus réflexes et que nous manions intuitivement. Les tensions qui naissent sur le chemin de tels éloignements comportent explicitement des contraintes *méthodologiques* que les processus réflexes de construction de connaissances ne suscitent pas, ou dont ils cachent l'action possible. Quant aux représentations engendrées sous l'empire de buts que la pensée naturelle ne nourrit pas, elles sont marquées d'un caractère d'étrangeté qui appellent une modélisation subséquente. Celle-ci, si elle est accomplie et si elle a 'réussi', par définition même de cette affirmation, incorpore les représentations mentionnées dans les formes familières de la pensée naturelle ou de la science classique[40].

### *4.7.2. La source commune des problèmes suscités par la forme descriptionnelle $D_M/G,me_G,V_M/$.*

Ce qui est resté insuffisamment compris à travers l'ensemble des problèmes qui émergent relativement à la forme descriptionnelle $D_M/G,me_G,V_M/$ est – uniformément – la spécificité majeure d'une description de microétat, à savoir le fait qu'il s'agit d'une description qui est toute première, primordiale. Malgré plus de 70 années d'existence de la mécanique quantique et d'utilisation du formalisme mathématique de cette théorie, la forme épistémologique *intégrée* des descriptions de microétats symbolisée ici $D_M/G,me_G,V_M/$, avec sa genèse et ses conséquences, sont restées foncièrement étrangères à nos esprits éduqués par des interactions cognitives avec le réel macroscopique. La base de cette affirmation est que, dès que la forme descriptionnelle $D_M/G,me_G,V_M/$ est connue et comprise *via* sa genèse, les conséquences des impératifs méthodologiques que l'émergence de cette forme comporte, ainsi que l'indépendance mutuelle des étapes descriptionnelles et les relativités que cette forme incorpore, dissolvent d'ores et déjà le problème de complétude, le problème ontologique, le problème de la spécificité de l'indéterminisme quantique face au 'hasard' classique, et le problème de la définition de 'la coupure quantique-classique'.

On peut penser que ces acquis se montreront utiles pour libérer également le formalisme mathématique de la mécanique quantique des problèmes d'interprétation qui l'entachent. Car on perçoit clairement que les problèmes examinés plus haut, tels qu'ils se manifestent face à la forme $D_M/G,me_G,V_M/$, concentrent en eux, distillée, l'essence des problèmes qui portent les mêmes dénominations et qui ont d'abord été formulés relativement au formalisme quantique parce que ce formalisme s'est constitué et a été utilisé avant que sa manière de signifier ait été élucidée.

---

[40] En ce qui concerne les théorie de la relativité d'Einstein, il est difficile d'imaginer une modélisation réussie, parce que ces théories *mélangent* le niveau de description transférée (qui se manifeste dans les lois de transformation des coordonnées d'espace-temps) et le niveau classique des modèles macroscopiques (qui se manifeste dans la manière d'introduire les entités-objet de description). Cependant que le formalisme de la mécanique quantique, d'une façon cryptique mais parfaitement pure, ne concerne *que* des descriptions transférées : c'est en cela que consiste l'importance épistémologique paradigmatique de ce formalisme.

### 4.8. L'organisation d'espace-temps assignable à la description d'un microétat : l'*arbre de probabilité* de *un* microétat

Nous voilà donc orientés en ce qui concerne le contenu d'une description de microétat et la nature spécifique des probabilités primordiales qu'une telle description comporte. Cherchons maintenant pour une description transférée de microétat – avec sa genèse – une expression plus synthétique que celle qui s'est déjà constituée. Nous atteindrons ce but en examinant plus en détail en quoi, exactement, l'organisation qualitative du concept de probabilité primordiale qui s'est constitué, se distingue de celle qu'avait dans son esprit Kolmogorov lorsqu'il a formulé sa théorie classique des probabilités[41].

#### *4.8.1. Construction*

Ce qui suit immédiatement concerne le cas fondamental d'un micro-état d'*un* seul micro-système (cf. 3.4). Les autres cas seront considérés ensuite.

Soit une opération de génération $G$ qui engendre un microétat $me_G$. Dénotons ici $B$, $C$, $D$, etc. les grandeurs mécaniques redéfinies pour des microétats qui engendrent la description $D_M/G,me_G,V_M/$ de $me_G$ via des mesures *Mes(B)*, *Mes(C)*, *Mes(D)*, etc. Tout ce qui est essentiel concernant $D_M/G,me_G,V_M/$ – avec son entière genèse – peut être représenté d'une manière intuitive à l'aide d'un schéma d'espace-temps qui *globalise* la réalisation et les résultats d'un grand nombre de répétitions de la succession d'opérations *[G.Mes(B)]*, ainsi que de la succession d'opérations *[G.Mes(C)]*, ainsi que de la succession d'opérations *[G.Mes(D)]*, etc. Ce schéma est intuitif parce qu'il met en évidence visuellement les domaines d'espace-temps couverts par la réunion de tous les processus physiques impliqués dans la description $D_M/G,me_G,V_M/$[42]. Détaillons, mais en simplifiant au cas paradigmatique de seulement deux grandeurs dynamiques mutuellement incompatibles, $B$ et $C$.

Considérons d'abord le processus de génération $G$ du microétat $me_G$. Ce processus commence à un moment initial, disons $t_o$, et il finit à un moment ultérieur, disons $t_G$. Il possède donc une durée $t_G-t_o$. Il balaye aussi un certain domaine d'espace, disons $d_G$. Donc il couvre un domaine global d'espace-temps $d_G.(t_G-t_o)$. Au moment $t_G$ quand le processus de génération $G$ est accompli – donc le microétat $me_G$ peut déjà être supposé exister – on commence aussitôt un acte de *Mes(B)*. Ainsi l'on accomplit une succession *[G.Mes(B)]*. Cette succession finit avec l'enregistrement par l'appareil *A(B)*, d'un certain groupe de manifestations physiques observables. Au moment où l'enregistrement de ce groupe de manifestations est accompli, l'opération physique de *Mes(B)* est terminée et l'entière succession *[G.Mes(B)]* est elle aussi accomplie. Notons $t_B$ ce moment final. Le processus physique *Mes(B)* aura donc couvert un durée $(t_B-t_G)$. Il aura également balayé un certain domaine d'espace, disons $d_B$. Il se sera donc produit sur un domaine d'espace-temps *[$d_B.(t_B-t_G)$]*. Quant à l'entière succession *[G.Mes(B)]*, elle aura couvert le domaine total d'espace-temps *[$d_G.(t_G-t_o)+d_B.(t_B-t_G)$]*.

---

[41] L'exposé qui suit est très simplifié, faute des guidages complexes qu'offre la méthode générale de conceptualisation relativisée. Dans Mugur-Schächter [2006], pp.193-257 on peut trouver un exposé accompli.
[42] Il ne s'agit pas d'un modèle, mais juste d'une *représentation*.

Après avoir accompli le processus physique *Mes(B)*, on accomplit en outre une opération supplémentaire, *conceptuelle* cette fois : le groupe de marques physiques observables enregistré par l'appareil *A(B)* doit être *codé* conformément à la règle de codage associée à la re-définition conceptuelle de la grandeur *B* pour le cas des microétats, et ce codage fournit une traduction du groupe de marques physiques enregistrées, en termes d'une valeur *Bj* du spectre *{B1, B2,... Bj,...Bk}*[43] de la grandeur *B*, comme l'exige la définition-cadre *gqX2(me)* d'une grille de qualification de $me_G$ par la grandeur *B* (cf. *3.3.2.2*).

Répétons un très grand nombre *N* de fois la réalisations de la même succession *[G.Mes(B)]*, en remettant à chaque fois le chronomètre à *0* comme pour des épreuves sportives. Idéalement, le domaine total d'espace-temps couvert sera à chaque fois le même : $[d_G.(t_G-t_o)$. Quant au domaine d'espace-temps $d_B.(t_B-t_G)]$, il variera en général car l'enregistrement final d'une manifestation observable aura des coordonnées d'espace-temps aléatoires (puisque la situation est en général probabiliste, donc à la base, statistique). L'on considérera donc le domaine d'espace-temps *global* couvert par l'ensemble des actes de mesure et on le désignera par la même notation. Si le nombre *N* est assez grand, l'ensemble de toutes les *N* répétitions de la succession *[G.Mes(B)]* aura progressivement matérialisé *toutes* les valeurs du spectre *{B1, B2, B3,....Bj,... Bk}*, *j=1,2,....k,* de la grandeur *B*, puisque aucune parmi ces valeurs n'a une probabilité *a priori* nulle. Chacune de ces valeurs aura été obtenue avec une certaine fréquence relative. Et si *N* est très grand, alors l'ensemble *{n(G,B1)/N, n(G,B2)/N,.... n(G,Bj)/N.....n(G,Bk)/N.}* des fréquences relatives obtenues sera assimilable à la loi correspondante de probabilité

$$\{p(G,Bj)\} = \{p(G,B1),p(G,B2),....p(G,Bj),...p(G,Bk)\}, \quad j=1,2,....k,$$

présupposée 'existante' (cf. 3.6). Bref, au bout de ces *N* réalisations d'une succession *[G.Mes(B)]* le 'plafond' du domaine d'espace-temps $[d_G.(t_G-t_o)+d_A(t_A t_G)]$ se trouvera finalement garni par toutes les valeurs du spectre *{B1, B2, B3,....Bj,... Bk }* de la grandeur *B*, et – sur un niveau descriptionnel supérieur – l'on pourra inscrire l'entière loi factuelle de probabilité *{p(G,Bj)}* constatée sur ce spectre. Or le couple

$$[(B1, B2, B3,..Bj ..Bk), \quad \{p(G,Bj)\}]$$

est l'essence[44] de ce qui, dans la théorie moderne des probabilités formulée par Kolmogorov ([1933]), est dénommé *un espace de probabilités* : le résultat qui s'est constitué est donc, en termes simplifiés, l'espace de probabilité qui inclut la loi de probabilité *{p(G,Bj)}*.

Considérons maintenant la grandeur *C*, qui par hypothèse est *in*compatible avec la grandeur *B*, et refaisons concernant *C* un chemin strictement analogue à celui qu'on vient d'indiquer concernant la grandeur *B*. Au bout d'un très grand nombre *N* de répétitions de la succession de deux opérations *[G.Mes(C)]* couvrant un domaine global d'espace-temps $[d_G.(t_G-t_o)+d_C.(t_C-t_G)]$ qui cette fois correspond à *Mes(C)*. Au dessus du plafond de ce nouveau domaine d'espace-temps l'on aura finalement inscrit un autre espace de probabilité

$$[(C1, C2, C3,....Cq,..Cm), \ \{ p(G,Cq \}]$$

---

[43] Nous supposons un spectre fini, pour simplifier.
[44] Pour simplifier, l'algèbre que l'on définit sur l'univers d'événements élémentaires n'est pas mentionnée ici. La définition complète d'un espace de probabilité sera rappelée dans le paragraphe *3.8.3* et une discussion très approfondie de ce concept se trouve dans Mugur-Schächter [2006].

Puisque les grandeurs *B* et *C* sont incompatibles, les deux domaines d'espace-temps $d_B.(t_B-t_G)$ et $d_C.(t_C-t_G)$ couverts respectivement par des *Mes(B)* et des *Mes(C)*, seront *différents* (cf. *3.6.1.2*). Mais le domaine d'espace-temps $d_G.(t_G-t_o)$ couvert par l'opération de génération *G* est le *même* dans les successions [*G.Mes(B)*] et les successions [*G.[Mes(C)*] parce que les mesures de la grandeur *B* et celles de la grandeur *C* ont été effectuées toutes sur des exemplaires du même microétat $me_G$ engendré par l'opération de génération *G*. Ainsi la structure globale d'espace-temps de tout l'ensemble de successions de mesure accompli, est arborescente, avec un tronc commun couvrant le domaine d'espace-temps $d_G.(t_G-t_o)$ et deux branches distinctes qui couvrent globalement deux domaines d'espace-temps $d_B.(t_B-t_G)$ et $d_C.(t_B-t_G)$ différents. Chaque branche est surmontée d'un espace de probabilité spécifique à la branche. C'est pour cette raison qu'il est adéquat d'appeler cette structure un *arbre de probabilité du microétat $me_G$* correspondant à $G$[45]. On peut désigner cet arbre par le symbole $T[G,(V_M(B)\cup V_M(C)]$  (*T* : tree ; $(V_M(B)\cup V_M(C)$ : la vue (grille) qualifiante constituée par *l'union* des grandeurs mécaniques redéfinies pour des microétats qui ont été dénotées par *B* et *C*).

---

[45] Attention! L'expression 'arbre de probabilité' est employée dans d'autres contextes avec des significations tout à fait différentes de celle qui lui est assignée ici.

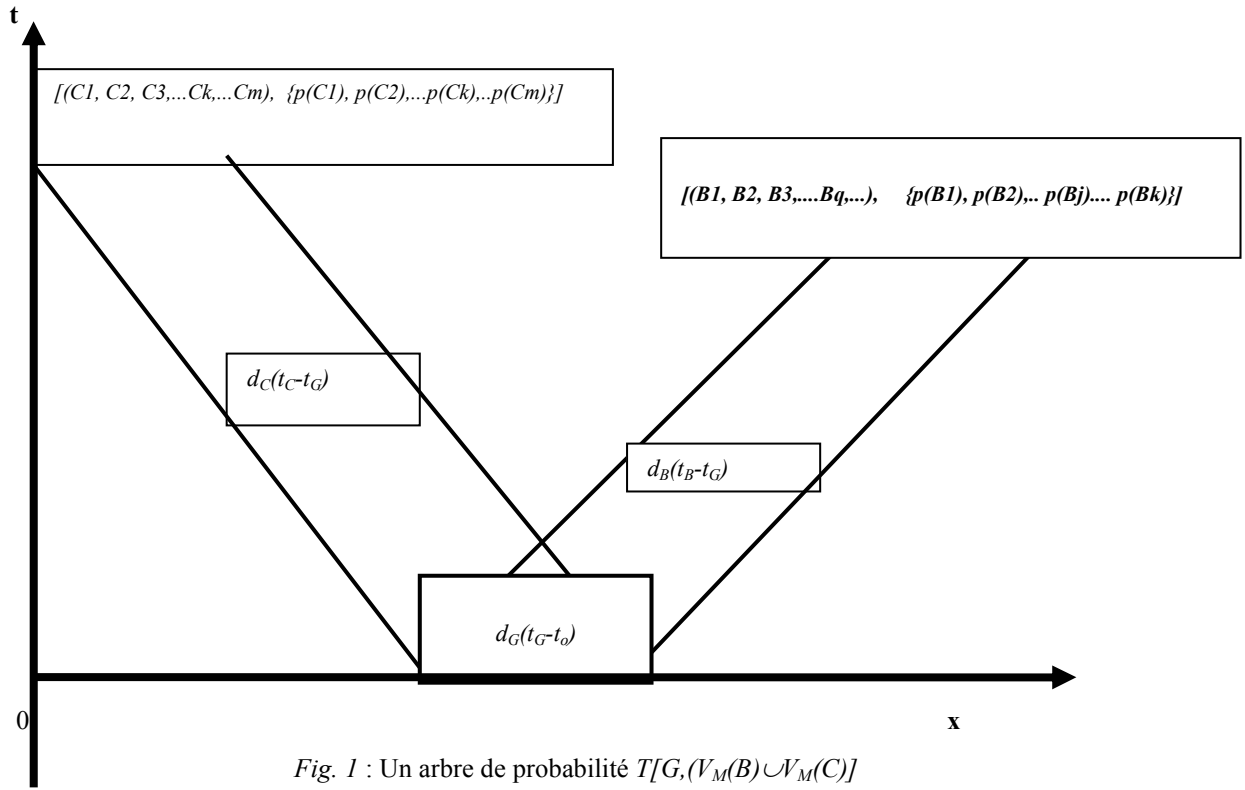

*Fig. 1* : Un arbre de probabilité $T[G,(V_M(B) \cup V_M(C)]$

Si l'on avait représenté *toutes* les branches possibles impliquées dans la description $D_M/G,me_G,V_M/$, correspondant à toutes les grandeurs mécaniques mutuellement incompatibles définies pour un microétat, on aurait dû employer l'article défini et dire *'l'arbre de probabilité du microétat $me_G$ correspondant à $G$'*. Dans ce cas le symbole – général – serait $T(G,V_M)$)[46].

Qu'on ne se laisse pas porter par des impressions inertielles de trivialité. *Le concept d'arbre de probabilité d'un microétat d'un seul microsystème est un concept probabiliste foncièrement nouveau qui* **déborde** *de plusieurs manières le concept d'un espace de probabilité de Kolmogorov*. Pour s'en rendre compte il suffit de considérer les précisions suivantes qu'il comporte concernant les notions de phénomène aléatoire, de dépendance probabiliste et de 'méta-loi de probabilité.

### *4.8.2. Les phénomènes aléatoires liés à un micro-état de un micro-système*

La théorie classique des probabilités est dépourvue d'une expression formalisée de la notion de phénomène aléatoire. Cette notion y est posée comme engendrant un espace de probabilité, mais elle n'intervient que dans le substrat *verbal* de la théorie. On ne symbolise mathématiquement que l'espace de probabilité engendré par le phénomène aléatoire, dont les éléments, par contre, sont exprimés et étudiés mathématiquement en détail. On peut caricaturer par le schéma :

[ **?** 'phénomène aléatoire' **?** ] → [espace de probabilité]

---

[46] Un physicien de la mécanique quantique perçoit tout de suite que les espaces de probabilité symbolisés au dessus les branches – à eux *seuls* et *tous ensemble* – sont l'équivalent qualitatif de la description du microétat $me_G$ offerte par le formalisme quantique *via* le vecteur d'état $|\Psi\rangle$ associé à $me_G$. En effet $|\Psi\rangle$ permet, à l'aide d'opérations mathématiques qui impliquent également les autres éléments de l'algorithme quantique, de calculer l'ensemble de lois de probabilité qui coiffent l'arbre $D_M/G,me_G,V_M/$. Mais l'arbre lui-même, tout ce qui, dans le schéma de la *Fig.1* est en dessous de la couronne de lois de probabilité, *n'est pas représenté*.

Les signes d'interrogation soulignent le vide de représentation laissé en ce qui concerne la genèse physique et conceptuelle d'un espace de probabilité.

Cependant que l'arbre de probabilité d'un microétat inclut au contraire une représentation minutieusement explicitée des phénomènes aléatoires qui engendrent les événements élémentaires à partir desquels on construit ensuite les espaces de probabilité qui coiffent les branches. *Tout* ce qui dans la figure 1 se trouve en dessous de l'espace de probabilités d'une branche, est phénomène aléatoire correspondant. On y voit, on comprend en détail, on sent toute la montée de la conceptualisation probabiliste : l'opération de génération $G$ *crée* l'entité-objet $me_G$ à étudier, de l'intérieur de la factualité physique a-conceptuelle ; l'acte de *Mes(X)*, où $X=B,C$, etc., porte cette entité-objet – en la *changeant* par interaction avec l'appareil $A(X)$ – jusqu'au *bord* de la conceptualisation, en produisant les marques physiques observables sur les enregistreurs de $A(X)$ ; la loi de codage qui traduit ces marques, en valeurs $Xn$ de $X$, constitue un prolongement – conceptuel – du processus physique de mesure, qui dépose *dans* la toute première couche de conceptualisation, une réalisation de l'univers *{X1, X2,...Xj....Xk}* des 'valeurs' possibles de $X$ dont est constitué le spectre des événements élémentaires ; et l'espace de probabilité correspondant *[{X1, X2....,Xj,...Xk}, {p(G,Xj)}]* (où l'algèbre est supprimée par simplification) est finalement construit sur cet univers d'événements élémentaires. Le phénomène aléatoire qui engendre cet espace est défini dans toutes ses phases.

Ainsi, à la faveur du cas particulier des microétats, le concept probabiliste *général* de phénomène aléatoire est en cours d'acquérir une représentation, qualitative mais à signifiance non-restreinte (Mugur-Schächter [2006]).

Or cette représentation est explicitement et foncièrement relative à l'opération de génération $G$ de l'entité-objet-d'étude $me_G$, à cette entité $me_G$ elle-même, et à la grille de qualification $V_M$ utilisée (ici $V_M \equiv V_M(B) \cup V_M(C)$) : des éléments qui dans la conceptualisation probabiliste classique n'interviennent pas.

### *4.8.3. Dépendance probabiliste*

Jusqu'ici, pour simplifier, nous avons fait abstraction d'un élément important d'un espace de probabilité au sens de Kolmogorov. Maintenant nous allons remédier à cette lacune afin de pouvoir faire des remarques importantes sur le concept de dépendance probabiliste.

Rappelons donc ce qui suit. Un espace de probabilité de Kolmogorov *[U, τ, p(τ)]* contient : un *univers* (ensemble) *U d'événements élémentaires*, une *algèbre τ d'événements définie sur U*, et une *mesure de probabilité p(τ) qui est définie sur l'algèbre d'événements τ*. Un événement au sens des probabilités est constitué par tout *ensemble* d'événements élémentaires. Une algèbre d'événements *e* définie sur l'univers $U$ d'événements élémentaires $e_i$, l'algèbre $\tau$ est un ensemble de sous-ensembles $e$ de $U$ – contenant $U$ lui-même ainsi que l'ensemble vide $\emptyset$ – et qui est tel que si les sous-ensembles d'événements élémentaires $A$ et $B$ sont contenus dans $U$ alors $U$ contient également la réunion $A \cup B$ et la différence $A-B$. Enfin, *une mesure de probabilité définie sur τ* consiste en un ensemble de nombres réels $p(A)$ dont chacun est associé à un événement $A$ de $\tau$ et qui satisfont aux conditions suivantes : *$0 \leq p(A) \leq 1$, $p(U)=1$* (normation), $p(\emptyset)=0$, et $p(A \cup B) \leq p(A)+p(B)$ où l'égalité se réalise ssi $A$ et $B$ sont disjoints (n'ont aucun événement élémentaire $e_i$ en commun $(A \cap B=\emptyset)$.

Deux événements $A$ et $B$ de l'algèbre $\tau$ sont posés par définition être mutuellement *'indépendants'* si le produit numérique $p(A)p(B)$ de leurs probabilités est égal à la probabilité $p(A \cap B)$ de l'événement-produit ensembliste $A \cap B$. (Cette définition est étendue au cas de deux (ou plusieurs) algèbres distinctes $\tau$ et $\tau' \neq \tau$, mais en présupposant la possibilité d'une réalisation *conjointe* de $A \in \tau$ et $B \in \tau'$).

Ces définitions restent pertinentes à l'intérieur, isolément, de *chaque* espace de probabilité qui coiffe une branche de l'arbre de probabilité $T(G,V_M)$ du microétat $me_G$.

Mais considérons maintenant un événement $A(X)$ de l'algèbre d'événements $\tau(U(Xj))$ posée sur l'univers $U(Xj)$ d'événements élémentaires de l'espace de probabilité

$$[U(Xj),\ \tau(U(Xj)),\ p[\tau(U(Xj))]]$$

qui coiffe la branche de $T(G,V_M)$ correspondant à des *Mes(X)*. Et soit un événement $B(Y)$ de l'algèbre d'événements $\tau(U(Yk))$ posée sur l'univers $U(Yk)$ d'événements élémentaires de l'espace de probabilité

$$[U(Yk)\},\ \tau(U(Yk)),\ p[\tau(U(Yk))]]$$

qui coiffe une *autre* branche de $T(G,V_Q)$ correspondant à des *Mes(Y)* incompatibles avec les *Mes(X)*.

Comment associer une définition mathématique à une dépendance entre deux événements de ce type ? Au sens de la pensée naturelle, deux tels événements ne sont certainement pas 'in-dépendants', puisqu'ils concernent le *même* microétat $me_G$ produit par une et même opération de génération $G$. Pourtant la notion d'occurrence 'conjointe' qui intervient dans la définition probabiliste classique de la dépendance entre deux événements, n'a pas de sens dans ce cas, car – par construction de $T(G,V_M)$ – les deux événements $A(Xj)$ et $B(Yk)$ ne se produisent *jamais* 'simultanément', i.e. pour un même exemplaire du microétat $me_G$. : deux branches distinctes de $T(G,V_M)$ sont mutuellement incompatibles dans tous leurs éléments.

On pourrait se dire que l'incompatibilité entre $A(Xj)$ et $B(Yk)$ est une forme limite de dépendance maximale, car ces deux événements s'excluent systématiquement, et donc, puisque les probabilités séparées $p(A(Xj))$ et $p(B(Yk))$ ne sont pas nulles, on *peut* exprimer la dépendance en question en écrivant $p(A(Xj).B(Yk))=0$, ce qui est différent de $p(A(Xj))$ x $p(B(Yk))$, comme il est exigé par la définition classique. Mais on flaire un glissement, un abus d'extension de la syntaxe probabiliste classique. Et la réticence se confirme lorsqu'on considère deux événements de deux *arbres* différents, fondés sur deux opérations de génération différentes $G$ et $G'$, donc correspondant à deux microétats différents : dans ce cas *aussi* la probabilité conjointe respective est toujours nulle, si par '*simultané*' on entend 'pour un même exemplaire d'un microétat *donné*'. Mais d'autre part, pourquoi, dans ce nouveau cas, y aurait-il systématiquement *dépendance* ?

Il semble clair que le domaine de validité de la conceptualisation probabiliste classique est confiné à un seul espace de probabilité. Or l'ensemble des espaces de probabilité appartenant à l'arbre de probabilité d'un microétat *ne peut pas être incorporé à un seul espace de probabilité, parce qu'ils comportent des événements mutuellement incompatibles*. Pourtant – en conséquence de l'existence du tronc commun de l'arbre – une certaine notion de 'dépendance probabiliste' entre les événements de ces espaces distincts, semble s'imposer intuitivement, mais cette notion tout simplement dépasse la notion classique de dépendance probabiliste. *Il faudra élargir la conceptualisation probabiliste classique.* Cette conclusion se trouve renforcée juste ci-dessous.

### 4.8.4. *Le méta-phénomène aléatoire lié à un micro-état de un micro-système et méta-dépendance probabiliste*

Chaque loi de probabilité $\{p(G,Xj),\ j=1,2,...n\}$ de l'arbre $T(G,V_M)$ d'un microétat est contenue dans l'espace de probabilité qui coiffe une seule branche et elle est relative à la triade $(G,me_G,V_M(X))$ spécifique de cette branche-là. Mais le fait qu'un et même couple *(G,me$_G$)* est impliqué dans le tronc commun de tous les phénomènes aléatoires de toutes les branches de l'arbre, ainsi que dans toutes les lois de probabilité que ceux-ci produisent, conduit irrépressiblement à poser (par exemple, afin de commencer avec le cas le plus élémentaire) qu'entre la probabilité *p(Yk)* d'un événement élémentaire de la branche des *Mes(Y)* de *T(G,V$_M$)* et la loi de probabilité d'une autre branche de *T(G,V$_M$)* – considérée *globalement*, car comment détailler ? – il existe une certaine relation 'méta-probabilistes'. On pose donc que l'on a

$$p(Yk) = \mathbf{F}\{p(G,Xj)\} \quad {}^{47}$$

Ici *'**F**'* est une relation fonctionnelle dont la structure reste à être spécifiée. (Dans le cas particulier de la figure 1 on écrirait donc *p(Ck) = **F**{p(Bj)}* où *{p(Bj)}* désigne l'entière loi de probabilité sur les valeurs observables *Bj* de la grandeur *B*, qui est incompatible avec la grandeur *C*).

Il existe un *fait d'expérience* dont ici on ne peut énoncer que l'essence qualitative et qui fonde plus concrètement le postulat avancé plus haut : Si la dispersion[48] des valeurs observables *Yk* de la grandeur *Y* distribuées selon la loi *{p(Yk)}*, est grande, alors la dispersion des valeurs observables *Xj* de la grandeur *X* – incompatible avec *Y* – distribuées selon la loi *{p(Xn)}*, est petite. Et *vice versa* [49]. Quelle que soit la forme mathématique de la relation fonctionnelle *'**F**'*, ce fait implique qu'elle *existe* et d'une façon qui comporte des manifestations physiques observables.

Selon le postulat posé, l'ensemble de tous les phénomènes aléatoires liés à toutes les grandeurs mécaniques *X* redéfinies pour des microétats et corespondant à *un* même couple *(G,me$_G$)*, peut être conçu comme *un méta-phénomène aléatoire global spécifique du microétat me$_G$* [50] : l'opération de génération *G* commune qui peuple le tronc de l'arbre fonde factuellement ce méta-phénomène aléatoire global.

Considérons maintenant *exclusivement* l'ensemble de toutes les lois de probabilité qui coiffent les branches de l'arbre d'un microétat *me$_G$* – séparément du méta-phénomène aléatoire global que l'on vient de définir et qui engendre cet ensemble de lois – muni de larelation méta-probabiliste inconnue *p(Yk)=**F**{p(G,Xj)}* posées entre ses éléments. Cet ensemble de lois de probabilité reliées peut être regardé comme constituant *une*, *'la' méta-loi de probabilité spécifique de me$_G$*[51,52].

La suite des remarques qui viennent d'être faites peut être synthétisée en disant que l'arbre de probabilité *T(G,V$_M$)* d'un microétat *me$_G$* constitue un *tout* probabiliste nouveau où les différents phénomènes aléatoires et les différents espaces de probabilité qui interviennent, s'unissent de manière organique.

---

[47] Un physicien de la mécanique quantique reconnaîtra dans l'assertion posée le trait fondamental de la *théorie des transformations de base* de Dirac. Il reconnaîtra également que dans le formalisme on n'associe pas à ce trait une signification probabiliste définie, et qui en outre soit non classique.
[48] L'éparpillement statistique.
[49] Un physicien reconnaît là tout de suite l'assise qualitative du *théorème* dit 'd'incertitude' établi dans le formalisme mathématique de la mécanique quantique (à ne pas identifier au *principe* d''incertitude' de Heisenberg).
[50] Les logiciens emploient quelquefois le mot 'méta' pour désigner le langage *d'immersion* du langage considéré. J'attire l'attention qu'ici, au contraire, ce mot est employé comme signifiant *'au dessus'*, 'qui se place à un niveau conceptuel plus élevé que celui où émerge le concept classique d'un seul phénomène aléatoire'.
[51] Le singulier souligne l'unité instillée par l'unicité de la paire *(G,me$_G$)* qui intervient dans toutes ces lois de probabilité.
[52] Mackey [1963], Gudder [1976], puis Suppes, Beltrametti, et d'autres (probablement à ce jour même), ont recherché – par des voies purement mathématiques – une formulation satisfaisante pour une méta-loi de probabilité associable à l'entier vecteur d'état |Ψ> correspondant à un microétat *me$_G$*. Ici le fondement qualitatif d'une telle formulation se fait jour naturellement et il permettra peut-être d'identifier la forme pertinente de la fonctionnelle dénotée *F*.

On pourrait avoir tendance à subsumer les relations méta-probabilistes posées plus haut, au concept probabiliste classique de corrélation probabiliste. Mais en fait les caractéristiques spécifiques du type de relations signalées ici, n'ont aucun correspondant explicitement élaboré dans le calcul classique des probabilités. Ici on est en présence de méta-relations probabilistes d'un type bien défini, foncièrement liée au fait que les lois de probabilité *{p(Yk)} et {p(Xn)}* concernent le même microétat. La conceptualisation probabiliste classique n'individualise pas cette sorte particulière de corrélations et la condition spécifique de leur émergence.

### *4.8.5. Conclusion sur le concept d'arbre de probabilté d'un micro-état d'un seul micro-système*

Le concept qualitatif d'arbre de probabilité d'un micro-état $me_G$ de un micro-système, qui déploie la structure globale d'espace-temps liée à la forme descriptionnelle $D_M/G,me_G,V_M/$, met au grand jour des implications importantes de cette forme descriptionnelle. Celles-ci, avec notamment la signification méta-probabiliste propre de l'ensemble des lois de probabilité qui coiffent les branches d'un arbre $T(G,V_M)$ – qu'il n'est pas possible d'incorporer à un espace unique – constituent un apport nouveau au concept général de probabilité (cf. [Mugur-Schächter 2006]).

## 4.9. Deux généralisations du concept d'arbre de probabilté d'un microétat de un microsystème

Dans 3.8 nous avons introduit seulement le cas fondamental de l'arbre de probabilité d'un microétat à génération 'simple' (pas composée) et comportant un seul microsystème. Que devient le concept d'arbre de probabilité dans le cas d'un micro-état à génération composée, ou de plusieurs micro-systèmes, ou les deux à la fois ?

### *4.9.1. L'arbre de probabilité d'un micro-état de deux ou plusieurs micro-systèmes*

Le concept d'arbre de probabilité établi pour un micro-état d'un seul micro-système se transpose d'une manière évidente à un micro-état de deux ou plusieurs micro-systèmes.

Pour fixer les idées considérons un micro-état de *2* micro-systèmes, *S1* et *S2*, engendré par une opération de génération d'état $G_{12}$ qui a impliqué à la fois les deux microsystèmes *S1* et *S2*. Symbolisons ce micro-état par $me_{G12}$ et soit $T(G_{12},V_M)$ l'arbre de probabilité de ce micro-état. Dans ce cas, une opération de mesure opérée sur $me_{G12}$, si elle est complète, comporte par définition (cf. 3.4) une mesure $Mes_1(X)$ d'une grandeur *X* opérée sur *S1*, et une mesure $Mes_2(Y)$ d'une grandeur *Y* opérée sur *S2*. Dénotons par $Mes_{12}(XY)$ une telle mesure complète, où l'ordre d'écriture *XY* indique, par référence à l'ordre des indices *1* et *2*, que *X* concerne le microsystème *S1* et *Y* concerne le microsystème *S2*. Soit *{X1,X2,...Xj,...Xϕ}* les valeurs de *X* qui sont prises en considération comme pouvant apparaître par une mesure de *X* (qu'elle ait été faites sur *S1* ou sur *S2*, n'importe) où $\phi$ est un entier 'final' qui dénote le nombre de ces valeurs de *X*. Soit *{Y1,Y2,...Yk,...Yφ}* les 'valeurs' de *Y* qui peuvent apparaître par une mesure de *Y* (qu'elle ait été faites sur *S1* ou sur *S2*, n'importe) où $\varphi$ est un autre entier 'final' qui dénote le nombre des valeurs de *Y* qui sont sont prises en considération comme pouvant apparaître par une mesure de *Y*. Les événements élémentaires de l'espace de probabilité qui coiffe la branche de l'arbre de probabilité $T(G_{12},V_M)$ du micro-état $me_{G12}$ correspondant à une mesure complète $Mes_{12}(XY)$ (une $Mes_1(X)$ et une $Mes_2(Y))$ consistent en toutes les paires possibles $Xj1Yk2$ d'une valeur $X$ enregistrée pour un exemplaire de micro-système $S1$, dénotée $Xj1$, et une valeur de $Y$ enregistrée pour un exemplaire de micro-système $S2$, dénotée $Yk2$ ; c'est donc l'ensemble de paires $\{Xj1Yk2\}$, $j=1,2,.....$ $\phi$, $k=1,2,.....$ $\varphi$. On peut avoir en particulier $X \equiv Y$. Mais en général les deux grandeurs considérées sont différentes. Elles peuvent même être non compatibles, dans le cas d'un micro-état de *un* micro-système. Ces distinctions n'ont pas de conséquences sur le fait suivant.

Pour un micro-état de *deux* micro-systèmes les grandeurs $X$ et $Y$ intervenant dans une $Mes_{12}(XY)$ sont toujours *compatibles* parce qu'elles sont effectuées sur deux micro-systèmes *distincts*, c'est à dire, parce qu'elles s'incorporent, respectivement, à une mesure d'indice *1* et une mesure d'indice *2* que l'on peut toujours réaliser *simultanément* sur un seul *exemplaire* du micro-état $me_{G12}$.

Il en résulte que les deux mesures de la paire $(Mes_1(X), Mes_2(Y))$ qui constitue l'acte de mesure complète $Mes_{12}(XY)$, appartiennent toujours à une *même* branche de l'arbre de probabilité $T(G_{12},V_M)$ de $me_{G12}$, celle des $Mes_{12}(XY)$, qui est coiffée par l'espace de probabilité fondé sur les événements élémentaires de l'ensemble de paires $\{X1jY2k\}$, $j=1,2,.....$ $\phi$, $k=1,2,.....$ $\varphi$. Donc les deux événements $Xj1$ et $Yk2$ produits par les mesures $Mes_1(X)$ et $Mes_2(Y)$ qui constituent un acte de $Mes_{12}(XY)$ – OBSERVABLES SÉPARÉMENT – appartiennent toujours à un *même* événement élémentaire d'un *même* espace de probabilités de l'arbre de probabilité de $me_{G12}$. Cela *quelle que soit la distance d'espace-temps* qui sépare les deux événements physiques 'enregistrement de la valeur $Xj$ concernant $S1$' et 'enregistrement de la valeur $Yk$ concernant $S2$' (i.e. la distance qui sépare leurs deux ici-maintenant respectifs). Or à l'intérieur d'un seul espace de probabilité, le concept *classique* de dépendance probabiliste est *valide en dehors de toute restriction d'espace-temps*[53].

On peut montrer que les magistrales analyses de Michel Bitbol ([1996] et [2000]) concernant la 'contextualité' des descriptions de microétats et les expressions verbales de celle-ci se trouvent en strict accord avec le concept d'arbre de probabilité d'un microétat de deux ou plusieurs microsystèmes. En lisant ces analyses attentivement on comprend mieux à quel point ce concept permet de comprendre la complexité surprenante de la situation examinée.

### *4.9.2. L'arbre de probabilité d'un micro-état à génération composée à un ou plusieurs micro-systèmes (micro-état quelconque)*

Considérons d'abord un microétat $me_{G(G1,G2)}$ à génération composée, d'un seul microsystème $S$, engendré par l'action sur $S$ d'une opération de génération $G(G1,G2)$ où se composent seulement deux opérations de génération $G1$ et $G2$ (cf. *3.2.4*). Le fait suivant est à noter soigneusement. Le microétat à génération composée $me_{G(G1,G2)}$ engendré pour le micro-système $S$ par l'opération de génération composée $G(G1,G2)$ – comme *tout* microétat donné – est lié à *un seul* arbre de probabilité $T(G(G1,G2),V_M)$ (où $V_M$ est une vue mécanique définie pour des microétats). Toutefois on s'attend irrépressiblement à ce qu'il y ait une relation définie entre le microétat à génération composée $me_{G(G1,G2)}$ effectivement produit par $G(G1,G2)$ et les deux microétats $me_{G1}$ et $me_{G2}$ qui se *seraient* réalisés, respectivement, *si* soit $G1$ seul soit $G2$ seul avait agi sur $S$. Et l'on s'attend également à ce que cette relation actuel-virtuel se reflète de quelque manière dans une relation entre l'arbre

---

[53] Le 'problème de localité' concerne un micro-état de deux micro-systèmes : on imagine l'étonnement que peuvent susciter ses caractéristiques lorsqu'on ignore la structure d'arbre de probabilité. Dans Mugur-Schächter [2008] les relations entre mécanique quantique et relativité, et notamment l'assertion de 'non localité' du formalisme quantique, pourront être précisées.

*T(G(G1,G2),V$_M$)* du microétat effectivement réalisé *me$_{G(G1,G2)}$* et les arbres *T(G1,V$_M$)* et *T(G2,V$_M$)* des deux microétats *me$_{G1}$* et *me$_{G2}$* qui auraient pu se réaliser pour *S* si *G1* seul ou *G2* seul, respectivement, avait agi sur *S*. Toutefois, dans l'approche présente qui ignore le formalisme mathématique de la mécanique quantique et est exigée strictement qualitative, on ne peut répondre à cette attente qu'en signalant un fait d'observation directe exprimé dans les termes négatifs suivants.

Imaginons que tous les trois arbres *T(G1,V$_M$)*, *T(G2,V$_M$)*, *T(G(G1,G2),V$_M$)* aient été réalisés physiquement et que les lois de probabilité qu'ils comportent peuvent être comparées expérimentalement. Soit *X* une grandeur qui contribue à la vue mécanique *V$_M$*. Soit *p12(Xj)* la probabilité trouvée dans *T(G(G1,G2),V$_M$)* pour l'événement qui consiste en l'enregistrement de la valeur *Xj* de *X*, et soient *p1(Xj)* et *p2(Xj)*, respectivement, les probabilités trouvées pour ce même événement dans les arbres *T(G1,V$_M$)* et *T(G,V$_M$)* des microétats *me$_{G1}$* et *me$_{G2}$*. La probabilité *p12(G(G1,G2),Xj)* n'est en général *pas* la somme des probabilités *p1(G1,Xj)* et *p2(G2,Xj)* : en général on trouve que

$$p12(G(G1,G2),Xj) \neq p1(G1,Xj) + p2(G2,Xj)$$

En *ce* sens, le microétat *me$_{G(G1,G2)}$* ne peut pas non plus être regardé comme la 'somme' des deux microétats *me$_{G1}$* et *me$_{G2}$*. On peut exprimer cette situation en disant que les microétats *me$_{G1}$* et *me$_{G2}$* se 'réalisent' tous les deux à l'intérieur du microétat à génération composée *me$_{G(G1,G2)}$*, mais qu'ils y 'interagissent' ou 'interfèrent' en modifiant mutuellement les effets probabilistes observables que chacun produit lorsqu'il se réalise séparément[54].

Les caractères d'un arbre de probabilité correspondant à un micro-état à génération composée qui, en plus, est aussi un micro-état de deux ou plusieurs micro-systèmes, découlent facilement des spécifications précédentes.

### 4.10. Conclusion sur le concept général d'arbre de probabilité d'un microétat

Les généralisation du concept d'arbre de probabilité d'un seul microétat qui viennent d'être spécifiées dans 3.8 et 3.9, confrontées aux définitions de 3.4, entraînent que le concept d'arbre de probabilité s'étend à *toutes* les sortes possibles de microétats, tout en acquérant dans chacun des cas généralisés certaines caractéristiques nouvelles, spécifiques du cas considéré[55]. Donc le symbole *D$_M$/G,me$_G$,V$_M$/* – où désormais *me$_G$* désigne un microétat absolument quelconque, impliquant un seul microsystème ou plusieurs ou une opération de génération simple ou composée – est doté d'une signifiance non restreinte.

Le contenu de chacun des symboles qui interviennent dans la l'écriture *D$_M$/G,me$_G$,V$_M$/*, a déjà été amplement explicité. Pourtant le désigné de cette notation restait abstrait, non intuitif. Tandis que la représentation d'espace-temps *T(G,V$_M$)* de la description *D$_M$/G,me$_G$,V$_M$/* en rend immédiatement présents à l'intuition tous les contenus, aussi bien que leurs relations. Le trait le plus marquant qui, sur le schéma d'un arbre de probabilité *T(G,V$_M$)* saute aux yeux, est l'importance cruciale du processus de génération *G* d'un microétat. Ce processus régit des dépendances observables qui souvent sont inattendues – surtout dans le cas d'un micro-état de deux micro-systèmes – et il guide pour représenter et comprendre ces dépendances.

---

[54] Lors de la construction du formalisme mathématique de la mécanique quantique, cette situation physique-représentationnelle a suggéré (plus ou moins explicitement) l'utilité de l'introduction d'une représentation *vectorielle* caractérisée par un axiome de 'superposition' *au sens mathématique*. En outre, cette même situation suggère aussi la possibilité et l'intérêt de la définition explicite d'un calcul avec des arbres de probabilités considérés *globalement* ([Mugur-Schächter [1991] et [2008]).
[55] Pour un physicien cela exprime que le concept d'arbre de probabilité sous-tend l'entier formalisme quantique.

Ce fait est chargé de conséquences qui, une fois perçues dans le cas particulier des microétats, peuvent s'étendre à l'entière conceptualisation du réel. Notamment, le concept d'arbre de probabilité se transpose à la conceptualisation probabiliste en général, où il *groupe* des phénomènes aléatoires distincts mais qui introduisent tous une même entité-objet, en reliant les effets observables de ces phénomènes aléatoires dans un 'pattern' non trivial de méta-dépendances probabilistes qui dans la théorie classique des probabilités ne sont pas définies spécifiquement. Ce 'pattern' revient à spécifier explicitement un type de corrélations probabilistes qui, bien que particulier, est doté d'une grande importance parce qu'il singularise ce qu'on peut regarder comme *les caractéristiques probabilistes observables d'une genèse commune* (Mugur-Schächter [2006] pp. 250-256).

### 4.11. Remarque sur l'évolution d'un microétat

On peut se demander si, en respectant les contraintes imposées ici, il est possible de dire quelque chose sur l'évolution d'un microétat dans des conditions extérieures données[56]. La réponse est positive.

Imaginons une opération $G$ de génération du microétat étiqueté $me_G$. Soit $t_o$ le moment que l'on assigne à la fin de l'opération $G$, donc au début de l'époque d'existence de l'exemplaire considéré du microétat $me_G$. Supposons aussi qu'au moment $t_o$ on ne démarre aucune opération de mesure sur cet exemplaire de l'état $me_G$, mais qu'on le laisse subsister pendant une durée $\Delta t = t - t_o$ dans des conditions extérieures, disons $\mathcal{CE}$, que l'on aura choisi de mettre en place (champs macroscopiques). Rien, dans la démarche élaborée ici, ne s'oppose à ce qu'on considère que l'association de l'opération de génération $G$, avec les conditions extérieures $\mathcal{CE}$ et le passage de la durée $\Delta t = t - t_o$, constituent une nouvelle opération de génération $G'=f(G, \mathcal{CE}, \Delta t)$, qui produit un nouveau microétat $me_G'$ que l'on peut étudier par des mesures exactement de la même manière que $me_G$. En outre, rien ne s'oppose non plus à ce que la durée $\Delta t$ soit choisie aussi petite (ou grande) qu'on veut. On peut donc étudier un ensemble de microétats correspondant à un ensemble d'opérations de génération $G'=f(G, \mathcal{CE}, \Delta t)$ où $G$ et $\mathcal{CE}$ *restent les mêmes*, cependant que $\Delta t$ change de la façon suivante. L'on accomplit un *ensemble* de mesures qui permet de spécifier l'entier arbre de probabilité $T(G', V_M)$ avec $G'=f(G, \mathcal{CE}, \Delta t1)$ et $t1$ dans $\Delta t1 = t1 - t_o$ très proche de $t_o$ ; puis on accomplit un ensemble de mesures avec $G'=f(G, \mathcal{CE}, \Delta t2)$ et $t2$ dans $\Delta t2 = t2 - t_o$ plus éloigné de $t_o$ ; et ainsi de suite, jusqu'à un arbre final correspondant à $G'=f(G, \mathcal{CE}, \Delta tf)$ où $\Delta tf = (tf - t_o) \equiv t - t_o$. De cette manière on peut constituer concernant le microétat *de départ* $me_G$, des connaissances qui sont en principe équivalentes à celles qu'offre une loi d'évolution.

Evidemment, jamais une démarche qualitative ne pourra atteindre l'efficacité de représentation d'un système bien construit d'algorithmes mathématiques. Mais ici il ne s'agit que d'une remarque de principe qui fonde la possibilité de parler de l'évolution du microétat $me_G$.

---

[56] Dans le formalisme quantique l'équation d'évolution de Schrödinger ne s'applique pas directement au microétat étudié, mais seulement au descripteur mathématique |$\Psi$> à l'aide duquel (avec les autres algorithmes aussi) l'on calcule l'ensemble des lois de probabilité qui coiffent ce qui, ici, a été dénommé l'arbre de probabilité de ce microétat. Et cette loi fait intervenir le hamiltonien d'évolution, i.e. le descripteur de la grandeur mécanique 'énergie totale'. Tout cela, qui est mathématique et à signification spécifiquement mécanique, échappe au domaine de pertinence de l'approche développée ici.

### 4.12. L'infra-mécanique quantique

Le processus qualitatif de construction de descriptions de microétats qui vient de s'accomplir ici, et son résultat, constituent ce que j'appelle *l'infra-mécanique quantique*.

L'infra-mécanique quantique met en évidence une forme descriptionnelle *'transférée', primordialement probabiliste*, marquée par trois inamovibles relativités qui sont rappelés dans le symbole $D_M/G,me_G,V_M/$ assigné à cette forme : les relativités à l'opération de génération $G$ du microétat $me_G$ à étudier, à ce microétat $me_G$ lui-même, et à la grille de qualification (la vue mécanique globale $V_M$ composée de l'ensemble des vues-aspect $V_M(X)$ correspondant aux grandeurs mécaniques $X$ re-définies pour un microétat).

Le concept de description transférée primordiale $D_M/G,me_G,V_M/$ est foncièrement distinct du concept de description au sens classique. Selon le concept classique désigné par le mot 'description' l'entité-objet-de-description préexiste en tant qu'un 'objet' au sens du langage courant, et elle est simplement *sélectionnée* afin d'être décrite, *via* une propriété (prédicat) préexistante elle aussi et qui qualifie l'entité-objet-de-description en même temps qu'elle la sélectionne. *Rien*, dans la conceptualisation classique, ne conduit à concevoir une description du type $D_M/G,me_G,V_M/$, ni la pensée et les langages courants, ni les grammaires, ni la logique, ni l'entière science classique avec ses mathématiques et notamment ses probabilités.

Quant au formalisme quantique, il n'en émane au sujet du type descriptionnel $D_M/G,me_G,V_M/$ que des considérations partielles et entremêlées.

L'arbre de probabilité $T(G,V_M)$ associé à la forme descriptionnelle $D_M/G,me_G,V_M/$ en étale l'entière structure d'espace-temps synthétisée dans un schéma géométrisé. Ce schéma met en évidence des significations probabilistes nouvelles qui s'inscrivent dans un développement à venir de la théorie générale des probabilité ; en outre ce schéma admet deux sortes de généralisations en conséquence desquelles le concept d'arbre de probabilité recouvre l'entier domaine de pertinence du formalisme quantique.

Telle est l'essence de l'infra-mécanique quantique.

Un œil averti discerne clairement dans l'infra-mécanique quantique, l'essence du contenu épistémologique de la mécanique quantique fondamentale, épurée quant à son statut descriptionnel : toute insertion de modélisations qui feraient coalescence avec la strate classique de la conceptualisation, subséquente selon l'ordre génétique des types descriptionnels, est éliminée. C'est la condition-cadre générale *CCG* qui accomplit cette épuration, en démarquant un vide – intéressant – laissé par l'éradication des éléments implicites de modélisation instillés dans le formalisme mathématique de la mécanique quantique à partir de la mécanique classique, afin de pouvoir associer aux microétats des qualifications spécifiquement mécaniques.

La dénomination 'infra-mécanique quantique' veut donc dire *dans le substrat de la mécanique quantique,* **en dessous du formalisme quantique**.

Elle n'indique *pas* une mécanique mais un "en dessous de [la mécanique quantique]" qui, elle, *est* une 'mécanique'. Il faudrait écrire infra-[mécanique quantique]. Mais ce serait fastidieux.

L'infra-mécanique quantique est un tout cohérent développé *indépendamment* du formalisme quantique. C'est une sorte de *théorie épistémo-physique*, si l'on peut dire, qui possède des contours propres et un contenu spécifique doté d'un mode conceptuel-descriptionnel de fonctionner qui est défini exhaustivement et de manière explicite, avec détail et rigueur. Cela fait contraste avec la façon de laquelle se laissent discerner les aspects épistémologiques impliqués dans les descriptions de microétats lorsqu'on veut les appréhender en n'examinant que le formalisme mathématique de la mécanique quantique, sans prendre appui direct sur les contraintes qui émanent de la situation cognitive dans laquelle on se trouve.

Par exemple, lorsqu'on cherche dans le formalisme mathématique de la mécanique quantique des reflets du concept d'arbre de probabilité d'un microétat tel qu'il s'est constitué dans l'infra-mécanique quantique, on peut en trouver, bien entendu. Mais le concept lui-même n'est *pas* formé dans le formalisme mathématique. Donc dans la mécanique quantique, les aperçus intuitifs que ce concept offre lorsqu'on en examine la structure intégrée à l'intérieur de l'infra-mécanique quantique, restent cachés. Quant aux aspects probabilistes non-classiques que la structure d'arbre de probabilité d'un microétat met en évidence – notamment l'organisation non triviale des dépendances et des indépendances probabilistes et la notion d'une méta-loi de probabilité globalement caractéristique d'un microétat donné – ils restent dépourvus de définition dans le formalisme quantique, bien qu'ils y soient implicitement incorporés. Les algorithmes quantiques ne contiennent que des bribes du concept d'arbre de probabilité d'un microétat, éparpillées dans des façons de dire plus ou moins adéquates et des écritures plus ou moins cryptiques[57]. Ce concept s'y trouve comme dans un état paradoxal de traces d'un jamais fait. L'identification *a posteriori* de ce concept à l'intérieur des algorithmes mathématiques du formalisme quantique – sous le guidage de la représentation qualitative construite ici – produit l'impression curieuse d'une sorte de reconstitution à la Cuvier à partir de fragments tirés d'une roche. Mais une reconstitution réfléchie dans un miroir de temps : par des fouilles dans un formalisme qui existe depuis longtemps, on met au jour des éléments d'une représentation des microétats qui vient juste de naître, et d'une organisation appartenant à un calcul des probabilités *futur*.

Quoi qu'il en soit, on se trouve désormais en possession d'une structure *de référence* non mathématisée pour attaquer globalement l'ensemble des questions d'interprétation soulevés par le formalisme quantique. Construite indépendamment du formalisme quantique, cette structure de référence incorpore les réponses méthodologiques explicites à toutes les contraintes épistémologiques explicitées elles aussi, qu'imposent la situation cognitive mise en jeu et les modes humains de conceptualiser. Il paraît donc probable que l'on réussira à sortir de ce piège dans lequel on se débat depuis plus de 70 ans, qui consiste à aborder les questions d'interprétation soulevées par le formalisme quantique d'une manière entachée de circularité, en utilisant à chaque fois (plus ou moins) le formalisme quantique lui-même parce que la mécanique quantique n'est rien de plus que ce formalisme et donc n'offre aucun autre terrain d'appui que ce formalisme, que l'on est en outre forcé de mêler à la pensée courante, qui est classique, puisqu'on doit aussi parler et écrire avec des mots courants. Cette situation hybride et immobilisante du point de vue logique, pourra enfin être brisée.

Les conséquences de cette nouvelle possibilité sont à construire (on pourra les jauger dans [Mugur-Schächter 2008]). Mais les élucidations qui d'ores et déjà se sont produites à l'intérieur même de l'infra-mécanique quantique considérée isolément, concernant quelques problèmes majeurs d'interprétation qui y émergent dans des termes quasi-identiques à ceux dans lesquels ils émergent aussi face au formalisme quantique, permettent de l'optimisme à l'égard de cette entreprise. Ces élucidations font penser à l'ombre jetée sur une plaine, vers midi, par un nuage opaque mais local: lorsqu'on constate l'ombre et on lève les yeux, on l'explique par la présence du nuage.

---

[57] Nous l'avons déjà beaucoup souligné, mais c'est tellement important que je répète une fois de plus : Le statut descriptionnel des algorithmes quantiques est notablement obscurci par le fait qu'ils *mélangent* les caractères qui proviennent des conditions cognitives et des modes humains généraux de conceptualisation, avec des caractères qui sont l'effet de choix de représentations mathématiques *modélisantes* tirées des formulations analytiques de la mécanique classique afin de pouvoir atteindre le but de définir des opérations de mesure dont on puisse penser qu'elles qualifient des microétats en termes *mécaniques*. Tandis que l'infra-mécanique quantique reste rigoureusement pure de tout mélange de cette sorte. Elle reste liée *exclusivement* au niveau descriptionnel primordial des descriptions transférées. Les spécificités descriptionnelles de ce tout premier niveau de conceptualisation, sont préservées de l'insertion subreptice d'éléments descriptionnels qui ont pris naissance dans le niveau foncièrement distinct de la pensée classique où l'on manipule des 'objets'-modèles nés – nécessairement – *sur la base* de descriptions primordiales transférées ( réflexes ou seulement implicites). La rançon de la préservation – méthodologique – de cet état de 'pureté' descriptionnelle, est l'impossibilité de spécifier des mesures de grandeurs 'mécaniques'.

## 5. Commentaire global

L'infra-mécanique quantique étant maintenant délinéée, il est possible de l'examiner de l'extérieur, dans sa globalité, afin de repérer la place qu'elle occupe face à trois aspects d'importance majeure : la façon de laquelle y interviennent l'espace, le temps et les géométries d'espace-temps ; le type de consensus qu'elle permet ; ce qu'elle implique sur les relations entre microphysique actuelle et les théories de la relativité d'Einstein.

Par exception, les considérations qui suivent s'adressent plus spécialement aux physiciens. Elles n'expriment ni des vues actuellement consensuelles, ni même des vues personnelles qui seraient toutes bien stabilisées. Il s'agit d'une très brève exploration informelle du terrain conceptuel introduit par l'infra-mécanique. Cette exploration est soumise aux lecteurs sous forme de remarques et de *questions*.

### 5.1. Remarques générales sur espace, temps et géométrie

Un individu humain normal ne peut percevoir, ni même seulement concevoir, une entité *physique* – objet, événement, substance – sans la placer dans l'espace et le temps. Ceci est un fait psychique qu'il paraît difficile de contester. Kant a exprimé ce fait en posant que l'espace et le temps sont deux formes *a priori* de l'intuition.

Dans ce qui suit j'adopte ce postulat.

J'ajouterai un aveu épistémologique : réciproquement, je n'arrive pas à concevoir de l'espace ou du temps en l'absence – strictement – de *toute* existence physique, ou au moins d'une émanation d'une existence physique, comme mon attention cachée quelque part pour surveiller, et mon souffle qui en quelque sorte dénombre qualitativement du passage de temps.

Le postulat kantien rappelé plus haut n'implique aucune *structure* d'espace, ou de temps, ou d'espace-temps. Il n'affirme qu'un fait concernant le psychisme des individus humains. Comment, alors, s'engendrent des 'géométries' d'espace, ou de temps, ou d'espace-temps ?

Henri Poincaré [1898] a notamment élaboré l'idée que l'assignation à l'espace – considéré comme une donnée première non structurée[58] – d'une structure *géométrique* euclidienne, émerge par l'intégration dans un système unique, stable et cohérent, de toute la diversité des aspects spatiaux qui se manifestent dans les *interactions* naturelles kinésiques et sensorielles entre les individus humains et du réel physique. Cette intégration, la géométrie euclidienne, s'exprime par un système de relations entre, exclusivement, des concepts spatiaux abstraits, points, lignes, figures (cf *1.5.3*), où les relativités à tel ou tel 'point de vue' sont *évacuées* [59], comme aussi les éléments physiques et biologiques sensoriels qui ont participé aux interactions.

Einstein (peu après Poincaré, mais sans faire référence à lui) a fait une assertion similaire dans son exposé de la relativité restreinte. Il y affirme que les interactions *de mesure* que des observateurs inertiels réalisent avec des mobiles macroscopiques, *via* des signaux lumineux, conduisent à la géométrie d'espace-*temps* 'Minkowski-Einstein'. Mais notons que, à la différence de la géométrie d'espace euclidienne, la géométrie d'espace-temps de Minkowski-Einstein n'est *pas* intégrative jusqu'au bout. Elle n'intègre (dans un schéma à deux cons de lumière et deux ailleurs) que l'ensemble des interactions de mesure d'*une seule classe* d'observateurs inertiels ayant tous des états identiques de mouvement inertiel. Une méta-intégration de toutes ces intégrations dans une synthèse unique de toutes les interactions de mesure de tous les observateurs inertiels, n'est pas opérée dans la relativité restreinte. C'est pour cette raison que la géométrie Minkowski-Einstein ne permet pas de définir une causalité générale cohérente.

---

[58] Poincaré ne fait pas référence à Kant, pour autant que je sache.
[59] Elles constituent un autre système, annexe, la géométrie euclidienne projective.

Quelques trente années plus tard Husserl a développé sa célèbre phénoménologie où il décrit les processus de *constitution transcendantale des 'objets' physiques*[60]. Une entité physique donnée est toujours perçue exclusivement de tel ou tel point de vue particulier, elle n'est jamais perçue de tous les points de vue possibles. Une perception relative à un point de vue donné, conduit à une description correspondante *particulière* de cette entité. Mais l'ensemble de toutes les différentes descriptions particulières possibles d'une entité donnée, est intégré par l'esprit dans un concept abstrait d'un 'objet' conçu comme l'invariant de cet ensemble (c'est ce qui, ici, a été à plusieurs reprise désigné par l'expression 'objet'-modèle).

Il paraît naturel (sinon même inévitable) de regarder le concept de géométrie euclidienne de Poincaré comme *un objet-cadre général* construit comme l'invariant de *tous* les ensembles d'interactions humaines courantes avec des entités physiques, dont chacun engendre un 'objet' au sens de Husserl. Dans les sciences cognitives actuelles une vue de cette sorte commence à poindre chez des neurobiologistes et elle est étayée par des philosophes (Berthoz et Petit [2007])[61].

Quant à la relativité générale, ce qu'elle introduit n'est *pas* ce que j'accepte d'appeler une géométrie d'espace-temps. C'est une *représentation géométrique* où une certaine géométrie d'espace-temps qui n'est pas explicitée de façon isolée, est d'emblée mise en coalescence avec un codage géométrique de distributions variables de masses et de champs, cela sous la contrainte d'un but : le but que la loi de mouvement de tout mobile macroscopique donné, observé par tout ensemble d'observateurs – *via* des signaux lumineux –, soit construite par tous ces observateurs comme une géodésique de la représentation géométrique mentionnée.

De ces considérations, retenons ce qui suit.

Une géométrie d'espace ou d'espace-temps émerge comme une 'constitution transcendantale' d'une structure intégrative abstraite fondée sur un ensemble *d'interactions d'un type donné* (des interactions naturelles, ou de mesures macroscopiques opérées par des observateurs physiciens *via* des signaux lumineux, ou encore, sans doute, des interactions d'autres catégories possibles).

*Lorsque le type d'interactions considéré change, la géométrie qui en émerge change elle aussi.*

---

[60] Rappelons que chez Husserl 'transcendantal' veut simplement dire 'par interaction'.
[61] On pourrait regarder la description d'une entité physique macroscopique donnée, fondée exclusivement sur telle ou telle structure perceptive particulière éprouvée par un individu humain à la suite d'interactions sensorielles avec cette entité physique, comme une description de l'entité physique qui est *transférée* sur les enregistreurs d'appareils sensoriels biologiques de l'individu humain. Mais dans la mesure où l'entité-objet-de-description qui intervient est connue à l'avance, il s'agirait d'une hybridation du concept de description transférée primordiale, avec le concept d'objet'-modèle-macroscopique.

## 5.2. Infra-mécanique quantique et mécanique quantique
### *versus*
### espace, temps et géométrie

Comment interviennent l'espace et le temps dans l'infra-mécanique quantique? Ils y interviennent d'abord fondamentalement, en tant que formes *a priori* de l'intuition des concepteurs humains qui ne peuvent concevoir des entités physiques sans les loger dans de l'espace et du temps. Ils y interviennent également dans les assignations de coordonnées d'espace et de temps impliquées dans les définitions des opérations $G$ de génération d'un microétat et des opérations de qualification par des *Mes(X)*. Ces assignations de coordonnées dépassent l'appartenance aux formes *a priori* de l'intuition humaine. Elles appartiennent à une activité scientifique qui exige l'incorporation dans une structure géométrique. Mais *laquelle* ? La géométrie d'espace euclidienne associée au temps absolu de Galilée et Newton, ou bien des géométries d'espace-temps de Minkowski-Einstein ?

Examinons la situation de plus près. Selon l'infra-mécanique quantique, l'algorithme qualitatif de construction de connaissances concernant un microétat $me_G$ conduit à des descriptions primordiales transférées $D_M/G,me_G,V_M/$. La genèse et le résultat de cette sorte de descriptions sont explicités dans la structure $T(G,V_M/)$ d'arbre de probabilité d'un microétat. Le protocole de construction de cet arbre comporte la répétition un très grand nombre de fois, de la succession *[G.Mes(X)]* où $X$ varie sur l'ensemble des grandeurs mécaniques redéfinies pour des microétats. Chaque succession *[G.Mes(X)]* implique les données d'espace-temps $d_G.(t_G-t_o)$ et $d_X.(t_X-t_G)$ (voir la *Fig.1*) ainsi que – en général – les coordonnées d'espace et de temps des manifestations physiques $\mu_j$, $j=1,2,...m$ observées sur les enregistreurs de l'appareil $A(X)$ mis en jeu. Que peut-on dire concernant ces différentes qualifications d'espace et de temps qui interviennent ?

La première remarque qui vient à l'esprit est que toutes ces qualifications s'appliquent *directement*, non pas à des interactions entre l'expérimentateur et les microétats $me_G$ étudiés, mais aux interactions de l'expérimentateur avec les appareils macroscopiques utilisés pour réaliser les opérations $G$ et *Mes(X)*. Chaque expérimentateur accomplit des opérations $G$ et *Mes(X)* dans son référentiel propre lié à son laboratoire, sans observer rien d'autre que ses appareils, leur état et les marques physiques qui s'y affichent, sans interagir avec rien d'autre. Nulle part n'interviennent ni des mobiles observés par plusieurs observateurs à la fois, ni des signaux lumineux pour accomplir les observations et notamment pour assigner des coordonnées d'espace et de temps. En outre, le but n'est pas d'établir, pour un mobile donné qui est observé à partir d'états d'observation différents, une loi de mouvement dont la forme soit invariante aux changements d'états d'observation inertiels. Le but, dans ce cas, est d'établir des distributions de probabilité fondées sur des dénombrements de marques physiques amassées sur des enregistreurs d'appareils, où elles peuvent attendre leur lecture et leur dénombrement aussi longtemps qu'on veut. *Tout cela paraît entièrement étranger aux géométries Minkowski-Einstein. Tout ce qui concerne les interactions entre l'expérimentateur et des entités physiques, semble s'intégrer à la géométrie euclidienne et au temps social conventionnel utilisés dans la physique classique.*

Toutefois on peut encore hésiter. La forme descriptionnelle qualitative $D_M/G,me_G,V_M/$ mise en évidence par l'infra-mécanique quantique pourrait induire en erreur, peut-on se dire, par sa généralité, par l'abstraction qui y est faite des définitions des opérations de mesure. En effet *dans le formalisme de la mécanique quantique* on spécifie entièrement la définition de chaque opération de mesure et cette définition est construite sur la base de modélisations de prolongement de la mécanique classique. Celles-ci peuvent notamment introduire des mesures de durées et de distances que l'on assigne *au microétat étudié lui-même*, conçu subrepticement comme un mobile microscopique qui, du point de vue des déplacements et aux dimensions près, serait assimilable à un mobile macroscopique ; ou du moins, qui impliquerait un élément assimilable à un tel mobile. C'est effectivement le cas pour la méthode du temps de vol discutée dans *3.3.2.1*. En ces circonstances, est-il toujours tout à fait pertinent de dire que les données d'espace et de temps qui interviennent dans une description primordiale transférée concernent toujours exclusivement les interactions entre l'expérimentateur et ses appareils ?

Cette objection dépasse l'infra-mécanique quantique, elle concerne les descriptions primordiales transférées de la *mécanique quantique*. Examinons-la en prenant appui sur la vue exprimée concernant les géométries. Admettons que dans le cas des enregistrements d'une durée et d'une distance lors d'un acte de mesure par la méthode du temps de vol, il s'agirait d'une interaction entre l'expérimentateur et le microétat étudié. Quel sens y aurait-il de faire usage, pour cette raison, du formalisme de la relativité restreinte qui a été construit sur la base d'interactions où les signaux lumineux tiennent une place centrale? des interactions accomplies par plusieurs observateurs qui tous observent un même mobile à partir de référentiels inertiels distincts? des interactions dont les résultats sont ensuite organisés (*via* de nouvelles variances assignées valeurs des grandeurs mécaniques) de telle manière que tous les observateurs en tirent une loi de mouvement de la même forme? *Rien* de tout cela ne correspond au cas d'une mesure de la quantité de mouvement d'un microétat par la méthode du temps de vol. Il n'y a dans ce cas ni observation par plusieurs observateurs inertiels différents, ni signaux lumineux, ni le but de construire une loi de mouvement invariante aux changements de l'état d'observation.

### *5.2.1. Infra-mécanique quantique et mécanique quantique versus consensus*

Pour le moment donc, le seul type de consensus que l'on peut exiger dans le cas d'une théorie à descriptions *primordiales transférées* comme la mécanique quantique, est celui obtenu par la comparaison directe des lois de probabilité élaborées séparément dans les différents référentiels propres. Chacun des observateurs, dans son référentiel propre, n'opère que sur de l'*inobservable*, d'une manière qui, par construction, est strictement analogue à celle de laquelle l'autre observateur opère dans son référentiel propre. Or selon le principe de relativité restreinte elle même :

« Le déroulement de phénomènes analogues liés de la même façon à divers systèmes de référence, ne dépend pas du mouvement rectiligne et uniforme du système de référence »[62].

En ce qui concerne le type de consensus exigible, la situation est du même genre que celle qui se réalise dans l'entière physique classique : l'assertion d'un résultat expérimental donné doit être *vérifiable* dans chaque référentiel propre.

### *5.2.2. La microphysique actuelle et les relativités d'Einstein*

Mais alors, peut-on se dire, pourquoi dans le calcul célèbre de la thèse de Louis de Broglie, l'application des lois de transformation de Lorentz-Einstein a conduit à la relation fondatrice $p=h/\lambda$ ? Pourquoi dans le calcul des durées de vie de microsystèmes radioactifs les transformations de Lorentz-Einstein sont pertinentes ? Pourquoi l'équation de Dirac a-t-elle conduit à une découverte nouvelle ?

---

[62] Marie-Antoinette Tonnelat [1971], p. 152.

Ces questions, sans doute, méritent beaucoup de réflexion. Je n'essaierai pas de les examiner ici car cela dépasserait trop le but de ce livre. Néanmoins j'avance les remarques suivantes.

La thèse de Louis de Broglie a introduit un *MODELE* de microétat, un modèle *individuel* et particulier, celui d'un microétat d'*électron* libre. Ce n'est pas un modèle fantôme consistant dans une écriture mathématique de prolongement d'une qualification mécanique classique, mais en plein un modèle de microétat. Par cela *le traitement de Louis de Broglie s'est placée d'emblée sur le niveau de la conceptualisation modélisante de type classique*. (Cela explique sans doute l'impossibilité qui s'est manifestée par la suite, d'inclure ce traitement, tel quel, dans la construction subséquente de la mécanique quantique, qui, elle, s'est placée sur le plan de conceptualisation primordiale, transférée). Or on peut très bien concevoir que, pour des *modélisations* de microétats libres d'électrons *dotés de charge électrique*, qui donc interagissent *via* des champs électriques beaucoup plus fortement que *via* des champs gravitationnels, l'emploi des transformations Lorentz-Einstein pour les coordonnées d'espace et de temps se soit montré pertinent. En effet, même si les observateurs humains et leurs signaux lumineux sont absents au niveau microscopique, les modèles d'états d'électrons prennent le relais en quelque sorte, car ils interagissent *via* des champs qui se propagent avec la vitesse de la lumière.

Chaque exemple de clair succès d'un traitement relativiste appliqué à des microétats, devrait être examinée selon des lignes similaires, i.e. en explicitant le type d'interactions impliqué.

En tout cas, une théorie qui, comme la mécanique quantique, se place toute entière sur le niveau de conceptualisation primordiale transférée, ne peut *pas* incorporer une géométrie fondée sur des interactions entre des *modèles* qui sont élaborés sur la base de descriptions primordiales transférées, dans une phase de conceptualisation qui est subséquente dans l'ordre génétique des modes de conceptualisation. On ne peut pas réaliser une cohérence logique-formelle en mélangeant aveuglément des niveaux de conceptualisation distincts, comme on ne peut pas avoir une équation correcte si les deux membres ne sont pas identiques du point de vue des dimensions sémantiques qui interviennent. La résistance insurmontable de la mécanique quantique, à une 'unification' globale avec la relativité restreinte, est une manifestation de rejet conceptuel en conséquence de l'incompatibilité des niveaux de conceptualisation où se placent ces deux théories. Une unification ne pourrait être envisagée qu'entre une *modélisation* achevée de la mécanique quantique fondamentale, et la relativité restreinte. Et encore, dans ces conditions, il n'y aurait *lieu* d'envisager une telle unification que si *(a)* les interactions entre microétats modélisés *justifient* toutes, uniformément, l'emploi des transformations de Lorentz-Einstein dans la représentation des façons dont les microéts modélisés 'ressentent' les 'influences' qui leur parviennent *via* des champs, en fonction de leur état de mouvement mécanique ; *(b)* si l'on peut justifier *jusqu'au bout* une métaphore en termes microphysiques, des conditions d'observation qui sous-tendent la relativité restreinte (observateurs inertiels, but d'établir une loi de mouvement invariante aux changements de l'état de mouvement mécanique). Ce qui est loin de paraître évident.

Quant à une unification de la mécanique quantique *fondamentale*, à descriptions primordiales transférées, avec la relativité générale, on perçoit immédiatement toutes les réticences qui émanent du point de vue qui vient d'être esquissé, tout autant quant à la possibilité d'un tel concept, qu'en ce qui concerne sa valeur en tant qu'un but.

La mécanique quantique fondamentale est l'*unique* théorie mathématique transférée, primordialement probabiliste, construite à ce jour. Cette théorie a réussi à incorporer dans ses algorithmes les *RACINES de la connaissance* humaine et à les dominer formellement. C'est pour cette raison qu'elle apparaît à tous comme une révolution majeure dans la pensée scientifique. Et c'est parce que la percée cognitive impliquée dans la mécanique quantique s'est réalisée directement en termes mathématiques, sans que l'on ait pu discerner d'une manière *intégrée* les contenus épistémologiques et méthodologiques si radicalement nouveaux captés dans les algorithmes mathématiques, que cette théorie apparaît comme cryptique : *la méthodologie de prise en compte explicite des contraintes qui émanent de la situation cognitive à supposer et du but de chaque étape descriptionnelle, manque*. C'est un manque à la base qui empêche de doter les modélisations microphysiques, de fondations élaborées avec précision. Nous avons d'abord, presque miraculeusement, capté les racines de la conceptualisation dans la mécanique quantique fondamentale, et ensuite la liaison avec ces racines s'est brisée et nous sommes remontés vers le ciel conceptuel des modèles classiques, comme accrochés à des cerfs volants dont la ficelle a été coupée.

Le modèle standard a sans doute réussi dans une large mesure à compenser cette brisure d'une véritable continuité avec la mécanique fondamentale. Il serait intéressant d'analyser les causes, les modalités et le degré de cette réussite.

Mais la théorie des cordes s'est énormément éloignée de la mécanique quantique fondamentale. On n'y perçoit plus aucune connexion conceptuelle avec cette théorie, seulement quelques signes mathématiques qui la rappellent. On y perçoit par contre une véritable fascination pour les démarches relativistes qui se sont forgées dans le domaine de la physique macroscopique, sous l'empire de situations observationnelles et de buts descriptionnels définis dans la physique macroscopique. La transposition acritique de ces démarches afin de représenter des entités et des interactions microscopiques, semble vouée à rester non pertinente. Quelle chance a-t-on de tomber en accord avec les manières de se manifester à nous, de ces entités et interactions, lorsqu'on recherche des invariants formels sans se soucier à chaque pas de savoir pour quel but on recherche telle ou telle invariance et en cohérence avec *quelle* méthode de transposition à une représentation microscopique, de *quels* traits de *quelle* situation cognitive? Or en physique mathématique la non pertinence épistémologique et méthodologique se manifeste comme un rejet conceptuel d'une greffe d'un organe abstrait incompatible avec l'organisme du savoir constitué précédemment.

Lorsqu'on a *vu*, comme au cours de l'humble cheminement développé ici, à quel point chaque pas sur un bref trajet de conceptualisation épistémologique-méthodologique d'un domaine du réel physique, dépend des contraintes qui émanent de la situation cognitive impliquée, du but de la représentation recherchée, des *décisions* méthodologiques qui s'imposent face à *cette* situation cognitive et *ce* but, spécifiquement, mais des décisions qu'on ne perçoit que successivement lorsqu'on se trouve nez à nez avec la nécessité de l'une d'elles lors d'une impasse dans la progression, ce n'est qu'alors qu'on réalise vraiment l'importance des aspects épistémologiques et des aspects de *méthode*.

*La microphysique actuelle souffre d'une crise d'absence d'une méthode épistémologique. Elle souffre plus encore de l'ignorance de l'existence de cette crise*.

L'entière physique moderne est viscéralement travaillée par la nécessité d'engendrer à partir d'elle-même une méthode épistémologique élaborée en concordance avec son propre niveau de détail, de rigueur et de synthèse ; une méthode qui, dans chaque phase descriptionnelle d'un développement représentationnel, puisse dicter explicitement les règles à respecter afin d'achever cette phase là d'une manière consistante avec tous les présupposés qui y interviennent.

Il ne me semble pas exclu qu'une 'unification' de la physique actuelle ne puisse s'accomplir que dans un sens très différent de celui qu'on poursuit depuis des dizaines d'années. À savoir, par *une méthode unitaire de construction des représentations quelconques*. Une méthode qui, lors de *chaque* action descriptionnelle, distingue clairement tous les éléments auxquels cette action est relative, les éléments qui la conditionnent localement, et qui sépare nettement les uns des autres les actes successifs de conceptualisation, ainsi que, globalement, les niveaux de conceptualisation, selon une hiérarchie dictée par une certaine genèse naturelle des processus de conceptualisation.

C'est dans la méthode de décrire que pourrait s'accomplir une unification. Dans le cadre d'une méthode descriptionnelle générale, chaque description pourrait conserver pleinement les spécificités de son contenu, tout en se trouvant en relation explicite avec les autres descriptions.

## CONCLUSION GENERALE

Nous avons entrepris de spécifier en détail les contenus épistémologiques impliqués dans la mécanique quantique. Le résultat peut surprendre. Tout à fait indépendamment du formalisme de la mécanique quantique, il s'est constitué une discipline qualitative – *une discipline épistémo-physique* – l'infra-mécanique quantique, dont le contenu consiste en l'organisation du substrat d'opérations cognitives et de significations encryptées dans le formalisme quantique. Dans l'infra-mécanique quantique cette organisation sous-jacente se montre débarrassée de tout élément mathématique à proprement parler (les probabilités ne sont que dénombrements et logique), concentrée en elle-même et dotée d'un contour propre.

*Sa genèse, explicitée pas à pas, marque d'un caractère de nécessité cette discipline épistémo-physique indépendante*.

Le cœur de l'infra-méacanique quantique est un type descriptionnel 'transféré' sur des enregistreurs d'appareil, primordialement probabiliste, foncièrement différent du type descriptionnel classique et qui, auparavant, n'avait jamais encore atteint le niveau d'une connaissance exprimée et intégrée. Désormais – *pour le cas spécial des microétats* – ce type descriptionnel primordial est défini avec rigueur et détail et il élucide la nature de la fameuse 'coupure quantique-classique'.

Le brouillard lourd, comme solide, qui cachait la manière de signifier de la mécanique quantique, est entièrement dissipé.

———

## BIBLIOGRAPHIE

J.S. Bell [1964], "On the Einstein Podolsky Rosen Paradox", *Physics* p.195.
J.S. Bell [1966], "On the Problem of Hidden Variables in Quantum Mmechanics", *Rev. of Modern Physics, 38, 447-52*.

A. Berthoz [1997], *Le Sens du Mouvement* (English translation by Odile Jacob : Harvard University Press, 1997).

A. Berthoz et J-L. Petit [2006], *Physiologie de l'action et phenomenologie*, Odile Jacob.

M. Bitbol [1996], *Mécanique quantique: une introduction philosophique*, Flammarion.

M. Bitbol [2000], *Physique & philosophie de l'esprit*, Flammarion.

D. Bohm [1952], "À Suggested Interpretation of the Quantum Theory in Terms of "Hiden" Variables, I and II", *Phys. Rev. 85, 166-193*.

L. de Broglie [1924] *Recherches sur la théorie des quanta*, Thèse, Fac. Des Sciences de Paris.

L. de Broglie [1956], *Une tentative d'interprétation causale et non-linéaire de la mécanique ondulatoire (la théorie de la double solution)*, Gauthier-Villars.

A. Einstein [1949], in *Albert Einstein, Philosopher and Scientist,* P. A. Schlipp ed., p. 86, Library of Living Philosophers.

A. Einstein [1955*]*, "The Space and the Time in Pre-relativistic Physics", in *The Meaning of Relativity,* Princeton University Press.

Gudder [1976], "A generalized measure and probability theory", in *Foundations of Probability Theory, Statistical Inference and Statistical Theories of Science*, Reidel, Dordrecht.

P. Holland [1993], *The Quantum Theory of Motion* (réédité en 1995, 1997, 2000, 2004), Cambridge Univ. Press.

G. Longo, [2002], "Laplace, Turing et la géométrie impossible du «jeu d'imitation »", *Intellectica* 2002/2.

Mackey [1963], *Mathematical Foundations of Quantum Mechanics*, Benjamin.

M. Mugur-Schächter [1964], *Etude du caractère complet de la mécanique quantique*, Gauthier Villars.

M. Mugur-Schächter [1977] "The quantum mechanical one-system formalism, joint probabilities and locality", in *Quantum Mechanics a half Century Later*, J. L. Lopes and M.. Paty, eds., Reidel.

M. Mugur-Schächter [1979], "Study of Wigner's Theorem on Joint Probabilities", *Found. Phys.*, Vol. 9.

M. Mugur-Schächter [1984], "Esquisse d'une représentation générale et formalisée des descriptions et le statut descriptionnel de la mécanique quantique", *Epistemological Letters*, Lausanne, cahier 36.

M. Mugur-Schächter [1991], "Spacetime Quantum Probabilities…… Part I… : Relativized Descriptions and Popperian Propensities", *Founds. of Phys*., Vol. 21.

M. Mugur-Schächter [1992], "Spacetime Quantum Probabilities II : Relativized Descriptions and Popperian Propensities", *Founds. of Phys*., Vol. 22.

M. Mugur-Schächter [1995], Mugur-Schächter M., (1995), "Une méthode de conceptualisation relativisée... ", *Revue Int. de Systémique*, Vol. 9.

M. Mugur-Schächter [2002A], "Objectivity and Descriptional Relativities", *Fonds. of Science*.

M. Mugur-Schächter [2002B], "From Quantum Mechanics to a Method of Relativized Conceptualization", in *Quantum Mechanics, Mathematics, Cognition and Action*, Mugur-Schächter M. and Van Der Merwe A., Eds., Kluwer Academic.

M. Mugur-Schächter [2002C], "En marge de l'article de Giuseppe Longo sur Laplace, Turing et la géométrie impossible du "jeu d'imitation »", *Intellectica* 2002/2.

M. Mugur-Schächter [2006], *Sur le tissage des connaissances*, Hermès-Lavoisier.

M. Mugur-Schächter [2008], "Quantum Mechanics Freed of Interprétation Problems", *en cours d'élaboration*.

Orwell [1945] (en anglais, Animal Farm (éditeur ?), trad. Française La ferme des animaux, Champ Libre ? (ou Gallimard ? : pas clair)

H. Poincaré [1898] "On the Foundations of Geometry", in *The Monist, V, 9*

H. Poincaré [1965], in *Science et hypothèse*, Flammarion.

M. Schlosshauer [2003], "Decoherence, the Measurement Problem, and Interpretations of Quantum Mechanics", *arXiv:quant-ph/0312059* vl, 6 Déc. 2003 (proposé pour publication dans Rev. of Mod. Phys.).

J. von Neumann [1955], *Mathematical Foundations of Quantum Mechanics*, Princeton University Press.

P. Teilhard de Chardin [ 1956 ], *La place de l'homme dans la nature*, Albin Michel.

Ma-A. Tonnelat, *Histoire du principe de relativité*, Flammarion.

E.P. Wigner, E.P. [1971] in *Perspectives in Quantum Theory*, W. Yourgrau and A. van der Merwe eds., MIT Press.

─────────